\documentclass[12pt,a4paper]{article}
\usepackage{amssymb,amsmath,amsthm,euscript}
\usepackage{mathabx}
\usepackage{mathtools}
\usepackage[matrix,arrow,curve]{xy}

\usepackage[utf8]{inputenc}
\usepackage[english]{babel}

\usepackage[breaklinks]{hyperref}
\usepackage[hyphenbreaks]{breakurl}

\newcommand{\C}{\mathcal C}
\renewcommand{\P}{\mathcal P}
\newcommand{\D}{\mathcal D}
\newcommand{\ZZ}{\mathbb Z}

\newcommand{\CC}{\mathbb C}

\def\llambda{{\boldsymbol{\lambda}}}
\def\aalpha{{\boldsymbol{\alpha}}}

\newcommand{\Ad}{\mathop{\mathrm{Ad}}\nolimits}
\newcommand{\Hom}{\mathop{\mathrm{Hom}}\nolimits}

\newcommand{\Mod}{\mathrm{\text{\normalfont-Mod}}}
\newcommand{\End}{\mathop{\mathrm{End}}\nolimits}
\newcommand{\Vect}{\mathop{\mathrm{Vect}}\nolimits}
\newcommand{\Mat}{\mathop{\mathrm{Mat}}\nolimits}
\newcommand{\Rep}{\mathop{\mathrm{Rep}}\nolimits}
\newcommand{\Spec}{\mathrm{Spec}}
\newcommand{\id}{\mathop{\mathrm{id}}\nolimits}
\newcommand{\tr}{\mathop{\mathrm{tr}}\nolimits}
\newcommand{\diag}{\mathop{\mathrm{diag}}\nolimits}
\newcommand{\Supp}{\mathop{\mathrm{Supp}}\nolimits}
\renewcommand{\d}{{\mathrm{d}}}
\newcommand{\cl}{{\mathrm{cl}}}
\newcommand{\GL}{{\mathrm{GL}}}

\newcommand{\Ilf}{I_{lf}}
\newcommand{\Iinflf}{(I_\infty)_{lf}}
\renewcommand{\le}{\leqslant}
\renewcommand{\ge}{\geqslant}
\newcommand{\Ker}{\mathop{\mathrm{Ker}}\nolimits}
\renewcommand{\Im}{\mathop{\mathrm{Im}}\nolimits}

\newcommand{\wt}{\widetilde}

\newtheorem{Th}{Theorem}[section]
\newtheorem{Lem}[Th]{Lemma}
\newtheorem{Prop}[Th]{Proposition}
\newtheorem{Cor}{Corollary}[Th]
\newtheorem{Conj}{Conjecture}[section]
\theoremstyle{definition}
\newtheorem{primer}{Example}[section]
\newtheorem{opr}{Definition}[section]
\newtheorem{Rem}{Remark}[section]

\numberwithin{equation}{section}

\title{Application of the reflection functor to integrable systems on quiver varieties}
\author{Alexey Silantyev\thanks{aleksejsilantjev@gmail.com}}
\date{}

\begin{document}

\maketitle

\vspace{-5mm}
\begin{center}
{\it Joint Institute for Nuclear Research, 141980 Dubna, Moscow region, Russia} \\
{\it State University "Dubna"{}, 141980 Dubna, Moscow region, Russia } \\
\end{center}
\vspace{5mm}

\begin{abstract}
This article reviews the results of~\cite{ChS} and~\cite{S}, also the theorem on completeness of flows is proved here. More precisely, this review concerns the following subjects. The theory of representations of quivers and of their preprojective algebras are described, this includes moduli spaces of representations of these algebras, quiver varieties and reflection functor. The proof that the bijection between moduli spaces induced by the reflection functor is an isomorphism of symplectic affine varieties is presented. Hamiltonian systems on the quiver varieties and an application of the reflection functor to these systems are descried. The results on the case of the cyclic quiver are reviewed and a role of the reflection functor in this case is discussed. In particular, "spin"{} integrable generalisations of the Calogero--Moser systems and their application to the generalised KP hierarchies are considered.
\end{abstract}

\tableofcontents

\section{Introduction}

The reflection functors appeared in the frame of the theory of quiver representations in the work~\cite{BGP}, which is in turn based on ideas of~\cite{GP}. When Crawley-Boevey and Holland defined (deformed) preprojective algebra for a quiver, they also introduced reflection functors on the category of its representations~\cite{CBH}. Here we describe an application of these functors to integrable systems and to hierarchies of differential equations.

In~\cite{Nakajima} Nakajima introduced quiver varieties in terms of quiver representations. A quiver variety can be represented as a moduli space of representations of a preprojective algebra~\cite{CB01}. The preprojective algebra that defines a quiver variety is constructed by a framed quiver. The framing is the addition of one vertex and some number of edges from this new vertex to the old ones.

Some quiver varieties coincide with completions of phase spaces for integrable systems. These systems are rational Calogero--Moser systems and their generalisations. The (symmetrized) phase space for the Calogero--Moser system of type $A_{n-1}$, where $n$ is the number of particles, was first completed by Wilson~\cite{W}. This completion is called Calogero--Moser space. To present it as a quiver variety one needs to consider a one-loop quiver. More generally, the completed phase space of the Calogero--Moser system associated with the generalised symmetric group $S_n\ltimes(\ZZ/m\ZZ)^n$ is isomorphic to the quiver variety for the cyclic quiver consisting of $m$ vertex and $m$ edges oriented in the same direction~\cite{EG} (the cases $m=1$ and $m=2$ correspond to the Calogero--Moser systems of types $A_{n-1}$ and $B_n$ respectively). In this case the framing is done by one new edge and the representations, which the quiver variety consists of, have the fixed dimension defined by the number $n$ in a certain way.

A moduli space of representations of a preprojective algebra for a general quiver is not always a manifold. In order to make it at least an affine variety (possibly irreducible), one should include only semi-simple modules to the moduli space. The conditions guaranteeing the smoothness of the moduli spaces were found in~\cite{CB01}. The moduli spaces have symplectic structure, which can be obtained from the construction of Hamiltonian reduction. Therefore, they are candidates for construction of integrable Hamiltonian systems on them. A Hamiltonian system with complete flows is defined here on an arbitrary quiver variety.

The reflection functor defines an isomorphism between different moduli spaces and, in particular, between quiver varieties. It was proved in~\cite{N03} that in the case of quiver varieties it preserves their algebraic and symplectic structures. In~\cite{S} we proved this statement for general moduli spaces and we showed that this functor relates the Hamiltonian systems on them. In particular, this helps to generalise the results on integrability. This is demonstrated in the case of the cyclic quiver, for which the integrability of the Hamiltonian systems on some quiver varieties was proved in~\cite{ChS}.
The integrable systems found there can be interpreted as "spin"{} versions of the Calogero--Moser systems for $S_n\ltimes(\ZZ/m\ZZ)^n$. The spin variables arise when the number of the additional edges in the framed quiver is grater than one. In particular, the quiver variety for the one-loop quiver with a general framing is a completion of the phase space of the Gibbons--Hermsen system defined in~\cite{GH}.

This construction can be generalised by using a multiplicative version of the preprojective algebra for the cyclic quiver~\cite{ChF}. It leads to generalisations of the Ruijsenaars--Schneider systems.

The Hamiltonian systems on the quiver varieties are applied to construct solutions of some integrable PDEs and of their hierarchies. This method originates from the work~\cite{AMM}, where solutions of the Korteweg--de Vries (KdV) equation were constructed as rational functions with poles moving as particles of the Calogero--Moser systems of type $A_{n-1}$. In the papers~\cite{ChCh} and \cite{Kr} the same was done for the Kadomtsev--Petviashvili (KP) equation. Later the solutions of the KP equation were extended to solutions of the whole KP hierarchy~\cite{W}. In this work Wilson proved (by means of adelic Grassmannians) that the rational solutions of the KP hierarchy is in one-to-one correspondence with the points of the Calogero--Moser spaces (which are the quiver varieties for the one-loop quiver). In~\cite{ChS} generalisations of the KP hierarchy and of its matrix version were proposed. Rational solutions of these generalisations were obtained by using the cyclic quivers and Hamiltonian flows on the corresponding quiver varieties.

The present work is based on~\cite{S}. Its structure is as follows.

In Section~\ref{sec2} basic notions and constructions are introduced. The notions of quiver and of its representation are defined in Section~\ref{sec21}. Section~\ref{secRoots} is devoted to roots, to their reflections and to a role of the roots in the theory of quiver representations. In Section~\ref{secPi} we introduce double quiver, preprojective algebra and consider a moduli space of representations of the preprojective algebra. In Section~\ref{secHamRed} a general construction of Hamiltonian reduction and its application to the construction of the moduli spaces are described.

Section~\ref{secRF} is devoted to the reflection functor and its application to the moduli spaces. In Section~\ref{secRFEquiv} we define the reflection functor and prove that it is an equivalence of categories. Section~\ref{secRFIsom} is devoted to the induced isomorphism between moduli spaces. It is proved here that it preserves algebraic and symplectic structures.

In Section~\ref{secKolMnIntSys} quiver varieties and Hamiltonian systems on them are considered for a general quiver. In Section~\ref{secKolMn} the quiver varieties are defined as a particular case of the moduli spaces. Section~\ref{secHamKolMn} is devoted to a construction of some special Hamiltonian systems on the quiver varieties and, more generally, on the moduli spaces (for a framed quiver). In Section~\ref{secKolMnRF} the reflection functor is applied to these Hamiltonian systems.

Section~\ref{secCykl} is devoted to the Hamiltonian systems on the quiver varieties for the cyclic quiver (with $m$ vertices). It begins with the general case in Section~\ref{secHamSysCycl} and continues with particular cases in other sections. In Section~\ref{secCM} the case $m=1$ with the simplest framing is described. In Section~\ref{secGH} the consideration is generalised to the general framing. In Section~\ref{secde0} we consider the case of the general $m$ with some simple framing. Section~\ref{secddelta} is devoted to the case of a more complicated framing.

In Section~\ref{secApplKP} the results obtained in the previous section are applied to the KP hierarchy and its generalisations. In Section~\ref{secKP} we recall the definition of the KP hierarchy and explain how the Calogero--Moser spaces are applied to describe the rational solutions of this hierarchy. Section~\ref{secGKP} is devoted to generalisation of the KP hierarchy by means of the Cherednik algebra for the cyclic group of order $m$. This allows us to define two generalisations of the KP hierarchy. Their rational solutions are constructed by means of the quiver varieties for the cyclic quiver. In Section~\ref{secGMKP} this generalises to the case of the matrix KP hierarchy.

Finally, a list of open questions on this subject is given.

Additionally, two appendices are written. In Appendix~\ref{appPoisson} a method of calculation of the Poisson brackets in matrix form is described. It is applied to derivation of some formulae used in the main text. Appendix~\ref{appRRep} is devoted to an alternative proof of the fact that the reflection functor preserves the algebraic structure of the moduli spaces.
 
{\it Acknowledgements}. 
The author is grateful to Yury~Berest and Oleg~Chalykh for explanation of some results and useful discussions.
This paper is partially supported by EPSRC under grant EP/K004999/1.

\section{Preliminaries}
\label{sec2}

Here we introduce basic notions and formulate main theorems of the theory of representations of quivers and of preprojective algebras associated with quivers. In particular, we define root systems and describe their relationship with the representations. Finally, we describe moduli spaces of representations of preprojective algebra in terms of Hamiltonian reduction.

\subsection{Quivers, path algebras and their representations}
\label{sec21}

{\it Quiver} $Q$ is an oriented graph. Formally, this is a 4-tuple $(I,E,t,h)$ consisting of two sets $I$ and $E$, which we suppose to be finite, and two maps $t\colon E\to I$ and $h\colon E\to I$. The set $I$ is interpreted as the set of vertices of the graph, $E$ --- as the set of edges. The values $t(a)$ and $h(a)$ are interpreted as tail and head of an edge $a\in E$ respectively. We write $a\colon i\to j$ or $i\xrightarrow{a}j$ when  $t(a)=i$ and $h(a)=j$, and say that $a\in E$ is an edge from the vertex $i\in I$ to the vertex $j\in I$. Since this notations make unnecessary the explicit usage of the maps $t$ and $h$, we usually write $Q=(I,E)$, assuming that the unmarked maps "tail"{} and "head"{} are given. We also write $a\in Q$ instead of $a\in E$.

Let us fix the field of complex numbers $\CC$ for the basic field. All the vector spaces and algebras will be over $\CC$.

{\it Representation of the quiver} $Q=(I,E)$ is a family of vector spaces $V_i$, $i\in I$, with a family of linear maps $V_a\colon V_i\to V_j$ given for each edge $a\colon i\to j$.

Denote by $\CC I$ the finite-dimensional vector space $\bigoplus\limits_{i\in I}\CC1_i$ equipped with the multiplication
\begin{align} 1_i\cdot1_j=\delta_{ij}1_i. \label{1i1j}
\end{align}
Thus, $\CC I$ is a commutative semi-simple algebra with the unity $1=\sum\limits_{i\in I}1_i$.

Any $\CC I$-module $V$ is decomposed into the direct sum: $V=\bigoplus\limits_{i\in I}V_i$, where $V_i=1_iV$. On vectors $v=(v_i)_{i\in I}\in V$, $v_i\in V_i$, the algebra $\CC I$ acts as $1_jv=(\delta_{ij}v_i)_{i\in I}$. Conversely, any family of linear spaces $(V_i)_{i\in I}$ gives a representation of the algebra $\CC I$ on the space $V=\bigoplus\limits_{i\in I}V_i$ with the action described above. In particular, a representation of the quiver $Q=(I,E)$ gives a representation of $\CC I$.

If $V$ is finite-dimensional, then all $V_i=1_iV$ are finite-dimensional. In this case the vector $\alpha=(\alpha_i)_{i\in I}\in\ZZ_{\ge0}^I$ with components $\alpha_i=\dim_\CC V_i$ is called {\it dimension} of the $\CC I$-module $V$ and it is denoted by $\dim_{\CC I}V$. Note that $\dim_\CC V=|\alpha|$, where $|\alpha|:=\sum\limits_{i\in I}\alpha_i$. There exists a one-to-one correspondence between finite-dimensional $\CC I$-modules (up to an isomorphism) and vectors $\alpha\in\ZZ_{\ge0}^I$. We assign to each vector $\alpha$ with components $\alpha_ i\in\ZZ_{\ge0}$ the $\CC I$-module $V=\bigoplus\limits_{i\in I}V_i$, corresponding to the family $V_i=\CC^{\alpha_i}$; any $\CC I$-module of the dimension $\alpha$ is isomorphic to $V$.

A representation of the quiver $Q=(I,E)$ that has dimension $\alpha$ as a $\CC I$-module is called a representation of the quiver $Q$ of dimension $\alpha$. The set of all representations of $Q$ on $\CC I$-module $V=\bigoplus\limits_{i\in I}V_i$ with $V_i=\CC^{\alpha_i}$, which respect the structure of $\CC I$-module, is denoted by $\Rep(Q,\alpha)$. To give such a representation one can take an arbitrary family of linear operators from $\CC^{\alpha_i}$ to $\CC^{\alpha_j}$ for each edge $a\colon i\to j$. That is $\Rep(Q,\alpha)=\prod\limits_{a\colon i\to j}\Hom(\CC^{\alpha_i},\CC^{\alpha_j})$ (product over all $a\in Q$). Note that the set $\Rep(Q,\alpha)$ has a structure of a vector space and, in particular, of an affine variety.

Consider the vector space $\CC E=\bigoplus\limits_{a\in E}\CC a$. It has a natural structure of $\CC I$-bimodule:
\begin{align}
 &1_k\cdot a=\delta_{jk}a, && a\cdot 1_k =\delta_{ik}a &&\text{ for $a\colon i\to j$.} \label{1ia}
\end{align}

Define {\it path algebra} for the quiver $Q=(I,E)$ as $\CC Q=T_{\CC I}\CC E$, where $T_AM$ is the tensor algebra of the bimodule $M$ over the algebra $A$. It has the following $\ZZ$-grading:
\begin{align*}
 &\CC Q=\bigoplus_{\ell=0}^\infty (\CC Q)_\ell, \qquad\quad (\CC Q)_0=\CC I, \qquad\quad (\CC Q)_1=\CC E, \\
 &(\CC Q)_2=\CC E\otimes_{\CC I}\CC E, \qquad (\CC Q)_3=\CC E\otimes_{\CC I}\CC E\otimes_{\CC I}\CC E, \qquad \ldots
\end{align*}
If $\ell\ge1$ then the vector space $(\CC Q)_\ell$ is freely generated by the elements of the form $a_\ell\cdots a_2 a_1$, where $a_k\colon i_{k-1}\to i_k$ are edges going successively between the vertices $i_0,i_1,\ldots,i_\ell\in I$. Such elements are called {\it paths} of the length $\ell$ from the vertex $i_0$ to the vertex $i_\ell$. In particular, the paths of the length one are edges $a\in Q$. The elements $1_i$ generating the subalgebra $\CC I\subset\CC Q$ are called {\it trivial paths}. The trivial path $1_i$ is the path of the length zero from $i$ to $i$. The path algebra $\CC Q$ can be defined as an algebra generated by $1_i$ and $a\in Q$ with relations~\eqref{1i1j}, \eqref{1ia}. Note also that these relations imply
\begin{align*}
 &ba=0 &&\text{whenever $a\colon i\to j$,\quad $b\colon i'\to j'$,\qquad $j\ne i'$}.
\end{align*}

Let $(V_i,V_a)$ be a representation of the quiver $Q$. Define an action of $\CC Q$ on $V=\bigoplus\limits_{i\in I}V_i$. Let $1_i$ act as above. Then, for $a\colon i\to j$ and $v=(v_l)_{l\in I}$ define $av=V_av_i$, where $V_av_i$ is understood as an element of $V$ via the embedding $V_j\hookrightarrow V$. This is in agreement with the relations~\eqref{1ia} and hence we obtain a structure of $\CC Q$-module on $V$. Conversely, any $\CC Q$-module $V$ is a $\CC I$-module and the edges $a\colon i\to j$ act on $v=(v_l)_{l\in I}$ as $av=V_av_i$, for some linear maps $V_a\colon V_i\to V_j$. This means that the representations of the path algebra ($\CC Q$-modules) are in one-to-one correspondence with the representations of quiver $Q$. Moreover, there is an identification $\Rep(\CC Q,\alpha)=\Rep(Q,\alpha)$, where $\Rep(\CC Q,\alpha)$ is a set of $\CC$-algebra homomorphisms (representations) $\rho\colon\CC Q\to\End V$ on $V=\bigoplus_{i\in I}V_i$, $V_i=\CC^{\alpha_i}$, which respect the $\CC I$-module structure: $\rho(1_i)=1_i$ $\;\forall\, i\in I$. The variety of all $n$-dimensional representations $\rho\colon\CC Q\to\End(\CC^n)$ is decomposed into disconnected union: $\Rep(\CC Q,n)=\bigsqcup\limits_{|\alpha|=n}\Rep(\CC Q,\alpha)$.

Let $\GL(\alpha):=\prod\limits_{i\in I}\GL(\alpha_i,\CC)$. The action of an element $g=(g_i)_{i\in I}\in\GL(\alpha)$ on a vector $v=(v_i)_{i\in I}\in V=\bigoplus\limits_{i\in I}V_i$ is defined by the action of each component $g_i\in\GL(\alpha_i,\CC)$ on $V_i=\CC^{\alpha_i}$, that is $g.v=(g_i.v_i)_{i\in I}$. This induces an action of the group $\GL(\alpha)$ on $\Rep(Q,\alpha)$:
\begin{align}
 &V_a\mapsto g_jV_a g_i^{-1}, &&a\colon i\to j. \label{GLaction}
\end{align}

Let $V=(V_i,V_a)$ and $\wt V=(\wt V_i,\wt V_a)$ are representations of the quiver $Q$. Their morphism is a morphism of $\CC Q$-modules $\phi\colon V\to\wt V$. It has the form $\phi=(\phi_i)_{i\in I}:=\bigoplus\limits_{i\in I}\phi_i$, where $\phi_i\colon V_i\to\wt V_i$ are linear maps such that $\phi_jV_a=\wt V_a\phi_i$ for each $a\colon i\to j$.
Any element $g\in\GL(\alpha)$ maps a representation $V=(V_i,V_a)$ to an isomorphic representation: their isomorphism is given by the maps $\phi_i=g_i\colon\CC^{\alpha_i}\to\CC^{\alpha_i}$. Moreover, the representation $V,\wt V\in\Rep(Q,\alpha)$ are isomorphic if and only if $\wt V=g.V$ for some $g\in\GL(\alpha)$.

For an affine variety $M$ denote by $\CC[M]$ the algebra of regular functions $f\colon M\to\CC$. If a group $G$ acts on $M$, we define the subalgebra of invariant functions:
\begin{align*}
 \CC[M]^G:=\{f\in\CC[M]\mid f(gx)=f(x)\;\forall\, g\in G,\,x\in M\}.
\end{align*}

An element $p\in\CC Q$ is called a {\it cycle}, if it is a path from a vertex $i\in I$ to the same vertex. For each cycle $p=a_\ell\cdots a_2 a_11_{i}$ we define the regular function $\tr_\alpha(p)\in\CC\big[\Rep(Q,\alpha)\big]$, which maps a representation $V=(V_i,V_a)\in\Rep(Q,\alpha)$ to the number $\tr(V_{a_k}\cdots V_{a_2}V_{a_1}\id_{\CC^{\alpha_i}})\in\CC$, where the trace is taken over the space $V_i=\CC^{\alpha_i}$ (for trivial paths one obtains $\tr_\alpha(1_i)=\alpha_i$). The formula~\eqref{GLaction} implies that $\tr_\alpha(p)\in\CC\big[\Rep(Q,\alpha)\big]^{\GL(\alpha)}$.

The structure of the algebras of invariant functions on the space of representations is given by the following theorem~\cite{LeBPro}.

\begin{Th} \label{ThLeBP} {\normalfont (Le Bruyn--Procesi)}
The algebra $\CC\big[\Rep(Q,\alpha)\big]^{\GL(\alpha)}$ is generated by the functions $\tr_\alpha(p)$.
\end{Th}

\subsection{Root systems} 
\label{secRoots}

The theory of quiver representations is closely related to the notion of root. In order to define the root system we first define reflections by means of a bilinear form depending on a quiver. The roots play an important role in the theory of reflection functor developed in Section~\ref{secRF}.

For a quiver $Q=(I,E)$ define the numbers
\begin{align*}
 &n_{ij}=n_{ji}=|\{a\colon i\to j\}|+|\{a\colon j\to i\}|, &&i,j\in I.
\end{align*}
If $i\ne j$, then $n_{ij}$ is the number of edges between the vertices $i$ and $j$ (in both directions). At the same time, $n_{ii}$ is the double number of loops in the vertex $i$. Denote the subset of loop free vertices by $\Ilf=\{i\in I\mid n_{ii}=0\}$. Equip the lattice $\ZZ^I=\{\alpha=(\alpha_i)_{i\in I}\mid\alpha_i\in\ZZ\}$ with the $\ZZ$-valued symmetric from:
\begin{align} \label{sbf}
 (\alpha,\beta)=2\sum_{i\in I}\alpha_i\beta_i-\hspace{-4pt}\sum_{a\in Q\atop a\colon i\to j}(\alpha_i\beta_j+\alpha_j\beta_i)
\end{align}
(notations of the form $\sum\limits_{a\in Q\atop a\colon i\to j}$ or $\sum\limits_{a\colon i\to j}$ should be understood as $\sum\limits_{i,j\in I}\sum\limits_{a\in Q\atop a\colon i\to j}$).
Let $\varepsilon_i\in\ZZ^I$ are basis vectors: $(\varepsilon_i)_j=\delta_{ij}$. Then
\begin{align} \label{sbfe}
 (\varepsilon_i,\varepsilon_j)= 2\delta_{ij}-n_{ij}.
\end{align}
In particular, one has $(\varepsilon_i,\varepsilon_i)= 2$ for any $i\in\Ilf$ .

{\it Reflection} at a vertex $i\in\Ilf$ is the $\ZZ$-linear map
\begin{align} \label{si}
 &s_i\colon\ZZ^I\to\ZZ^I, &&s_i\alpha=\alpha-(\alpha,\varepsilon_i)\varepsilon_i.
\end{align}
Note that $s_i^2=1$ and $s_i\varepsilon_i=-\varepsilon_i$ for any $i\in\Ilf$. Denote the group generated by the reflections $s_i$, $i\in\Ilf$, by $W$.

If we forget the orientation of a quiver $Q$, we obtain a usual (nonoriented) graph. We call it the {\it graph of the quiver} $Q$. A quiver is said to be {\it connected} if its graph is connected. Any subset of vertices $I'\subset I$ induces the quiver $Q'=(I',E')$ with the set of edges $E'=\{a\in Q\mid a\colon i\to j,\;i,j\in I'\}$. The subset $\Supp(\alpha)=\{i\in I\mid\alpha_i\ne0\}\subset I$ is called the {\it support} of the vector $\alpha\in\ZZ^I$. We say that $\alpha$ has a connected support if the quiver induced by the subset $\Supp(\alpha)$ is connected.

A vector $\alpha\in\ZZ^I$ is called a {\it real root} if it belongs to an orbit $W\varepsilon_i$ for some $i\in\Ilf$. We denote the set of all real roots by $\Delta_{re}(Q)$. Then, let us define the following set called the {\it fundamental domain}:
\begin{align*}
 &F=\{\alpha\in\ZZ_{\ge0}^I\mid\alpha\ne0,\;(\alpha,\varepsilon_i)\le0\;\forall\,i\in I\text{ and $\alpha$ has a connected support}\}
\end{align*}
(the condition $(\alpha,\varepsilon_i)\le0$ for the vertices $i\notin\Ilf$ is satisfied automatically). A vector $\alpha\in\ZZ^I$ is called an {\it imaginary root}, if $\alpha\in WF$ or $-\alpha\in WF$. We denote the set of all imaginary roots by $\Delta_{im}(Q)$. An element $\alpha\in\ZZ^I$ is called a {\it root}, if it belongs to $\Delta(Q):=\Delta_{re}(Q)\cup\Delta_{im}(Q)$. The set $\Delta(Q)$ is called a {\it root system} of the quiver $Q$. 

Let $q(\alpha)=\frac12(\alpha,\alpha)=\sum\limits_{i\in I}\alpha_i^2-\hspace{-4pt}\sum\limits_{a\colon i\to j}\alpha_i\alpha_j$. This is a $\ZZ$-valued function on $\ZZ^I$. Since $(s_i\alpha,s_i\beta)=(\alpha,\beta)$ for any $\alpha,\beta\in\ZZ^I$, $i\in\Ilf$, the function $q(\alpha)$ is $W$-invariant in the sense that $q(w\alpha)=q(\alpha)$ for any $\alpha\in\ZZ^I$, $w\in W$. Note that $(\varepsilon_i,\varepsilon_i)=2$ $\;\forall\,i\in\Ilf$ and that $(\alpha,\alpha)=\sum\limits_{i\in I}\alpha_i(\alpha,\varepsilon_i)\le0$ $\;\forall\,\alpha\in F$. Hence $q(\alpha)=1$ for the real roots $\alpha$ and $q(\alpha)\le0$ for the imaginary roots $\alpha$. Consequently, the sets $\Delta_{re}(Q)$ and $\Delta_{im}(Q)$ do not intersect: $\Delta(Q)=\Delta_{re}(Q)\sqcup\Delta_{im}(Q)$.

Note also that the bilinear form~\eqref{sbf}, the numbers $n_{ij}$, the group $W$ and the root system $\Delta(Q)$ depend on the graph of the quiver $Q$ only, but not on the orientation of its edges. If this graph contains a loop, multiple lines or a cycle (nonoriented), then $\Delta_{im}(Q)$ is not empty. Indeed if there exists a cycle $i_1$--$i_2$--\ldots--$i_\ell$--$i_1$, then $\varepsilon_{i_1}+\varepsilon_{i_2}+\ldots+\varepsilon_{i_\ell}\in F$ (loops and multiple lines are particular cases of nonoriented cycle). But if even there are no loops, multiple lines and cycles, then $\Delta_{im}(Q)$ can be non-empty anyway.%
\footnote{For example, $\Delta_{im}(Q)\ne\emptyset$, if the graph of the quiver $Q$ is an {\it extended} Dynkin graph of type$D$ or $E$, that is a Dynkin graph of an affine Lie algebra of these types (see e.g.~\cite{Kac} or \cite{CBL}).
}
The cases, when all the roots are real, correspond exactly to so-called {\it finite} graphs, whose connected components are Dynkin diagrams for the simple Lie algebras of the types $A$, $D$, $E$. In this case the set $\Delta(Q)=\Delta_{re}(Q)$ is finite and coincides with the root system of the corresponding semi-simple Lie algebra. The group $W$ in this case is the corresponding Weyl group. In other cases the sets $\Delta(Q)$ and $W$ are infinite~\cite{Kac}.%
\footnote{Infiniteness of $\Delta(Q)$ follows from the fact that $n\alpha\in\Delta_{im}(Q)$ for any $n\in\ZZ_{\ge1}$ and $\alpha\in\Delta_{im}(Q)$. If the group $W$ was finite, then it would be a reflection group, but the finite reflection groups are classified (see e.g.~\cite{Hum}) and the conditions $(\varepsilon_i,\varepsilon_i)=2$, $(\varepsilon_i,\varepsilon_j)\in\ZZ$ are satisfied for the types $A$, $D$, $E$ only.
}

A root $\alpha\in\Delta(Q)$ is called {\it positive} ({\it negative}), if $\alpha_i\ge0$ $\;\forall\,i\in I$ ($\alpha_i\le0$ $\;\forall\,i\in I$). The positive and negative roots form the sets $\Delta^+(Q)$ and $\Delta^-(Q)$. As for the finite root systems one has the formula $\Delta(Q)=\Delta^+(Q)\sqcup\Delta^-(Q)$ in general case. It is proved by means of the theory of Lie algebras~\cite{Kac}. We use the notations $\Delta^+_{re}(Q)=\Delta_{re}(Q)\cap\Delta^+(Q)$ and $\Delta^+_{im}(Q)=\Delta_{im}(Q)\cap\Delta^+(Q)$. These sets have the following properties~\cite{Kac}: if $\alpha\in\Delta_{im}^+(Q)$, then $w\alpha\in\Delta_{im}^+(Q)$ $\;\forall\,w\in W$; if $\alpha\in\Delta_{re}^+(Q)$, $i\in\Ilf$ and $\alpha\ne\varepsilon_i$, then $s_i\alpha\in\Delta_{re}^+(Q)$.

The following theorem, which relates the notion of roots with the theory of quiver representations, was proved in~\cite{Kac} (see also~\cite{CBL}, \cite{CB01}).

\begin{Th} \label{ThKac} {\normalfont (Kac)}
\begin{enumerate}
 \item A quiver $Q$ has an indecomposable representation of the dimension $\alpha$ if and only if $\alpha\in\Delta^+(Q)$;
 \item for $\alpha\in\Delta^+_{re}(Q)$ there exists a unique indecomposable representation (up to an isomorphism);
 \item for $\alpha\in\Delta^+_{im}(Q)$ there are infinitely many pairwise non-isomorphic indecomposable representations.
\end{enumerate}
\end{Th}

This theorem implies that there exists a representation of a quiver $Q$ of the dimension $\alpha$, if $\alpha$ is decomposed into a sum of positive roots, and the number of the representations for each decomposition depends on the type of each summand (either it is real or imaginary root).

\subsection{Preprojective algebra and its representations}
\label{secPi}

Suppose we have a quiver $Q=(I,E)$. Now let us double the number of edges: for each edge $a\colon i\to j$ we add a new edge between the same vertices but in opposite direction, we denote it by $a^*\colon j\to i$. In this way we obtain the quiver $\overline Q=(I,E\sqcup E^*)$, where $E^*=\{a^*\mid a\in E\}$, it is called the {\it double} of $Q$ or {\it double quiver}. The edges $a$ and $a^*$ are said to be {\it dual} to each other: if $b=a^*\in\overline Q$ for some $a\in Q$, then we also write $b^*$ for $a$. The space of representations of the double quiver of the dimension $\alpha\in\ZZ_{\ge0}^I$ is
\begin{align*}
 \Rep(\overline Q,\alpha)= \prod_{a\in Q\atop a\colon i\to j}\Big(\Hom(\CC^{\alpha_i},\CC^{\alpha_j})\times\Hom(\CC^{\alpha_j},\CC^{\alpha_i})\Big)=T^*\Rep(Q,\alpha).
\end{align*}
As a cotangent bundle it is equipped with the symplectic form
\begin{align} \label{omega}
 \omega=\sum_{a\in Q} \tr(\d V_{a^*}\wedge\d V_a)=\sum_{a\in Q\atop a\colon i\to j}\sum_{k=1}^{\alpha_i}\sum_{l=1}^{\alpha_j}\d(V_{a^*})_{kl}\wedge\d(V_a)_{lk}.
\end{align}

Let $\lambda=(\lambda_i)\in\CC^I$ be a vector with components $\lambda_i\in\CC$, $i\in I$. Denote by $\Pi^\lambda(Q)$ the quotient algebra of the path algebra $\CC\overline Q$ over the relations
\begin{align}
 &\sum_{a\in Q, j\in I\atop a\colon j\to i}aa ^*-\hspace{-6pt}\sum_{a\in Q, j\in I\atop a\colon i\to j}a^*a=\lambda_i1_i, &&i\in I.
\end{align}
By using the notations
\begin{align}
 &(-1)^a=1, &&(-1)^{a^*}=-1 &&\text{for $a\in Q$} \label{m1aNotation}
\end{align}
one can rewrite these relations in more compact form:
\begin{align}
 &\sum_{a\in\overline Q, j\in I\atop a\colon j\to i}(-1)^a aa^*=\lambda_i1_i, &&i\in I.
\end{align}
The algebra $\Pi^\lambda(Q)$ is called the {\it deformed preprojective algebra} or simply {\it preprojective algebra} ($\lambda$ is a parameter of deformation)~\cite{CBH}.

For a representation of the algebra $\Pi^\lambda(Q)$ on a vector space its composition with the projection $\CC\overline Q\twoheadrightarrow\Pi^\lambda(Q)$ gives a representation of the double quiver $\overline Q$ on this space. Moreover, different representations of $\Pi^\lambda(Q)$ give different representations of $\overline Q$. A representation $V=(V_i,V_a,V_{a^*})$ of the quiver $\overline Q$ is obtained from a representation of the algebra $\Pi^\lambda(Q)$, if and only if
\begin{align}
 &\sum_{a\in\overline Q, j\in I\atop a\colon j\to i}(-1)^a V_a V_{a^*}=\lambda_i\id_{V_i}, &&i\in I. \label{VPi}
\end{align}
In finite-dimensional case this is a system of algebraic equations (one has $(\dim V_i)^2$ equations for each $i\in I$). All the $\Pi^\lambda(Q)$-modules of dimension $\alpha$ form an affine subvariety $\Rep\big(\Pi^\lambda(Q),\alpha\big)\subset\Rep(\overline Q,\alpha)$, defined by the equations~\eqref{VPi}.

Let $\lambda\cdot\alpha=\sum\limits_{i\in I}\lambda_i\alpha_i$. By taking trace of the equations~\eqref{VPi} and summing over $i\in I$, we derive $\lambda\cdot\alpha=0$. This is a necessary condition for existence of a representation of $\Pi^\lambda(Q)$ of dimension $\alpha$.

We say that the $\CC Q$-submodule $V'\subset V$ is a $\CC Q$-module {\it summand} of $V$, if the $\CC Q$-module $V$ is a $\CC Q$-module direct sum of $V'$ and a $\CC Q$-submodule $V''\subset V$.

\begin{Th} \label{ThLift} {\normalfont \cite{CBL}}
A representation $V=(V_i,V_a)$ of the quiver $Q$ can be extended to a representation $V=(V_i,V_a,V_{a^*})$ of the algebra $\Pi^\lambda(Q)$, if and only if $\lambda\cdot\dim_{\CC I}V'=0$ for any $\CC Q$-module summand $V'\subset V$.
\end{Th}

\begin{Cor} \label{ThLiftCor}
 The set $\Rep\big(\Pi^\lambda(Q), \alpha\big)$ is not empty if and only if $\alpha=\sum_l\alpha^{(l)}$, where $\alpha^{(l)}\in\Delta^+(Q)$ are such that $\lambda\cdot\alpha^{(l)}=0$. In this case there exists $V\in\Rep\big(\Pi^\lambda(Q),\alpha)$ being decomposed into a sum of indecomposable $\Pi^\lambda(Q)$-modules $V^{(l)}\in \Rep\big(\Pi^\lambda(Q), \alpha^{(l)}\big)$.
\end{Cor}

\noindent{\bf Proof.} Let $V\in\Rep\big(\Pi^\lambda(Q), \alpha\big)$. Decompose $V$ into a direct sum of indecomposable $\CC Q$-modules: $V=\bigoplus_{l} V^{(l)}$, $V^{(l)}\in \Rep\big(Q, \alpha^{(l)}\big)$. Then $\alpha=\sum_l\alpha^{(l)}$ and due to Kac Theorem~\ref{ThKac} we have $\alpha^{(l)}\in\Delta^+(Q)$. Since $V^{(l)}$ are $\CC Q$-module summands of $V$, Theorem~\ref{ThLift} implies that $\lambda\cdot\alpha^{(l)}=0$.

Conversely, if $\alpha^{(l)}\in\Delta^+(Q)$ are such that $\lambda\cdot\alpha^{(l)}=0$, then there exist indecomposable $\CC Q$-modules $V^{(l)}\in\Rep\big(Q, \alpha^{(l)}\big)$, and by virtue of Theorem~\ref{ThLift} they are extended to $\Pi^\lambda(Q)$-modules $V^{(l)}\in\Rep\big(\Pi^\lambda(Q),\alpha^{(l)}\big)$. Each $V^{(l)}$ is indecomposable as $\Pi^\lambda(Q)$-module, because otherwise it would be decomposable as $\CC Q$-module. The $\Pi^\lambda(Q)$-module $V=\bigoplus_{l} V^{(l)}$ belongs to $\Rep\big(\Pi^\lambda(Q),\alpha\big)$ for $\alpha=\sum_l\alpha^{(l)}$. \qed

The group $\GL(\alpha)$ acts on $\Rep(\overline Q,\alpha)$. Explicitly the action~\eqref{GLaction} of an element $g=(g_i)$ on $V=(V_i,V_a,V_{a^*})\in\Rep(\overline Q,\alpha)$ has the form
\begin{align} \label{GLactionD}
 &V_a\mapsto g_jV_a g_i^{-1}, & &V_{a^*}\mapsto g_iV_{a^*} g_j^{-1}, &&a\colon i\to j.
\end{align}
These formulae imply that this action preserves the relations~\eqref{VPi} for any $\lambda\in\CC^I$. Thus, we have an action of the group $\GL(\alpha)$ on $\Rep\big(\Pi^\lambda(Q),\alpha\big)$, which relates isomorphic (and only isomorphic) representations of the preprojective algebra $\Pi^\lambda(Q)$. Isomorphism classes of representations of the preprojective algebra (of $\Pi^\lambda(Q)$-modules) of a fixed dimension $\alpha\in\ZZ_{\ge0}^I$ form the orbit space $\Rep\big(\Pi^\lambda(Q), \alpha\big)/\GL(\alpha)$. This quotient not always has a structure of an algebraic variety.

Let $p(\alpha)=1-q(\alpha)=1+\hspace{-4pt}\sum\limits_{a\colon i\to j}\alpha_i\alpha_j-\hspace{-2pt}\sum\limits_{i\in I}\alpha_i\alpha_i$. The following theorem follows from~\cite[Corollary~1.4, Lemma~6.5]{CB01}.

\begin{Th} \label{ThSmooth}
If $\Rep\big(\Pi^\lambda(Q),\alpha\big)\ne\emptyset$ and all the modules $V\in\Rep\big(\Pi^\lambda(Q),\alpha\big)$ are simple, then $\alpha\in\Delta^+(Q)$ and the orbit space $\Rep\big(\Pi^\lambda(Q),\alpha\big)/\GL(\alpha)$ is a smooth connected affine variety of the dimension~$2p(\alpha)$.
\end{Th}

\begin{Rem}
In accordance with Corollary~\ref{ThLiftCor} the condition $\Rep\big(\Pi^\lambda(Q),\alpha\big)\ne\emptyset$ in Theorem~\ref{ThSmooth} can be replaced by the requirement that $\alpha$ is a positive root satisfying $\lambda\cdot\alpha=0$.
\end{Rem}

When the conditions of Theorem~\ref{ThSmooth} are satisfied $\Rep\big(\Pi^\lambda(Q),\alpha\big)/\GL(\alpha)$ is an affine variety called {\it moduli space}. Regular functions on $\Rep\big(\Pi^\lambda(Q),\alpha\big)/\GL(\alpha)$ are exactly $\GL(\alpha)$-invariant regular functions on $\Rep\big(\Pi^\lambda(Q),\alpha\big)$, that is
\begin{align} \label{CatQ}
\CC\big[\Rep\big(\Pi^\lambda(Q),\alpha\big)/\GL(\alpha)\big]=\CC\big[\Rep\big(\Pi^\lambda(Q),\alpha\big)\big]^{\GL(\alpha)}.
\end{align}

\begin{Rem} \label{RemCQ}
In general case one should consider closed orbits only, since the regular (and hence continuous) invariant functions do not distinguish unclosed orbits from adjacent closed orbits. The orbit of the point $V\in\Rep\big(\Pi^\lambda(Q),\alpha\big)$ is closed if and only if $\Pi^\lambda(Q)$-module $V$ is semi-simple~\cite[\S~3.3]{CBL}. In the general case the moduli space is defined as a space of orbits of semi-simple modules and is denoted by $\Rep\big(\Pi^\lambda(Q),\alpha\big)//\GL(\alpha)$. In this case the algebra of regular functions on the moduli space is also given by the right hand side of~\eqref{CatQ}.%
\footnote{The variety $M//G:=\Spec\big(\CC[M]^G\big)$ is called a {\it categorical quotient} of an affine variety $M$ by the action of a group $G$.
}
Theorem~\ref{ThSmooth} is generalised as follows: if $\Rep\big(\Pi^\lambda(Q),\alpha\big)\ne\emptyset$ and a generic point $V\in\Rep\big(\Pi^\lambda(Q),\alpha\big)$ is a simple module, then the moduli space $\Rep\big(\Pi^\lambda(Q),\alpha\big)//\GL(\alpha)$ is an irreducible affine variety, which is smooth in the points corresponding to the simple modules (see~\cite[Corollary~1.4, Lemma~6.5]{CB01}). Let $R^+_\lambda=\{\alpha\in\Delta^+(Q)\mid\lambda\cdot\alpha=0\}$. In accordance with~\cite[Theorem~1.2]{CB01} the condition that the set $\Rep\big(\Pi^\lambda(Q),\alpha\big)$ is not empty and a generic module from this set is simple is equivalent to $\alpha\in\Sigma_\lambda$, where $\Sigma_\lambda$ is a set of roots $\alpha\in R^+_\lambda$ satisfying $p(\alpha)>\sum\limits_{l=1}^rp(\alpha^{(l)})$ for any decomposition $\alpha=\sum\limits_{l=1}^r\alpha^{(l)}$ such that $r\ge2$ and $\alpha^{(l)}\in R^+_\lambda$ $\;\forall\,l$.
\end{Rem}

\subsection{Moduli space as a Hamiltonian reduction}
\label{secHamRed}

The moduli space has a symplectic structure inherited from the space $\Rep(\overline Q,\alpha)$. To describe it we use Hamiltonian reduction~\cite{ArnoldSG,ArnoldMech,Et}. First we give a general construction of the Hamiltonian reduction (in terms of algebraic geometry), and then we apply it to our case, obtaining the moduli spaces as symplectic varieties.

Let $M$ be a connected smooth variety equipped with a symplectic form $\omega_{M}$. Let $G$ be a connected (affine) Lie group $G$, which acts on the variety $M$, preserving its symplectic structure. Infinitesimally this action is given by vector fields $\mathcal V_\theta\in\Vect(M)$ enumerated by the elements $\theta\in\mathfrak g$, where $\mathfrak g$ is the Lie algebra of~$G$. This action is called a {\it Poisson action}, if there exist functions $H_\theta\in C^\infty(M)$ such that $\mathcal V_\theta f=\{H_\theta,f\}$ and $\{H_\theta,H_\eta\}=H_{[\theta,\eta]}$ for all $\theta,\eta\in\mathfrak g$, $f\in C^\infty(M)$. If, for example, $M_0$ is a smooth connected variety and $G$ is a connected group acting on $M$, then the action induced on $M=T^*M_0$ is Poisson with respect to the canonical symplectic structure of the cotangent bundle~\cite[Chatper~3, \S~3.1]{ArnoldSG},~\cite[Appendix~5, \S~A]{ArnoldMech}.

We will interested in the case when $M=T^*\CC^n=\CC^{2n}$ for some $n$ and the group $G$ acts on $\CC^n$ by linear homogeneous transformations. In the standard coordinates $x_1,\ldots,x_n,p_1,\ldots,p_n$ the symplectic form and the corresponding Poisson brackets have the form $\omega_M=\sum\limits_{i=1}^n\d p_i\wedge\d x_i$, $\{f,h\}=\sum\limits_{i=1}^n\Big(\dfrac{\partial f}{\partial x_i}\dfrac{\partial h}{\partial p_i}-\dfrac{\partial f}{\partial p_i}\dfrac{\partial h}{\partial x_i}\Big)$. One can suppose that $G\subset\GL(n,\CC)$ and $\mathfrak g\subset\mathfrak{gl}(n,\CC)$ (if the representation $G\to\GL(n,\CC)$ is not faithful then one can replace the group $G$ by its quotient by the kernel of this representation). Then the Hamiltonians of the action of $G$ on $M$ have the form $H_\theta=\sum\limits_{i,j=1}^n p_i\theta_{ij}x_j$, $\theta\in\mathfrak g$.

Let $\mathfrak g^*$ be the space dual to $\mathfrak g$. The $\mathfrak g^*$-valued function $P\colon M\to\mathfrak g^*$ defined by the formula $P(x)(\theta)=H_\theta(x)$, $x\in M$, $\theta\in\mathfrak g$ is called a {\it moment map}. Consider a point $\lambda\in\mathfrak g^*$ and let $G_\lambda=\{g\in G\mid\Ad^*_g(\lambda)=\lambda\}$ be its stabiliser with respect to the coadjoint action. Since $H_\theta$ are regular functions on $M=T^*\CC^n$, the full preimage $P^{-1}(\lambda)$ is an affine variety in $M$ (possibly reducible). The moment map relates the action of $G$ on $M$ with its coadjoint action as $P(gx)=\Ad^*_g\big(P(x)\big)$ (see~\cite[Chapter~3, \S~3.1]{ArnoldSG},~\cite[Appendix~5, \S~A]{ArnoldMech}). Therefore the preimage $P^{-1}(\lambda)\subset M$ is invariant with respect to the subgroup $G_\lambda\subset G$. Suppose that all the $G_\lambda$-orbits in $P^{-1}(\lambda)$ are closed, then the quotient $N_\lambda:=P^{-1}(\lambda)/G_\lambda$ is an affine variety. It is called a {\it Hamiltonian reduction}.

\begin{Th} \label{ThHR} {\normalfont(see e.g.~\cite[Appendix~5, \S~B]{ArnoldMech}).}
Suppose the variety $N_\lambda=P^{-1}(\lambda)/G_\lambda$ to be smooth. Then it has the symplectic structure $\omega_\lambda$: for $x\in P^{-1}(\lambda)$ and $\xi,\eta\in T_{[x]}N_\lambda$ one sets
\begin{align}
 \omega_\lambda(\xi,\eta)=\omega_{M}(\xi',\eta'), \label{omegaHR}
\end{align}
where $\xi',\eta'\in T_x\big(P^{-1}(\lambda)\big)\subset T_xM$ are any preimages of $\xi$, $\eta$ with respect to the differential of the projection $P^{-1}(\lambda)\twoheadrightarrow N_\lambda$.
\end{Th}

The restriction of an invariant function $h\in\CC[M]^G$ to $P^{-1}(\lambda)\subset M$ is $G_\lambda$-invariant function identified with a function on $N_\lambda$ and denoted by $h_\lambda\in\CC[N_\lambda]$.

\begin{Prop} \label{propPoisson}
The Poisson brackets $\{-,-\}_{N_\lambda}$ and $\{-,-\}_M$ associated with the symplectic forms $\omega_\lambda$ and $\omega_M$ respectively are related as $\{f_\lambda,h_\lambda\}_{N_\lambda}([x])=\{f,h\}_M(x)$, where $f,h\in\CC[M]^G$, $x\in P^{-1}(\lambda)$.
\end{Prop}

\noindent{\bf Proof.} Let $X_f$ be the Hamiltonian field with a Hamiltonian $f$, it is defined by means of $\omega$ as follows: $\omega_x(X_f(x),\xi)=(df)_x(\xi)$ at any point $x$ and for any tangent vector $\xi$ at $x$. Then the Poisson brackets associated with a symplectic form $\omega$ are $\{f,h\}=-\omega(X_f,X_h)$. The fields $X_f$ and $X_h$ (at the point $x$) are preimages of $X_{f_\lambda}$ and $X_{h_\lambda}$ (at the point $[x]$) respectively~\cite[Appendix~5, \S~C]{ArnoldMech}. Hence $\{f_\lambda,h_\lambda\}_{N_\lambda}=-\omega_\lambda(X_{f_\lambda},X_{h_\lambda})=-\omega_M(X_f,X_h)=\{f,h\}$. \qed

In our case $M_0=\Rep(Q,\alpha)$, $M=\Rep(\overline Q,\alpha)=T^*M_0$ and $\GL(\alpha)$ acts of the vector space $M_0$ by linear homogeneous transformations. The Lie algebra of $\GL(\alpha)$ is
\begin{align*}
 \End(\alpha):=\bigoplus\limits_{i\in I}\End(\CC^{\alpha_i})=\bigoplus\limits_{i\in I}\mathfrak{gl}(\alpha_i,\CC).
\end{align*}
Its dual vector space can be identified with $\End(\alpha)$ by means of the bilinear form $(\theta,\eta)=\sum\limits_{i\in I}\tr(\theta_i\eta_i)$, where $\theta=(\theta_i),\eta=(\eta_i)\in\End(\alpha)$.

The kernel of the representation $\GL(\alpha)\to\GL(M_0)$ is the subgroup $\CC^\times$, consisting of elements $g\in\GL(\alpha)$ with components $g_i=c\in\CC\backslash\{0\}$. The corresponding quotient group $G(\alpha):=\GL(\alpha)/\CC^\times$ acts effectively on $M=\Rep(\overline Q,\alpha)$, so it is embedded to $\GL(M_0)$. The action of $G(\alpha)$ on $M$ is Poisson, since it is induced by the action of $G(\alpha)$ on $M_0=\Rep(Q,\alpha)$. The Lie algebra of the quotient group $G=G(\alpha)$ is the quotient algebra $\mathfrak g=\End(\alpha)/(\CC\cdot1)$, where $1=(\id_{\CC^{\alpha_i}})_{i\in I}$. Its dual space is
\begin{align}
 \mathfrak g^*=\End(\alpha)_0:=\big\{u=(u_i)\in\End(\alpha)\mid\sum\limits_{i\in I}\tr u_i=0\big\}\subset\End(\alpha).
\end{align}
Let us identify $\lambda=(\lambda_i)_{i\in I}\in\CC^I$ with the element $\big(\lambda_i \id_{\CC^{\alpha_i}}\big)_{i\in I}\in\End(\alpha)$. Then the formula $\sum\limits_{i\in I}\tr(\lambda_i \id_{\CC^{\alpha_i}})=\lambda\cdot\alpha$ implies that the condition $\lambda\cdot\alpha=0$ is equivalent to $\lambda\in\End(\alpha)_0$. One can always suppose this condition to be satisfied: it is necessary for the non-emptiness of $\Rep\big(\Pi^\lambda(Q),\alpha\big)$. The moment map $P_\alpha\colon \Rep(\overline Q,\alpha)\to\mathfrak g^*$ has the form
\begin{align} \label{PalphaV}
 &P_\alpha(V)=\big(P_{\alpha,i}(V)\big)_{i\in I}, && P_{\alpha,i}(V)= \sum_{a\in\overline Q, j\in I\atop a\colon j\to i}(-1)^{a}V_aV_{a^*},
\end{align}
(see Appendix~\ref{appPoisson}). This implies $P_\alpha^{-1}(\lambda)=\Rep\big(\Pi^\lambda(Q),\alpha\big)$. The stabiliser of $\lambda=(\lambda_i)\in\mathfrak g^*$ is $G_\lambda=G=G(\alpha)$. We will suppose that the conditions of Theorem~\ref{ThSmooth} are satisfied. Then the Hamiltonian reduction is exactly the moduli space:
\begin{align*}
 N_\lambda(\alpha):=\Rep\big(\Pi^\lambda(Q),\alpha\big)/G(\alpha)=\Rep\big(\Pi^\lambda(Q),\alpha\big)/\GL(\alpha).
\end{align*}
It is smooth connected affine variety. By virtue of Theorem~\ref{ThHR} it is equipped with the symplectic form~$\omega_{\lambda,\alpha}$ given by the formula~\eqref{omegaHR}.

This symplectic form on $N_\lambda(\alpha)$ can be understood as follows. Let $z_k$ be local coordinates in a domain of the variety $N_\lambda(\alpha)$. In this domain a point $z=(z_k)$ can be presented by a (semi-simple) module $V(z)=\big(V_i,V_a(z)\big)\in\Rep\big(\Pi^\lambda(Q),\alpha\big)$, so that we obtain smooth $\Hom(\CC^{\alpha_i},\CC^{\alpha_j})$-valued functions $V_a=V_a(z)$ ($a\in\overline Q$, $a\colon i\to j$, $i,j\in I$) satisfying~\eqref{VPi}. Then~\eqref{omegaHR} in this domain has the form
\begin{multline} \label{omegaF}
 \omega_{\lambda,\alpha} =\sum_{a\in Q} \tr(\d V_{a^*}\wedge\d V_a)= \sum_{a\in Q} \tr\big(\d V_{a^*}(z)\wedge\d V_a(z)\big)= \\
 =\sum_{a\in Q}\sum_{k,l} \tr\Big( \frac{\partial V_{a^*}}{\partial z_k} \frac{\partial V_a}{\partial z_l}\Big) \d z_k\wedge\d z_l.
\end{multline}

Consider the functions $\tr_\alpha(p)\in\Rep(\overline Q,\alpha)$, where $p\in\CC\overline Q$ are cycles in the double quiver (see Section~\ref{sec21}). Their restriction to the subvariety $\Rep\big(\Pi^\lambda(Q),\alpha\big)$ gives the functions $\tr_\alpha(p)_\lambda\in\CC[N_\lambda(\alpha)]$. Theorem~\ref{ThLeBP} implies the following fact (see~\cite[Lemma~2.2]{CB02}).

\begin{Prop} \label{PropTr}
 The algebra $\CC\big[N_\lambda(\alpha)\big]=\CC\big[\Rep\big(\Pi^\lambda(Q),\alpha\big)\big]^{G(\alpha)}$ is generated by the restrictions of the functions $\tr_\alpha(p)$ to $\Rep\big(\Pi^\lambda(Q),\alpha\big)$, i.e. by $\tr_\alpha(p)_\lambda$, where $p$ runs over all cycles in $\overline Q$.
\end{Prop}

\begin{Rem} \label{RemCQHR}
If not all the $G_\lambda$-orbits in $P^{-1}(\lambda)$ are closed, the Hamiltonian reduction is defined as $N_\lambda=P^{-1}(\lambda)//G_\lambda$, this is a variety consisting of the closed orbits. If it is smooth, it has symplectic structure defined by  the formula~\eqref{omegaHR}, where $x\in P^{-1}(\lambda)$ is such that the orbit $Gx$ is closed. In the case $M=\Rep(\overline Q,\alpha)$, $G=G(\alpha)$ and $\lambda=(\lambda_i)$ the Hamiltonian reduction is the moduli space $N_\lambda(\alpha)=\Rep\big(\Pi^\lambda(Q),\alpha\big)//\GL(\alpha)$ (see Remark~\ref{RemCQ}). Proposition~\ref{PropTr} is also valid for this variety.
\end{Rem}

\section{Reflection functor for the preprojective algebras}
\label{secRF}

First we define the reflection functor on the representations of the preprojective algebras associated with a quiver~\cite{CBH} (see also~\cite{CBL}). We show that it gives an equivalence of categories. Then we consider how it acts on the moduli spaces and prove that it preserves their algebraic and symplectic structures.

\subsection{Definition of the reflection functor}
\label{secRFEquiv}

Let $i\in\Ilf$. Recall that the reflection $s_i$ acts on the lattice $\ZZ^I$ by the formula~\eqref{si}. Define a {\it dual reflection} as the linear transformation $r_i\colon\CC^I\to\CC^I$ by the formula
\begin{align}
 &(r_i\lambda)_j=\lambda_j-(\varepsilon_i,\varepsilon_j)\lambda_i.
\end{align}
The chain of the equalities
\begin{align*}
 r_i\lambda\cdot\alpha= \sum_j \big(\lambda_j-(\varepsilon_i,\varepsilon_j) \lambda_i\big)\alpha_j= \lambda\cdot\alpha-(\varepsilon_i,\alpha)\lambda_i= \lambda\cdot s_i\alpha 
\end{align*}
and the involution property $s_i^2=1$ imply
\begin{align}
 &r_i^2=1, &&\lambda\cdot\alpha= r_i\lambda\cdot s_i\alpha, &&&&i\in\Ilf.
\end{align}

Fix a vertex $k\in\Ilf$ and a vector $\lambda\in\CC^I$. We say that a reflection is {\it admissible} at the vertex $k\in\Ilf$ for given $\lambda\in\CC^I$, if $\lambda_k\ne0$. In this case we will construct the functor $F^\lambda_k$ from the category of $\Pi^\lambda(Q)$-modules to the category of $\Pi^{r_k\lambda}(Q)$-modules, which maps representations of the dimension $\alpha$ to representations of the dimension~$s_k\alpha$.

Let $V=(V_i,V_a)$, $i\in I$, $a\in\overline Q$, be a representation of the preprojective algebra $\Pi^\lambda(Q)$. To define the value of the functor $F^\lambda_k$ on $V$, we construct a representation $V'$ of the algebra $\Pi^{r_k\lambda}(Q)$ as follows.

Introduce the notations%
\footnote{Usually to define the reflection functor one supposes for simplicity that $a\in Q$ for any edge $a\colon j\to k$ (see e.g.~\cite{CB01}). To reach this condition one needs to reorient the quiver $Q$ and use an isomorphism $\Pi^\lambda(Q)\simeq\Pi^\lambda(Q')$, where $Q'$ is the reoriented quiver (reorientation depends on $k$). We will not suppose that, since the usage of the notation~\eqref{m1aNotation} makes this simplification to be unnecessary.
}
\begin{align}
 H=\{a\in\overline Q\mid a\colon j\to k\text{ for some }j\}\subset E\sqcup E^* 
\end{align}
and $V_\oplus=\bigoplus\limits_{a\in H\atop a\colon j\to k}V_j$. For each $a\colon j\to k$ we have the canonical embedding $\mu_a\colon V_j\hookrightarrow V_\oplus$ and projection $\pi_a\colon V_\oplus\twoheadrightarrow V_j$. Define linear maps $\mu\colon V_k\to V_\oplus$ and $\pi\colon V_\oplus\to V_k$ as
\begin{align}
 &\mu=\sum\limits_{a\in H}\mu_aV_{a^*}, &&\pi=\frac1{\lambda_k}\sum\limits_{a\in H}(-1)^aV_a\pi_a. \label{mupi}
\end{align}
Then the relation~\eqref{VPi} for $i=k$ takes the form $\pi\mu=1$. In particular, it follows that the map $\mu$ is injective and $\pi$ is surjective. Moreover the exact sequence
\begin{align}
 \xymatrix{0\ar[r]&\Ker\pi\ar[r]&V_\oplus\ar[r]^\pi &V_k\ar[r]&0}
\end{align}
splits by $\mu\colon V_k\to V_\oplus$, that is $V_\oplus=\Ker\pi\oplus V_k$. Let $V'_k=\Ker\pi$ and $V'_j=V_j$ for $j\ne k$. Denote by $\mu'\colon V'_k\hookrightarrow V_\oplus$ and $\pi'\colon V_\oplus\twoheadrightarrow V'_k$ the canonical embedding and projection of the direct sum:
\begin{gather}
 \xymatrix{V'_k\ar@<0.5ex>[r]^{\mu'}&V_\oplus\ar@<0.5ex>[l]^{\pi'}\ar@<-0.5ex>[r]_\pi &V_k\ar@<-0.5ex>[l]_\mu} \label{directsum} \\
 \pi\mu=1, \qquad\qquad\pi\mu'=0, \label{directsum1} \\
\pi'\mu'=1, \qquad\qquad\pi'\mu=0, \label{directsum2}\\
\mu\pi+\mu'\pi'=1.\label{directsum3} 
\end{gather}
Define maps $V'_a\colon V'_i\to V'_j$, $a\in\overline Q$, $a\colon i\to j$, as follows: if $a\in H$, then $a\colon j\to k$, $a^*\colon k\to j$ for some $j\in I$ and then we define $V'_a\colon V_j\to V'_k$ and $V'_{a^*}\colon V'_k\to V_j$ by the formulae
\begin{align}
 &V'_a=-\lambda_k(-1)^a\pi'\mu_a, && V'_{a^*}=\pi_a\mu'; \label{Vanew}
\end{align}
if $a\notin H$ and $a^*\notin H$, then we set $V'_a=V_a$.

\begin{Lem} \label{LemVbVa}
 For any $a,b\in H$ one has
\begin{align}
 &V'_{b^*}V'_a=-\lambda_k(-1)^a\delta_{ab}+V_{b^*}V_a,  \label{LemVbVa1} \\
 &\sum_{c\in H}(-1)^cV'_cV_{c^*}=0,  \label{LemVbVa2} \\
 &\sum_{c\in H}(-1)^cV_cV'_{c^*}=0. \label{LemVbVa3}
\end{align}
\end{Lem}

\noindent{\bf Proof.} By using~\eqref{directsum3} and $\pi_b\mu_a=\delta_{ab}$ we obtain
\begin{multline*}
 V'_{b^*}V'_a=-\lambda_k(-1)^a\pi_b\mu'\pi'\mu_a=-\lambda_k(-1)^a\pi_b(1-\mu\pi)\mu_a=-\lambda_k(-1)^a\delta_{ab}+\lambda_k(-1)^a\pi_b\mu\pi\mu_a=\\
=-\lambda_k(-1)^a\delta_{ab}+(-1)^a\sum_{c,d\in H}(-1)^d\pi_b\mu_cV_{c^*}V_d\pi_d\mu_a=-\lambda_k(-1)^a\delta_{ab}+V_{b^*}V_a.
\end{multline*}
Further, the right formulae in~\eqref{directsum2} and \eqref{directsum1} imply
\begin{align}
 &\sum_{c\in H}(-1)^cV'_cV_{c^*}=-\lambda_k\sum_{c\in H}\pi'\mu_cV_{c^*}=-\lambda_k\pi'\mu=0, \\
 &\sum_{c\in H}(-1)^cV_cV'_{c^*}=\sum_{c\in H}(-1)^cV_c\pi_c\mu'=\lambda_k\pi\mu'=0
\end{align}
respectively. \qed

Now let us show that $V'=(V'_i,V'_a)$ is a representation of the algebra $\Pi^{r_k\lambda}(Q)$. We need to check the relations~\eqref{VPi} for all $i\in I$. Note that $n_{kk}=0$ implies $(\varepsilon_k,\varepsilon_k)=2$, so that $(r_k\lambda)_k=\lambda_k-2\lambda_k=-\lambda_k$. The relation~\eqref{VPi} for $i=k$ is proven as follows:
\begin{align*}
 \sum_{a\colon j\to k}(-1)^aV'_aV'_{a^*}=\sum_{a\in H}(-1)^aV'_aV'_{a^*}=-\lambda_k\sum_{a\in H}\pi'\mu_a\pi_a\mu'=-\lambda_k\pi'\mu'=-\lambda_k=(r_k\lambda)_k,
\end{align*}
where we used $\sum\limits_{a\in H}\mu_a\pi_a=1$ and the left formula in~\eqref{directsum2}.
In the case $i\ne k$ we obtain $(\varepsilon_i,\varepsilon_k)=-n_{ik}$, $(r_k\lambda)_i=\lambda_i+n_{ik}\lambda_k$ and
\begin{align} \label{inek}
 \sum_{a\colon j\to i}(-1)^aV'_aV'_{a^*}=\sum_{a\in\overline Q,j\ne k\atop a\colon j\to i}(-1)^aV_aV_{a^*}+\hspace{-4pt}\sum_{a^*\in H\atop a\colon k\to i}(-1)^aV'_aV'_{a^*}.
\end{align}
The second term in~\eqref{inek} has the form
\begin{align*}
 \sum_{a^*\in H\atop a\colon k\to i}(-1)^aV'_aV'_{a^*}=-\hspace{-4pt}\sum_{a\in H\atop a\colon i\to k}(-1)^aV'_{a^*}V'_a=-\hspace{-4pt}\sum\limits_{a\in H\atop a\colon i\to k}\big(-\lambda_k+(-1)^aV_{a^*}V_a\big),
\end{align*}
where we used the formula~\eqref{LemVbVa1} for $a=b$. By taking into account
\begin{align*}
 \big|\{a\in H\mid a\colon i\to k\}\big|=\big|\{a\in Q\mid a\colon i\to k\}\big|+\big|\{a^*\in Q\mid a^*\colon i\to k\}\big|=n_{ik} 
\end{align*}
we continue:
\begin{align*}
 \sum_{a^*\in H\atop a\colon k\to i}(-1)^aV'_aV'_{a^*}=n_{ik}\lambda_k-\hspace{-4pt}\sum\limits_{a\in H\atop a\colon i\to k}(-1)^aV_{a^*}V_a=n_{ik}\lambda_k+\hspace{-4pt}\sum\limits_{a\in\overline Q\atop a\colon k\to i}(-1)^aV_aV_{a^*}.
\end{align*}
By summing one yields
\begin{align*}
 \sum_{a\colon j\to i}(-1)^aV'_aV'_{a^*}=\sum_{a\in\overline Q\atop a\colon j\to i}(-1)^aV_aV_{a^*}+n_{ik}\lambda_k=\lambda_i+n_{ik}\lambda_k=(r_k\lambda)_i.
\end{align*}
Hence, $V'$ is a $\Pi^{r_k\lambda}(Q)$-module.

To define the functor on morphisms, consider a morphism $\phi\colon V\to\wt V$ between two $\Pi^\lambda(Q)$-modules $V=(V_i,V_a)$ and $\wt V=(\wt V_i,\wt V_a)$. It is given by the maps $\phi_i\colon V_i\to\wt V_i$ such that $\phi_jV_a=\wt V_a\phi_i$ $\;\forall\,a\colon i\to j$. The construction described above gives us $\Pi^{r_k\lambda}(Q)$-modules $V'=(V'_i,V'_a)$ and $\wt V'=(\wt V'_i,\wt V'_a)$.
Set $\phi'_i=\phi_i$ for $i\ne k$ and $\phi'_k=-\frac1{\lambda_k}\sum\limits_{b\colon j\to k}(-1)^b\wt V'_b\phi_j V'_{b^*}$. Let us check that $\phi'=(\phi'_i)$ is a morphism $V'\to\wt V'$.

If $a\colon i\to j$ and $i\ne k\ne j$, then $\phi'_jV'_a=\phi_jV_a=\wt V_a\phi_i=\wt V'_a\phi'_i$. Let $a\colon i\to k$, then by using the formula~\eqref{LemVbVa1}, we obtain
\begin{align*}
 \phi'_kV'_a=-\frac1{\lambda_k}\sum_{b\colon j\to k}(-1)^b\wt V'_b\phi_jV'_{b^*}V'_a=\wt V'_a\phi_i-\frac1{\lambda_k}\sum_{b\colon j\to k}(-1)^b\wt V'_b\phi_jV_{b^*}V_a.
\end{align*}
The second term vanishes:
\begin{align*}
 \sum_{b\colon j\to k}(-1)^b\wt V'_b\phi_jV_{b^*}=\sum_{b\in H}(-1)^b\wt V'_b\wt V_{b^*}\phi_k=0
\end{align*}
due to~\eqref{LemVbVa2}.
Analogously, for $a^*\colon k\to i$ we have
\begin{align*}
 \wt V'_{a^*}\phi'_k=-\frac1{\lambda_k}\sum_{b\colon j\to k}(-1)^b\wt V'_{a^*}\wt V'_b\phi_j V'_{b^*}=\phi_iV'_{a^*}-\frac1{\lambda_k}\sum_{b\colon j\to k}(-1)^b\wt V_{a^*}\wt V_b\phi_j V'_{b^*}.
\end{align*}
And again:
\begin{align*}
 \sum_{b\colon j\to k}(-1)^b\wt V_b\phi_j V'_{b^*}=\sum_{b\in H}(-1)^b\phi_k V_b V'_{b^*}=0
\end{align*}
due to~\eqref{LemVbVa3}. Thus $\phi'$ is a morphism.

Denote by $\Pi^\lambda(Q)\Mod$ the category of the (left) $\Pi^\lambda(Q)$-modules.

\begin{opr}
Let $k\in \Ilf$ and $\lambda\in\CC^I$ be such that $\lambda_k\ne0$. The functor
\begin{align}
 F_k^\lambda\colon\Pi^\lambda(Q)\Mod\to\Pi^{r_k\lambda}(Q)\Mod
\end{align}
that maps a $\Pi^\lambda(Q)$-module $V$ to the $\Pi^{r_k\lambda}(Q)$-module $V'$ and a $\Pi^\lambda(Q)$-module morphism~$\phi$ to the $\Pi^{r_k\lambda}(Q)$-module morphism $\phi'$ is called the {\it reflection functor}.
\end{opr}

Recall (see e.g.~\cite{MacLane}) that two categories $\C$ and $\D$ are called {\it equivalent}, if there exist functors $F\colon\C\to\D$ and $G\colon\D\to\C$ such that the objects $X\in\C$ and $Y\in\D$ are naturally isomorphic to the objects $GF(X)$ and $FG(Y)$ respectively (the term {\it naturally} means that one has $\phi=GF(\phi)$ and $\psi=FG(\psi)$ for any morphisms $\phi$ in $\C$ and $\psi$ in $\D$ with respect to the identification $GF(X)=X$ and $FG(Y)=Y$). In this case the objects $X\in\C$ and $F(X)\in\D$ has the same properties such as simplicity, semi-simplicity, indecomposability etc. In particular, isomorphic objects are mapped to isomorphic ones.

\begin{Th} \label{REquiv}
 For any $k\in\Ilf$ and $\lambda\in\CC^I$ such that $\lambda_k\ne0$ the functor $F^\lambda_k$ gives an equivalence of categories $\Pi^\lambda(Q)\Mod$ and $\Pi^{r_k\lambda}(Q)\Mod$.
\end{Th}

\noindent{\bf Proof.} Since $(r_k\lambda)_k=-\lambda_k\ne0$ and $r_k^2(\lambda)=\lambda$, there is another reflection functor $F_k^{r_k\lambda}\colon\Pi^{r_k\lambda}(Q)\Mod\to\Pi^\lambda(Q)\Mod$. Let $V'=F^\lambda_k(V)$ and $V''=F_k^{r_k\lambda}(V')$. Consider elements of the construction of $\Pi^\lambda(Q)$-module $V''$ from $\Pi^{r_k\lambda}(Q)$-module $V'$. The space $V_\oplus$, the embeddings $\mu_a$ and the projections $\pi_a$ for this construction are the same as for the construction of $V'$ from $V$. By replacing $\lambda_k$, $V_a$ and $V_{a^*}$ in the definition~\eqref{mupi} by $-\lambda_k$, $V'_a$ and $V'_{a^*}$, we see that the maps~\eqref{mupi} for the construction of $V''$ from $V'$ are equal to
\begin{align*}
 &\sum_{a\in H}\mu_a V'_{a^*}=\sum_{a\in H}\mu_a \pi_a\mu'=\mu', &&-\frac1{\lambda_k}\sum_{a\in H}(-1)^a V'_a\pi_a=\sum_{a\in H}\pi'\mu_a \pi_a=\pi'.
\end{align*}
This means that the diagram of the direct sum~\eqref{directsum} for the construction of $V''$ from $V'$ has the form
\begin{gather}
 \xymatrix{V''_k=V_k\ar@<0.5ex>[r]^{\mu}&V_\oplus\ar@<0.5ex>[l]^{\pi}\ar@<-0.5ex>[r]_{\pi'} &V'_k\ar@<-0.5ex>[l]_{\mu'}}. \label{directsumEquiv}
\end{gather}
By substituting $\lambda_k\to-\lambda_k$, $\mu'\to\mu$ and $\pi'\to\pi$ in the formulae~\eqref{Vanew}, one yields
\begin{align}
 &V''_a=\lambda_k(-1)^a\pi\mu_a=(-1)^a\sum_{b\in H}(-1)^b V_a\pi_b\mu_a=V_a, \\
 &V''_{a^*}=\pi_a\mu=\pi_a\sum_{b\in H}\mu_b V_{b^*}=V_{a^*}.
\end{align}
Thus, $V''$ coincides with $V$ as $\Pi^\lambda(Q)$-module. Now, let $\phi\colon V\to\wt V$ be a morphism of $\Pi^\lambda(Q)$-modules, $F^\lambda_k(\phi)=\phi'\colon V'\to\wt V'$ and $F^{r_k\lambda}_k(\phi')=\phi''\colon V\to\wt V$. Then $\phi''_i=\phi'_i=\phi_i$ for any $i\ne k$ and
\begin{align*}
 \phi''_k=-\frac1{(r_k\lambda)_k}\sum_{b\colon j\to k}(-1)^b\wt V''_b\phi'_j V''_{b^*}=\frac1{\lambda_k}\sum_{b\colon j\to k}(-1)^b\wt V_b\phi_j V_{b^*}
=\frac1{\lambda_k}\sum_{b\in H}(-1)^b\wt V_b\wt V_{b^*}\phi_k=\phi_k.
\end{align*}
Hence the identification $F^{r_k\lambda}_k F^\lambda_k(V)=V$ is a natural isomorphism for $\Pi^\lambda(Q)$-modules $V$. By replacing $\lambda$ with $r_k\lambda$ we obtain a natural isomorphism $F^\lambda_k F^{r_k\lambda}_k(V)=V$ for $\Pi^{r_k\lambda}(Q)$-modules $V$. \qed

\subsection{Symplectic isomorphism of the moduli spaces}
\label{secRFIsom}

The reflection functor $F^\lambda_k$ maps a finite-dimensional module to a finite-dimensional one. More precisely, if $V$ is a $\Pi^\lambda(Q)$-module of the dimension $\alpha\in\ZZ_{\ge0}^I$, then $V'=F_k^\lambda(V)$ is a $\Pi^{r_k\lambda}(Q)$-module of the dimension $s_k\alpha$. Indeed, $\dim V'_i=\dim V_i=\alpha_i$ for $i\ne k$, and
\begin{align*}
 \dim V'_k=\dim V_\oplus-\dim V_k=\sum_{a\in H\atop a\colon j\to k}\dim V_j-\dim V_k=\sum_{j\ne k}n_{jk}\alpha_j-\alpha_k.
\end{align*}
On the other hand,
\begin{align}
s_k\alpha=\alpha-(\alpha,\varepsilon_k)\varepsilon_k=\sum_{i\in I}\alpha_i\varepsilon_i-\big(2\alpha_k-\hspace{-2pt}\sum_{j\ne k}n_{jk}\alpha_j\big)\varepsilon_k,
\end{align}
or, componentwise: $(s_k\alpha)_i=\alpha_i$ for $i\ne k$ and $(s_k\alpha)_k=\sum\limits_{j\ne k}n_{jk}\alpha_j-\alpha_k$.

Form the construction of the module $V'$ we can not extract a canonical basis in $V'_k$ by the bases of the spaces $V_i$, $i\in I$. Hence, the functor $F^\lambda_k$ does not give a (regular) map from $\Rep\big(\Pi^\lambda(Q),\alpha\big)$ to $\Rep\big(\Pi^{r_k\lambda}(Q),s_k\alpha\big)$. However, Theorem~\ref{REquiv} implies that $F^\lambda_k$ defines a bijection between the orbit spaces
\begin{align} \label{RFBij}
 \Rep\big(\Pi^\lambda(Q),\alpha\big)/G(\alpha)\leftrightarrow\Rep\big(\Pi^{r_k\lambda}(Q),s_k\alpha\big)/G(s_k\alpha).
\end{align}
This bijection can not be considered as regular, if one of these spaces is not a variety. Note also that $s_k\alpha\notin\ZZ_{\ge0}^I$ for some $\alpha\in\ZZ_{\ge0}^I$. The following statement is a consequence of Theorem~\ref{REquiv}.

\begin{Prop} \label{PropEquivCor}
Let $\alpha\in\ZZ_{\ge0}^I$ and $\lambda\in\CC^I$ such that $\lambda_k\ne0$.
\begin{itemize}
 \item[(a)] If $\Rep\big(\Pi^\lambda(Q),\alpha\big)\ne\emptyset$, then $\Rep\big(\Pi^{r_k\lambda}(Q),s_k\alpha\big)\ne\emptyset$, and, in particular, $s_k\alpha\in\ZZ_{\ge0}^I$.
\footnote{This can be shown by using Corollary~\ref{ThLiftCor} and the theory of root systems.
}
 \item[(b)] If all the modules $V\in\Rep\big(\Pi^\lambda(Q),\alpha\big)$ are simple, then all $V\in\Rep\big(\Pi^{r_k\lambda}(Q),s_k\alpha\big)$ are also simple.
\end{itemize}
\end{Prop}

The Proposition~\ref{PropEquivCor} implies that in the case $\lambda_k\ne0$ the condition of Theorem~\ref{ThSmooth} is satisfied for $\lambda$ and $\alpha$ if and only if it is satisfied for $r_k\lambda$ and $s_k\alpha$. In this case the varieties $N_\lambda(\alpha)$ and $N_{r_k\lambda}(s_k\alpha)$ are smooth and connected.

We will suppose below that $\lambda_k\ne0$ and the condition of Theorem~\ref{ThSmooth} is satisfied. Then the bijection~\eqref{RFBij} is a map of the varieties
\begin{align} \label{varrho}
 \varrho^{\lambda,\alpha}_k\colon N_\lambda(\alpha)\to N_{r_k\lambda}(s_k\alpha).
\end{align}
Let us clarify how the functions $\tr_\alpha(p)$ change under this map. By a {\it subcycle} of a cycle $p=a_\ell\cdots a_2a_1$ we call any cycle of the form $a_{j_m}\cdots a_{j_2}a_{j_1}$, where $1\le j_1<j_2<\ldots<j_m\le\ell$ and $0\le m\le\ell$ (subcycle of the length $m=0$ is a trivial cycle $1_i$ where $i\in I$ is such that $a_j\colon i\to i'$ for some $j=1,\ldots,\ell$ and $i'\in I$).

\begin{Lem} \label{LemTr}
Let $p=a_\ell\cdots a_2a_1$ be a cycle in $\overline Q$, namely $a_j\colon i_{j-1}\to i_j$, where $j=1,\ldots,\ell$ and $i_\ell=i_0$. Denote $a_{\ell+1}=a_1$.
\begin{itemize}
 \item[(a)] The function $\tr_{s_k\alpha}(p)_{r_k\lambda}\circ\varrho^{\lambda,\alpha}_k$ is a linear combination of $\tr_\alpha(p')_\lambda$, where $p'$ are subcycles of $p$. In particular, $\tr_{s_k\alpha}(p)_{r_k\lambda}\circ\varrho^{\lambda,\alpha}_k\in\CC[N_\lambda(\alpha)]$.
 \item[(b)] Suppose that $a_j\ne a_{j+1}^*$ whenever $i_j= k$. In this case we have $\tr_{s_k\alpha}(p)_{r_k\lambda}\circ\varrho^{\lambda,\alpha}_k=\tr_\alpha(p)_\lambda$. In particular, it is so, if $a_j\in Q$ for all $j$ or if $a_j^*\in Q$ for all~$j$.
\end{itemize} 
\end{Lem}

\noindent{\bf Proof.} Consider modules $V=(V_i,V_a)\in\Rep\big(\Pi^\lambda(Q),\alpha\big)$ and $V'=(V'_i,V'_a)=F^\lambda_k(V)$. The value of the function $\tr_{s_k\alpha}(p)_{r_k\lambda}\circ\varrho^{\lambda,\alpha}_k$ at the point $V$ is $\tr_{s_k\alpha}(p)_{r_k\lambda}(V')=\tr(V'_{a_\ell}\cdots V'_{a_2}V'_{a_1})$. If $i_{j-1}\ne k$ and $i_{j}\ne k$, then $V'_{a_j}=V_{a_j}$. If $i_j=k$ then due to formula~\eqref{LemVbVa1} we derive
\begin{align}
 V'_{a_{j+1}}V'_{a_j}=V_{a_{j+1}}V_{a_j}-\lambda_k(-1)^{a_j}\delta_{a_j,a_{j+1}^*}. \label{VaVaInTr}
\end{align}
Moreover, for such $j$ we have $i_{j-1}\ne k$ and $i_{j+1}\ne k$ (where $i_{\ell+1}=i_1$). Hence, by substituting~\eqref{VaVaInTr} for such $j$ to $\tr(V'_{a_\ell}\cdots V'_{a_2}V'_{a_1})$ we obtain a linear combination of functions $\tr(V_{a_{j_m}}\cdots V_{a_{j_2}}V_{a_{j_1}})$ with coefficients independent of $V$. If the condition of the item~(b) is satisfied, then the second term in the right hand side of~\eqref{VaVaInTr} vanishes and we obtain $\tr(V_{a_\ell}\cdots V_{a_2}V_{a_1})$. \qed

Let us recall that the affine variety $N_{\lambda}(\alpha)$ is equipped by the symplectic structure~\eqref{omegaF}.

\begin{Th} \label{ThIsom}
 The map $\varrho^{\lambda,\alpha}_k\colon N_{\lambda}(\alpha)\to N_{r_k\lambda}(s_k\alpha)$ is an isomorphism of symplectic varieties.
\end{Th}

\noindent{\bf Proof.} Since the map $\varrho^{\lambda,\alpha}_k$ is bijective we have only to prove that it is regular and preserves the symplectic structure. The regularity of the map $\varrho^{\lambda,\alpha}_k$ is equivalent to $f\circ\varrho^{\lambda,\alpha}_k\in\CC\big[N_{\lambda}(\alpha)\big]$ for all $f\in\CC\big[N_{r_k\lambda}(s_k\alpha)\big]$. But this follows from the item~(a) of Lemma~\ref{LemTr}, since the algebra $\CC\big[N_{r_k\lambda}(s_k\alpha)\big]$ is generated by the functions $\tr_{s_k\alpha}(p)_{r_k\lambda}$ due to Proposition~\ref{PropTr} (in Appendix~\ref{appRRep} we give a direct proof of the regularity of $\varrho^{\lambda,\alpha}_k$).

Now we need to establish the equality $(\varrho^{\lambda,\alpha}_k)^*\omega_{r_k\lambda,s_k\alpha}=\omega_{\lambda,\alpha}$. Let us write its right hand side as
\begin{multline*}
 \omega_{\lambda,\alpha}=\sum_{a\in Q\atop a,a^*\notin H} \tr(\d V_{a^*}\wedge\d V_a)+\hspace{-2pt}\sum_{a\in Q\atop a\in H} \tr(\d V_{a^*}\wedge\d V_a)+\hspace{-2pt}\sum_{a^*\in Q\atop a\in H} \tr(\d V_{a}\wedge\d V_{a^*})=\\
=\sum_{a\in Q\atop a,a^*\notin H} \tr(\d V_{a^*}\wedge\d V_a)+\hspace{-2pt}\sum_{a\in H}(-1)^a \tr(\d V_{a^*}\wedge\d V_a).
\end{multline*}
In the same way, by denoting $V'=F^\lambda_k(V)$, we obtain
\begin{align*}
 (\varrho^{\lambda,\alpha}_k)^*\omega_{r_k\lambda,s_k\alpha}
=\sum_{a\in Q\atop a,a^*\notin H} \tr(\d V'_{a^*}\wedge\d V'_a)+\hspace{-2pt}\sum_{a\in H}(-1)^a \tr(\d V'_{a^*}\wedge\d V'_a).
\end{align*}
Since the first terms in the right hand sides coincide, it is enough to check the equality of the second terms of the right hand sides:
\begin{multline*}
 \sum_{a\in H}(-1)^a \tr(\d V'_{a^*}\wedge\d V'_a)=
 -\lambda_k\sum_{a\in H} \tr\big(\pi_a\d\mu'\wedge\d(\pi')\mu_a\big)= \\
 =-\lambda_k \tr\big(\d\mu'\wedge\d(\pi')\sum_{a\in H}\mu_a\pi_a\big)
 =-\lambda_k \tr\big(\d\mu'\wedge\d\pi'\big).
\end{multline*}
By substituting~\eqref{directsum3} we obtain
\begin{align*}
 \tr\big(\d\mu'\wedge\d\pi'\big)=\tr\big((\mu\pi+\mu'\pi')\d\mu'\wedge\d\pi'\big)
=\tr\big(\pi\d\mu'\wedge\d(\pi')\mu\big)+\tr\big(\pi'\d\mu'\wedge\d(\pi')\mu'\big).
\end{align*}
By differentiating~\eqref{directsum1} and \eqref{directsum2}, we derive
\begin{align} \label{directsum23d}
\begin{aligned}
 \d(\pi)\mu&=-\pi\d\mu, &&&&&\d(\pi)\mu'&=-\pi\d\mu', \\
 \d(\pi')\mu'&=-\pi'\d\mu', &&&&&\d(\pi')\mu&=-\pi'\d\mu. \\
\end{aligned}
\end{align}
By using~\eqref{directsum23d},~\eqref{directsum3} and the general formulae
\begin{align*}
 &\tr\big(x\d y\wedge x\d y\big)=0, &&\tr\big(\d(y)x\wedge\d(y)x\big)=0
\end{align*}
(where $x,y$ are matrix-valued functions), one yields
\begin{multline*}
 \tr\big(\d\mu'\wedge\d\pi'\big)=\tr\big(\d(\pi)\mu'\pi'\wedge\d\mu\big)-\tr\big(\d(\pi')\mu'\wedge\d(\pi')\mu'\big)= \\
 =\tr\big(\d(\pi)(1-\mu\pi)\wedge\d\mu\big)
 =\tr\big(\d\pi\wedge\d\mu\big)+\tr\big(\pi\d\mu\wedge\pi\d\mu\big)
 =\tr\big(\d\pi\wedge\d\mu\big).
\end{multline*}
Thus, by substituting~\eqref{mupi} we obtain
\begin{multline*}
 \sum_{a\in H}(-1)^a \tr(\d V'_{a^*}\wedge\d V'_a)
  =-\lambda_k \tr\big(\d\mu'\wedge\d\pi'\big)=-\lambda_k \tr\big(\d\pi\wedge\d\mu\big)= \\
=-\hspace{-6pt}\sum_{a,b\in H}(-1)^a\tr\big(\d(V_a)\pi_a\wedge\mu_b\d V_{b^*}\big)
=-\hspace{-2pt}\sum_{a\in H}(-1)^a\tr\big(\d V_a\wedge\d V_{a^*}\big)
=\sum_{a\in H}(-1)^a\tr\big(\d V_{a^*}\wedge\d V_a\big),
\end{multline*}
where we used $\pi_a\mu_b=\delta_{ab}$. \qed

\begin{Rem}
 In the case of quiver varieties (see Section~\ref{secKolMn}) the statement of the theorem~\ref{ThIsom} was proved by Nakajima~\cite{N03}.
\end{Rem}

\begin{Rem} \label{RemCQIsom}
The results obtained in this section can be generalised to the case $N_\lambda(\alpha)=\Rep\big(\Pi^\lambda(Q),\alpha\big)//G(\alpha)$ described in Remark~\ref{RemCQHR}. Theorem~\ref{REquiv} implies that the functor $F^\lambda_k$ maps semi-simple modules and only them to semi-simple ones. Hence, even if the condition of Theorem~\ref{ThSmooth} is not satisfied then $F^\lambda_k$ induces the bijection $\varrho_k^{\lambda,\alpha}\colon N_\lambda(\alpha)\to N_{r_k\lambda}(s_k\alpha)$. The fact that this bijection is an isomorphism of affine varieties is proven in the same way (see also Proposition~\ref{PropIsomApp}). In this case the variety $N_\lambda(\alpha)$ is smooth if and only if $N_{r_k\lambda}(s_k\alpha)$ is smooth (if $\lambda_k\ne0$), and the isomorphism $\varrho_k^{\lambda,\alpha}$ preserves the symplectic structure.
\end{Rem}

\begin{Rem}
The proof of the regularity of $\varrho^{\lambda,\alpha}_k$, given above, uses non-trivial facts: Theorem~\ref{ThLeBP} and Proposition~\ref{PropTr}, based on this theorem. In Appendix~\ref{appRRep} an elementary proof of the regularity of $\varrho^{\lambda,\alpha}_k$ is given (it is given in more general case described in Remark~\ref{RemCQIsom}).
\end{Rem}

\section{Hamiltonian systems on quiver varieties}
\label{secKolMnIntSys}

Quiver varieties are particular case of the moduli spaces of representations of the preprojective algebras. By using their special form we can define Hamiltonian systems on them. The reflection functor is applied to the quiver varieties as to the moduli spaces and transforms the Hamiltonian systems to each other.

\subsection{Quiver varieties}
\label{secKolMn}

We define quiver varieties by following~\cite{CB01}.

Consider first the framing construction: for a given quiver $Q=(I,E)$ and a vector $\zeta=(\zeta_i)\in\ZZ_{\ge0}^I$ we define the {\it framed quiver} $Q_\zeta$ as follows. Add a new vertex, which we denote by $\infty$. For each vertex $i\in I$ we also add $\zeta_i$ new edges from $\infty$ to $i$. In this way we obtain a new quiver $Q_\zeta=(I_\infty,E_\zeta)$, where
\begin{align}
 &I_\infty=\{\infty\}\sqcup I,  &&E_\zeta=E\sqcup\{b_{ir}\colon\infty\to i\mid i\in I,\;r=1,\ldots,\zeta_i\}.
\end{align}
Given vectors $\alpha\in\ZZ_{\ge0}^I$ and $\lambda\in\CC^I$ can be uniquely extended to the vectors $\aalpha\in\ZZ_{\ge0}^{I_\infty}$ and $\llambda\in\CC^{I_\infty}$ such that $\aalpha_\infty=1$ and $\llambda\cdot\aalpha=0$ as follows:
\begin{align} \label{aalphallambda}
 &\aalpha=(1,\alpha), &&\llambda=(-\lambda\cdot\alpha,\lambda).
\end{align}
Since $\aalpha_\infty=1$, one has $\GL(\aalpha)\simeq\GL(\alpha)\times\CC^\times$. This isomorphism implies $G(\aalpha)\simeq\GL(\alpha)$. The {\it quiver variety} $M_\lambda(\alpha,\zeta)$ is defined as the moduli space $N_\llambda(\aalpha)$ for the quiver~$Q_\zeta$:
\begin{align}
 M_\lambda(\alpha,\zeta)=\Rep\big(\Pi^\llambda(Q_\zeta),\aalpha\big)/G(\aalpha)=\Rep\big(\Pi^\llambda(Q_\zeta),\aalpha\big)/\GL(\alpha).
\end{align}

A module $V\in\Rep(\overline Q_\zeta,\aalpha)$ belongs to $\Rep\big(\Pi^\llambda(Q_\zeta),\aalpha\big)$ if and only if
\begin{align} \label{VPiI}
 &\sum_{a\in\overline Q_\zeta,j\in I\atop a\colon j\to i}(-1)^a V_aV_{a^*}\equiv\sum_{a\in\overline Q,j\in I\atop a\colon j\to i}(-1)^a V_aV_{a^*}+\hspace{-2pt}\sum_{r=1}^{\zeta_i}V_{b_{i,r}}V_{b_{i,r}^*}=\lambda_i\id_{V_i}, && i\in I.
\end{align}
For $i=\infty$ the corresponding condition
\begin{align} \label{VPiinf}
 \sum_{a\in\overline Q_\zeta,j\in I\atop a\colon j\to\infty}(-1)^a V_aV_{a^*}\equiv-\hspace{-2pt}\sum_{j\in I}\sum_{r=1}^{\zeta_j}V_{b_{j,r}^*}V_{b_{j,r}}=\llambda_\infty\equiv-\lambda\cdot\alpha
\end{align}
follows from~\eqref{VPiI}: one needs to take trace of~\eqref{VPiI} and to sum over $i\in I$.

We call a vector $\lambda\in\CC^I$ to be {\it regular}, if $\lambda\cdot\alpha\ne0$ for any $\alpha\in\Delta(Q)$ (see~\cite{BCE}).

\begin{Lem}
 If the vector $\lambda\in\CC^I$ is regular, then every module $V\in\Rep\big(\Pi^\llambda(Q_\zeta),\aalpha\big)$ is simple.
\end{Lem}

\noindent{\bf Proof.} Since $\dim_\CC V=|\aalpha|\ge1$, the module $V$ is not zero. If $V$ is not simple, then there exists a non-trivial $\Pi^\llambda(Q_\zeta)$-submodule $V^{(1)}\subset V$. Let $V^{(2)}=V/V^{(1)}$ is the quotient module, $\aalpha^{(1)}=\dim_{\CC I_\infty}V^{(1)}$, $\aalpha^{(2)}=\dim_{\CC I_\infty}V^{(2)}$. Then $\aalpha^{(l)}=(0,\alpha^{(l)})$ for some $l\in\{1,2\}$ and $V^{(l)}$ is a $\Pi^\lambda(Q)$-module. Since $\alpha^{(l)}\ne0$ there exists an indecomposable $\CC Q$-module summand $V'$ of $ V^{(l)}$. Theorem~\ref{ThKac} implies $\alpha':=\dim_{\CC I}V'\in\Delta(Q)$ and by virtue of Theorem~\ref{ThLift} one has $\lambda\cdot\alpha'=0$, but this contradicts to the regularity of $\lambda$. \qed

Now let us apply Theorem~\ref{ThSmooth}.

\begin{Th} \label{ThReg}
 If the vector $\lambda\in\CC^I$ is regular and $\Rep\big(\Pi^\llambda(Q_\zeta),\aalpha\big)\ne\emptyset$, then the quiver variety $M_\lambda(\alpha,\zeta)=\Rep\big(\Pi^\llambda(Q_\zeta),\aalpha\big)/\GL(\alpha)$ is a smooth connected affine variety of the dimension 
\begin{align} \label{dimKolMn}
2p(\aalpha)=2\zeta\cdot\alpha-2q(\alpha)=2\sum\limits_{i\in I}\zeta_i\alpha_i+2\hspace{-5pt}\sum\limits_{a\colon i\to j}\alpha_i\alpha_j-2\sum\limits_{i\in I}\alpha_i^2.
\end{align}
\end{Th}

Note that the condition of this theorem implies $\aalpha\in\Delta^+(Q)$. Below we suppose that $\lambda$ is regular and $\aalpha=(1,\alpha)$ is a positive root unless otherwise indicated. In accordance to Corollary~\ref{ThLiftCor} this implies non-emptiness of $\Rep\big(\Pi^\llambda(Q_\zeta),\aalpha\big)$ and, hence validity of the condition of Theorem~\ref{ThReg}. Note also that $\dim M_\lambda(\alpha,\zeta)\ge1$ if and only if $\aalpha\in\Delta_{im}^+(Q_\zeta)$.

\subsection{Hamiltonians and integrals of motion}
\label{secHamKolMn}

Now we define Hamiltonian systems on the quiver varieties by generalising ideas proposed in~\cite{ChS}. These systems are given by Poisson commuting Hamiltonians. We also consider more general integrals of motions.

Let $p=a_\ell\cdots a_2a_11_{i_0}\in\CC\overline Q$, where $a_k\colon i_{k-1}\to i_k$. We say that $p$ is a path in $Q^*$ if $a_k^*\in Q$ $\;\forall\,k=1,\ldots,\ell$ and that $p$ is a cycle in $Q^*$ if, moreover, $i_\ell=i_0$. Denote by $\mathsf P_{i_0,i_\ell}$ the set of all paths in $Q^*$ from a fixed vertex $i_0$ to a fixed vertex $i_\ell$. Then the set of all cycles $Q^*$ from $i_0=i_\ell$ is $\mathsf P_{i_0}:=\mathsf P_{i_0,i_0}$ and the set of all paths in $Q^*$ is $\mathsf P:=\bigsqcup\limits_{i,j\in I}\mathsf P_{ij}$.

Consider a module $V\in\Rep(\Pi^\llambda(Q_\zeta),\aalpha)$; for $p=a_\ell\cdots a_2a_11_i\in\mathsf P_{ij}$ we use the notation $V_p:=V_{a_\ell}\cdots V_{a_2}V_{a_1}\id_{\CC^{\alpha_i}}\in\Hom(\CC^{\alpha_i},\CC^{\alpha_j})$. For cycles $p$ in $Q^*$ define functions $H_p\in\CC\big[M_\lambda(\alpha,\zeta)\big]=\CC\big[\Rep(\Pi^\llambda(Q_\zeta),\aalpha)\big]^{\GL(\alpha)}$ by the formula
\begin{align} \label{Hp}
 &H_p(V)=\tr(V_p), &&p\in\bigsqcup_{i\in I}\mathsf P_i.
\end{align}
If $p$ and $p'$ are two cycles in $Q^*$ that differ by a cyclic permutation of edges, then $H_p=H_{p'}$. Due to Proposition~\ref{propPoisson} the Poisson brackets of functions on $M_\lambda(\alpha,\zeta)$ coincides with the Poisson brackets of the corresponding (invariant) functions on $\Rep(\Pi^\llambda(Q_\zeta),\aalpha)$. Since the entries of the matrix $V_{a^*}$ Poisson commute with entries of $V_{b*}$ for any $a,b\in Q$, we obtain $\{H_p,H_{p'}\}=0$ for any cycles $p$ and $p'$ in $Q^*$. In this way one yields a Poisson commuting family of the Hamiltonians on $M_\lambda(\alpha,\zeta)$, but sometimes it is not enough for integrability.

On the space $\mathsf L_\zeta:=\oplus_{i,j\in I\atop p\in\mathsf P_{ij}}\Hom(\CC^{\zeta_j},\CC^{\zeta_i})$ we define a structure of Lie algebra:
\begin{align}
 &[A,B]_p=\sum_{p',p''\in\mathsf P \atop p''p'=p} \big(A_{p''}B_{p'}-B_{p''}A_{p'}\big), &&p\in\mathsf P,
\end{align}
where $A_p$ means the corresponding component of an element $A\in\mathsf L_\zeta$. Further, for a given module $V\in\Rep(\Pi^\llambda(Q_\zeta),\aalpha)$ introduce the matrices $v_i\in\Hom(\CC^{\zeta_i},\CC^{\alpha_i})$, $w_i\in\Hom(\CC^{\alpha_i},\CC^{\zeta_i})$, $i\in I$, with entries $(v_i)_{l,r}=(V_{b_{ir}})_l$ and $(w_i)_{r,l}=(V_{b_{ir}^*})_l$. We write in this case $V=(V_a,v_i,w_i)$, where $a\in\overline Q$ and $i\in I$. In these terms we define functions $I_A\in\CC\big[M_\lambda(\alpha,\zeta)\big]$ as
\begin{align} \label{IA}
 &I_A(V)=-\hspace{-4pt}\sum_{i,j\in I} \sum_{\;\;p\in\mathsf P_{ij}}\tr(A_pw_jV_pv_i), && A\in\mathsf L_\zeta.
\end{align}
Their Poisson brackets are calculated in Appendix~\ref{appPoisson}:
\begin{align}
 &\{I_A,I_B\}=I_{[A,B]}, &&A,B\in\mathsf L_\zeta. \label{IAIB}
\end{align}
Thus, we obtain a Lie algebra of integrals of motion for the Hamiltonians~\eqref{Hp}:
\begin{align}
\{H_p,I_A\}=0
\end{align}
(this formula is proven in the same way as the Poisson commutativity of $H_p$). Below we will be interested in commuting subalgebras in the Lie algebra spanned by~\eqref{Hp}, \eqref{IA}.

Let $E_r\in\Hom(\CC^n,\CC^m)$ be a rectangular matrix with the entries $(E_r)_{st}=\delta_{rs}\delta_{rt}$ (we assume that $E_r=0$ whenever $r>\min(n,m)$). Define the elements $E^{(\ell)}_r\in\mathsf L_\zeta$, $\ell\in\ZZ_{\ge0}$, $r\in\ZZ_{\ge1}$, by setting $(E^{(\ell)}_r)_p=\delta_{\ell,|p|}E_r$, where $|p|$ is the length of the path $p$. These elements pairwise commute in $\mathsf L_\zeta$:
\begin{multline*}
 [E^{(k)}_r,E^{(\ell)}_s]_p=\sum_{p',p''\in\mathsf P\atop p''p'=p}\big((E^{(k)}_r)_{p''}(E^{(\ell)}_s)_{p'}-(E^{(\ell)}_s)_{p''}(E^{(k)}_r)_{p'}\big)= \\
=\delta_{rs}\sum_{p',p''\in\mathsf P\atop p''p'=p}(\delta_{k,|p''|}\delta_{\ell,|p'|}-\delta_{\ell,|p''|}\delta_{k,|p'|})E_r=\delta_{rs}(\delta_{k+\ell,|p|}-\delta_{k+\ell,|p|})E_r=0
\end{multline*}
(here we used the fact that a path of the length $k+\ell$ can be uniquely divided into a product of two paths of the lengths $k$ and $\ell$ respectively). Thus we derive the following statement.

\begin{Th} \label{ThHamil}
 The functions~\eqref{Hp} and
\begin{multline} \label{Hkr}
H_{\ell,r}(V)=I_{E^{(\ell)}_r}(V)=-\hspace{-4pt}\sum_{i,j\in I}\sum_{p\in\mathsf P_{ij}\atop |p|=\ell}\tr(E_rw_jV_pv_i), \\
\ell,r\in\ZZ,\quad \ell\ge0,\quad 1\le r\le\max\limits_{i\in I}\zeta_i,\qquad
\end{multline}
form a family of Poisson commuting Hamiltonians on the quiver variety $M_\lambda(\alpha,\zeta)$:
\begin{align}
 &\{H_p,H_{p'}\}=0, &&\{H_p,H_{k,r}\}=0, &&\{H_{k,r},H_{\ell,s}\}=0.
\end{align}
\end{Th}

\begin{Rem} \label{RemHamil0}
 For $\ell=0$ the Hamiltonians~\eqref{Hkr} are $H_{0,r}=-\hspace{-2pt}\sum\limits_{i\in I}\tr(E_rw_iv_i)$. Due to the equation~\eqref{VPiinf} they  are related as $\sum_r H_{0,r}=-\lambda\cdot\alpha$.
\end{Rem}

\begin{Rem}
 The choice of the Hamiltonians~\eqref{Hkr} is not unique. We define them in this way to obtain integrable systems in the case of cyclic quiver, but sometimes it does not give integrability, for example, for the trees (when $Q$ does not have cycles). In the case of trees there is another commuting subalgebra in $\mathsf L_\zeta$, which gives an integrable family of Hamiltonians of the form~\eqref{IA}.
\end{Rem}

For general $\llambda\in\CC^{I_\infty}$ and $\aalpha\in\ZZ_{\ge0}^{I_\infty}$ such that $\llambda\cdot\aalpha=0$ we obtain the moduli space $N_\llambda(\aalpha)$ for the quiver $Q_\zeta$. One can define $H_p$ and $I_A$ on $N_\llambda(\aalpha)$ by the formulae~\eqref{Hp} and \eqref{IA}, but in the latter one, one should consider $v_i\in\Hom(\CC^{\aalpha_\infty}\otimes\CC^{\zeta_i},\CC^{\alpha_i})$ and $w_i\in\Hom(\CC^{\alpha_i},\CC^{\aalpha_\infty}\otimes\CC^{\zeta_i})$ as matrices with entries $(v_i)_{l,kr}=(V_{b_{ir}})_{lk}$ and $(w_i)_{kr,l}=(V_{b_{ir}^*})_{kl}$, where $i\in I$, $l=1,\ldots,\alpha_i$, $k=1,\ldots,\aalpha_\infty$, $r=1,\ldots,\zeta_i$.  In this case the formula~\eqref{IAIB} is also valid (see Appendix~\ref{appPoisson}).

Let $M$ be a Poisson complex manifold and $z_i$ be local complex coordinates. Each holomorphic function $H\colon M\to\CC$ defines local flows $z_i=z_i(t)$ by the system of equations $\dfrac{\d z_i}{\d t}=\{H,z_i\}$. These flows differ by the initial values $z_i(0)$. Remind that a function $H$ {\it defines complete flows} if each local flow for this function can be extended to a global flow. The latter is a holomorphic map $\varphi\colon\CC\to M$ such that for any chart $U\subset M$ with coordinates $z_i\colon U\to\CC$ the functions $z_i(t)=z_i\big(\varphi(t)\big)$ satisfy the equations above (we denote by $z_i(p)$ the value of the coordinate $z_i$ at the point $p\in M$).

\begin{Th}
 The functions~\eqref{Hp}, \eqref{IA} define complete flows on $M_\lambda(\alpha,\zeta)$ (and on a general $N_\llambda(\aalpha)$ for $Q_\zeta$).
\end{Th}

\noindent{\bf Proof.} The functions $H_p$ and $I_A$ are defined on $N_\llambda(\aalpha)$ as restrictions of corresponding functions on $\Rep(\overline Q_\zeta,\aalpha)$ to the subvariety $\Rep\big(\Pi^\llambda(Q_\zeta),\aalpha\big)=P_\aalpha^{-1}(\llambda)$. Due to $G(\alpha)$-invariance of $H_p$ and $I_A$ they Poisson commute with the moment map $P_\aalpha$. Hence the flows they define preserve this subvariety and induce the corresponding flows on $N_\llambda(\aalpha)$. This means that it is enough to prove that the Hamiltonians~\eqref{Hp}, \eqref{IA} define complete flows on the affine space $\Rep(\overline Q_\zeta,\aalpha)$. Consider first the case of the function $H_p$. Since $\{H_p,w_i\}$, $\{H_p,v_i\}$ and $\{H_p,V_{a^*}\}$ vanish for any $i\in I$ and $a\in Q$, the coordinates $w_i=w_i(t)$, $v_i=v_i(t)$, $V_{a^*}=V_{a^*}(t)$ are constant. This implies that $\{H_p,V_a\}$ is constant for any $a\in Q$, hence $V_a(t)=V_a(0)+t\{H_p,V_a\}$. Now consider the function $I_A$. The coordinates $V_{a^*}$ are constant in this case as well. Let $W\in\CC^{\aalpha_\infty(\alpha\cdot\xi)}$ be a vector with components $W_{ikrl}=(w_i)_{kr,l}$. The system of equations for $w_i=w_i(t)$ has the form
\begin{align} \label{wiIAFlow}
 \frac{\d w_i}{\d t}=\{I_A,w_i\}=-\hspace{-2pt}\sum_{j\in I}\sum_{\;\;p\in\mathsf P_{ij}}A_pw_jV_p
\end{align}
(see the end of Appendix~\ref{appPoisson}). The system~\eqref{wiIAFlow} can written as $\frac{\d W}{\d t}=\Omega W$ for some constant matrix $\Omega\in\Mat_{\aalpha_\infty(\alpha\cdot\xi)}(\CC)$. Its solution is $W(t)=\exp(t\Omega)W(0)$, so we obtain an entire solution $w_i=w_i(t)$ of the system~\eqref{wiIAFlow} for any initial values. In the same way we obtain an entire solution $v_i=v_i(t)$. Then we have the equations
\begin{align} \label{VaIAFlow}
 &\frac{\d(V_a)_{ll'}}{\d t}=\{I_A,(V_a)_{ll'}\}=-\hspace{-4pt}\sum_{i,j\in I} \sum_{\;\;p\in\mathsf P_{ij}}\tr\big(A_pw_j(t)\{V_p,(V_a)_{ll'}\}v_i(t)\big), &&a\in Q,
\end{align}
where $\{V_p,(V_a)_{ll'}\}=const$. The right hand side of~\eqref{VaIAFlow} is an entire function $f_{a,l,l'}(t)$, so we can write the solution as $(V_a)_{ll'}(t)=(V_a)_{ll'}(0)+\int\limits_0^tf_{a,l,l'}(t)\d t$. The integral does not depend on the path from $0$ to $t$ in the complex plane and it depends on $t$ as an entire function. \qed

\subsection{Isomorphisms of quiver varieties by the reflection functor }
\label{secKolMnRF}

Here we investigate how the reflection functor transforms the Hamiltonians $H_p$ and integrals~$I_A$, which are given on the quiver varieties as well as on more general moduli spaces associated with the framed quiver~$Q_\zeta$.

Note first that $\Iinflf=\{\infty\}\sqcup\Ilf$. Since $\varepsilon_i\in\Delta(Q)$, the regularity of $\lambda$ implies that $\lambda_i=\lambda\cdot\varepsilon_i\ne0$ for any $i\in\Ilf$. Hence the reflection is admissible at any vertex $i\in\Ilf$ for the parameter $\llambda=(-\lambda\cdot\alpha,\lambda)$. However, after the reflection at a vertex $i\in\Ilf$ we obtain $s_i\aalpha=(1,\alpha')$ and $r_i\llambda=(-\lambda'\cdot\alpha',\lambda')$, where the new vector $\lambda'\in\CC^I$ can be irregular (if $\zeta_i\ne0$). Moreover, there is no guarantee that the reflection at the vertex $\infty$ is admissible.

Let $\llambda',\llambda''\in\CC^{I_\infty}$ and $\aalpha',\aalpha''\in\ZZ_{\ge0}^{I_\infty}$. We will say that the pair $(\llambda'', \aalpha'')$ is obtained from $(\llambda', \aalpha')$ by a chain of admissible reflections, if $(\llambda'', \aalpha'')$ can be obtained from $(\llambda', \aalpha')$ by a finite number of transformations of the form
\begin{align} \label{preobr}
&(\llambda,\aalpha)\mapsto (r_k\llambda,s_k\aalpha), &&k\in\Iinflf, \quad \llambda_k\ne0.
\end{align}
We will be interested in the case of quiver varieties, i.e. the case when the initial and final pairs have the form~\eqref{aalphallambda} for regular vectors $\lambda$ (it is enough to require the regularity of this vector for the initial (or final) pair), but for the intermediate pairs the component $\aalpha_\infty$ may differ from $1$. Due to Proposition~\ref{PropEquivCor} all the modules $V\in\Rep\big(\Pi^\llambda(Q_\zeta),\aalpha\big)$ are simple for any intermediate pair $(\llambda,\aalpha)$ and, hence, $N_\llambda(\aalpha)$ is smooth and connected.

For every transformation~\eqref{preobr} we have a symplectic isomorphism $N_\llambda(\aalpha)\simeq N_{r_k\llambda}(s_k\aalpha)$ given by Theorem~\ref{ThIsom}. Let us find out how the functions $H_p$ and $I_A$ are transformed under this isomorphism. To do it we represent them as linear combinations of the functions $\tr_\aalpha(p)_\llambda$ (see the end of Section~\ref{secHamRed}):
\begin{align}
 &H_p=\tr_\aalpha(p)_\llambda, &&p\in\bigsqcup_{i\in I}\mathsf P_i; \\
 &I_A=-\hspace{-4pt}\sum_{i,j\in I}\sum_{p\in\mathsf P_{ij}}\sum_{r=1}^{\zeta_i}\sum_{s=1}^{\zeta_j}(A_p)_{rs}\tr_\aalpha(b_{js}^*pb_{ir})_\llambda, &&A\in\mathsf L_\zeta. \label{IAtr}
\end{align}
The function $f\in\CC\big[N_\llambda(\aalpha)\big]$ is mapped to $f\circ(\varrho^{\llambda,\aalpha}_k)^{-1}\in\CC\big[N_{r_k\llambda}(s_k\aalpha)\big]$. The item (b) of Lemma~\ref{LemTr} implies that the functions $H_p=\tr_\aalpha(p)_\llambda$ are always mapped to $H_p=\tr_{s_k\aalpha}(p)_{r_k\llambda}$.

Further, we need to obtain the image of the terms in the formula~\eqref{IAtr}. First consider the case $k\ne\infty$. For the paths $p\in\mathsf P_{ij}$ such that $|p|\ge1$ the functions $\tr_\aalpha(b_{js}^*pb_{ir})_\llambda$ are mapped to $\tr_{s_k\aalpha}(b_{js}^*pb_{ir})_{r_k\llambda}$. If $|p|=0$ then $p=1_i$ for some $i\in I$. In the case when $i\ne k$ or $r\ne s$ the functions $\tr_\aalpha(b_{is}^*1_ib_{ir})_\llambda=\tr_\aalpha(b_{is}^*b_{ir})_\llambda$ are mapped to $\tr_{s_k\aalpha}(b_{is}^*b_{ir})_{r_k\llambda}$. If $i=k$ and $r=s$ one has
\begin{align}
 \tr_\aalpha(b_{kr}^*b_{kr})_\llambda\circ(\varrho^{\llambda,\aalpha}_k)^{-1}(V')=\tr_\aalpha(b_{kr}^*b_{kr})_\llambda(V)=\tr(V_{b_{kr}^*}V_{b_{kr}}),
\end{align}
where $V'\simeq F^{\llambda}_k(V)$. By the formula~\eqref{VaVaInTr} we derive
\begin{align}
\tr(V_{b_{kr}^*}V_{b_{kr}})=\tr(V'_{b_{kr}^*}V'_{b_{kr}}+\llambda_k)=\tr_{s_k\aalpha}(b_{kr}^*b_{kr})_{r_k\llambda}(V')+\llambda_k\aalpha_\infty.
\end{align}
Thus, $\tr_\aalpha(b_{ir}^*b_{is})_\llambda\circ(\varrho^{\llambda,\aalpha}_k)^{-1}=\tr_{s_k\aalpha}(b_{ir}^*b_{is})_{r_k\llambda}+\delta_{ik}\delta_{rs}\llambda_k\aalpha_\infty$ for any $i\in I$, $r,s=1,\ldots,\zeta_i$.

Analogously, for $k=\infty$ and $p\in\mathsf P_i$ we obtain
\begin{align*}
 \tr_\aalpha(b_{ir}^*pb_{ir})_\llambda\circ(\varrho^{\llambda,\aalpha}_\infty)^{-1}(V')=\tr(V_pV_{b_{ir}}V_{b_{ir}^*})=\tr(V'_pV'_{b_{ir}}V'_{b_{ir}^*}-\llambda_\infty V'_p).
\end{align*}
Hence
\begin{align*}
 \tr_\aalpha(b_{js}^*pb_{ir})_\llambda\circ(\varrho^{\llambda,\aalpha}_\infty)^{-1}=\tr_{s_\infty\aalpha}(b_{js}^*pb_{ir})_{r_\infty\llambda}-\delta_{ij}\delta_{rs}\llambda_\infty\tr_{s_\infty\aalpha}(p)_{r_\infty\llambda}
\end{align*}
for any $p\in\mathsf P_{ij}$. In particular, for $p=1_i$ one yields
\begin{align}
 \tr_\aalpha(b_{is}^*b_{ir})_\llambda\circ(\varrho^{\llambda,\aalpha}_\infty)^{-1}=\tr_{s_\infty\aalpha}(b_{is}^*b_{ir})_{r_\infty\llambda}-\delta_{rs}\llambda_\infty\aalpha_i.
\end{align}

By taking into account these formulae we obtain the following proposition.

\begin{Prop}
 Under the map $f\mapsto f\circ(\varrho^{\llambda,\aalpha}_k)^{-1}$ the image of the algebra $\CC[H_p,I_A]$ generated by~\eqref{Hp}, \eqref{IA} and the image of its subalgebra $\CC[H_p,H_{\ell,r}]$ coincide with the algebras $\CC[H_p,I_A]$ and $\CC[H_p,H_{\ell,r}]$ respectively. More precisely,
\begin{align}
 &H_p\circ(\varrho^{\llambda,\aalpha}_k)^{-1}=H_p, &&p\in\bigsqcup_{i\in I}\mathsf P_i, &&k\in\Iinflf; \label{Hprho} \\
 &I_A\circ(\varrho^{\llambda,\aalpha}_k)^{-1}=I_A-\llambda_k\aalpha_\infty\tr(A_{1_k}), &&A\in\mathsf L_\zeta, &&k\in\Ilf; \\
 &I_A\circ(\varrho^{\llambda,\aalpha}_\infty)^{-1}=I_A+\llambda_\infty\sum_{i\in I}\sum_{p\in\mathsf P_i}\tr(A_p)H_p, &&A\in\mathsf L_\zeta, &&(k=\infty);
\end{align}
\begin{align}
 &H_{0,r}\circ(\varrho^{\llambda,\aalpha}_k)^{-1}=H_{0,r}-\llambda_k\aalpha_\infty, &&r\le\zeta_k &&k\in\Ilf; \\
 &H_{0,r}\circ(\varrho^{\llambda,\aalpha}_k)^{-1}=H_{0,r}, &&r>\zeta_k &&k\in\Ilf; \\
 &H_{0,r}\circ(\varrho^{\llambda,\aalpha}_\infty)^{-1}=H_{0,r}+\llambda_\infty\sum_{i\in I\atop r\le \zeta_i}\aalpha_i, &&\forall\,r &&(k=\infty); \\
 &H_{\ell,r}\circ(\varrho^{\llambda,\aalpha}_k)^{-1}=H_{\ell,r}, &&\ell\ge1, &&k\in\Ilf; \\
 &H_{\ell,r}\circ(\varrho^{\llambda,\aalpha}_\infty)^{-1}=H_{\ell,r}+\llambda_\infty\sum_{i\in I\atop r\le\zeta_i}\sum_{p\in\mathsf P_i\atop|p|=\ell}H_p, &&\ell\ge1, &&(k=\infty). \label{Hlrrhoinf}
\end{align}
\end{Prop}

Application of this proposition to the case we interested in leads us to the following theorem.

\begin{Th} \label{ThKolMnRF}
 Let $\lambda',\lambda''\in\CC^I$ be regular vectors (it is enough to suppose either $\lambda'$ or $\lambda''$ to be regular). Let $\zeta,\alpha',\alpha''\in\ZZ_{\ge0}^I$, $\llambda'=(-\lambda'\cdot\alpha',\lambda')$, $\llambda''=(-\lambda''\cdot\alpha'',\lambda'')$, $\aalpha'=(1,\alpha')$, $\aalpha''=(1,\alpha'')$. Suppose that the pair $(\llambda'', \aalpha'')$ is obtained from $(\llambda', \aalpha')$ by a chain of admissible reflections. The corresponding isomorphism of quiver varieties $\varrho\colon M_{\lambda'}(\alpha',\zeta)\simeq M_{\lambda''}(\alpha'',\zeta)$ induces an isomorphism of Poisson algebras of regular functions: $f\mapsto f\circ\varrho^{-1}$, $f\in\CC\big[M_{\lambda'}(\alpha',\zeta)\big]$. Subalgebras $\CC[H_p,H_{\ell,r}]\subset\CC[H_p,I_A]\subset\CC\big[M_{\lambda'}(\alpha',\zeta)\big]$ are mapped under this isomorphism to $\CC[H_p,H_{\ell,r}]\subset\CC[H_p,I_A]\subset\CC\big[M_{\lambda''}(\alpha'',\zeta)\big]$ respectively. More precisely, this map on these subalgebras is given by the formulae~\eqref{Hprho}--\eqref{Hlrrhoinf} for each reflection~\eqref{preobr}. In particular, the Hamiltonians $H_p$ are always mapped to $H_p$.
\end{Th}

\section{Cyclic quiver}
\label{secCykl}

In this section we consider the case of cyclic quiver, described in~\cite{ChS}. First we briefly consider the general framing. Then we describe cases of some particular framings in details.

\subsection{The quiver varieties and Hamiltonian systems for the cyclic quiver}
\label{secHamSysCycl}

The {\it cyclic quiver} with $m\ge1$ vertices is $Q=(I,E)$ where $I=\ZZ/m\ZZ=\{0,1,\ldots,m-1\}$ and $E=\{a_i\colon i\to i+1\mid i\in I\}$:
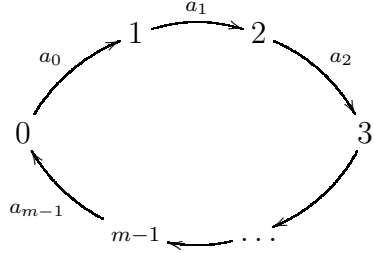
\begin{align*}
\xymatrix{& 1\ar@/^0.4pc/[r]^{a_1} & 2\ar@/^/[dr]^{a_2} & \\
0\ar@/^/[ur]^{a_0} &&&3\ar@/^/[dl] \\
&{\scriptstyle m-1}\ar@/^/[ul]^{a_{m-1}}&\ldots\ar@/^0.4pc/[l]&
}
\end{align*}
We use notations $\delta:=\sum\limits_{i\in I}\varepsilon_i$ and $|\lambda|:=\sum\limits_{i\in I}\lambda_i$ where $\lambda=(\lambda_0,\ldots,\lambda_{m-1})\in\ZZ_{\ge0}^I$. Since $\delta\in\Delta_{im}(Q)$, the regularity of $\lambda$ implies $|\lambda|\ne0$.

The framing is defined by a vector $\zeta=(\zeta_0,\ldots,\zeta_{m-1})\in\ZZ_{\ge0}^m$ and the framed quiver $Q_\zeta$ has the form
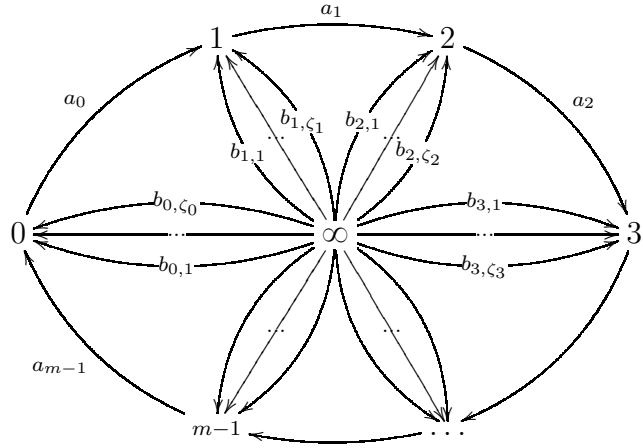
\begin{align*}
\xymatrix{&& 1\ar@/^/[rr]^{a_1} && 2\ar@/^1pc/[ddrr]^{a_2} && \\
{\phantom{\infty}}\\
0\ar@/^1pc/[uurr]^{a_0} &&&\infty\ar@/^1pc/[lll]|{b_{0,1}}\ar[lll]|{\ldots}\ar@/_1pc/[lll]|{b_{0,\zeta_0}} \ar@/^1pc/[uul]|{b_{1,1}}\ar[uul]|{\ldots}\ar@/_1pc/[uul]|{b_{1,\zeta_1}} \ar@/^1pc/[uur]|{b_{2,1}}\ar[uur]|{\ldots}\ar@/_1pc/[uur]|{b_{2,\zeta_2}} \ar@/^1pc/[rrr]|{b_{3,1}}\ar[rrr]|{\ldots}\ar@/_1pc/[rrr]|{b_{3,\zeta_3}} \ar@/^1pc/[ddr]\ar[ddr]|{\ldots}\ar@/_1pc/[ddr] \ar@/^1pc/[ddl]\ar[ddl]|{\ldots}\ar@/_1pc/[ddl] &&&3\ar@/^1pc/[ddll] \\
{\phantom{\infty}}\\
&&{\scriptstyle m-1}\ar@/^1pc/[uull]^{a_{m-1}}&&\ldots\ar@/^/[ll]&&
}
\end{align*}

For a representation $V\in\Rep\big(\Pi^\llambda(Q_\zeta),\aalpha\big)$ we denote
\begin{align}
 &X_i=V_{a_i}\colon V_i\to V_{i+1}, &&Y_i=V_{a_i^*}\colon V_{i+1}\to V_i, 
\end{align}
where $V_i=\CC^{\alpha_i}$, $i\in I$. Then the equations~\eqref{VPi} take the form
\begin{align}
 X_{i-1}Y_{i-1}-Y_iX_i+v_iw_i&=\lambda_i\id_{V_i}, && i=0,1,\ldots,m-1,  \label{VPiXY}\\
 \sum_{i=0}^{m-1}\tr(w_iv_i)&=\lambda\cdot\alpha, \label{VPiwv}
\end{align}
where~\eqref{VPiwv} follows from~\eqref{VPiXY}. The integrals introduced in Section~\ref{secHamKolMn} are expressed through the following operators:
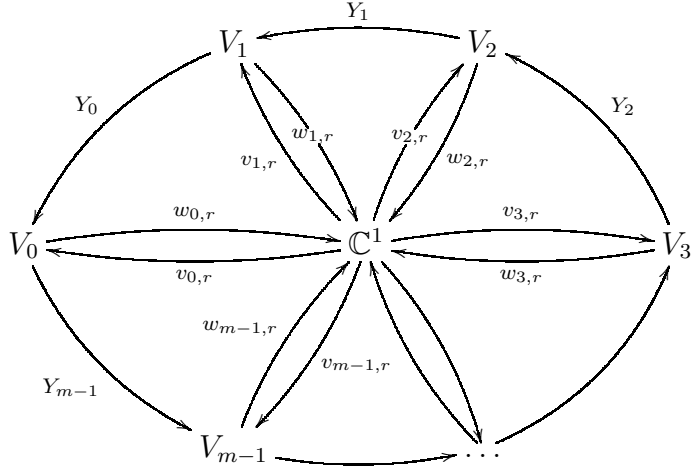
\begin{align*}
\xymatrix{&& V_1\ar@/_1pc/[ddll]_{Y_0}\ar@/^/[ddr]|{w_{1,r}} && V_2\ar@/_/[ll]_{Y_1}\ar@/^/[ddl]^{w_{2,r}} && \\
{\phantom{\infty}}\\
V_0\ar@/_1pc/[ddrr]_{Y_{m-1}}\ar@/^/[rrr]^{w_{0,r}} &&&\CC^1\ar@/^/[lll]^{v_{0,r}}\ar@/^/[uul]^{v_{1,r}}\ar@/^/[uur]|{v_{2,r}}\ar@/^/[rrr]^{v_{3,r}}\ar@/^/[ddr]\ar@/^/[ddl]^{v_{m-1,r}}&&&V_3\ar@/_1pc/[uull]_{Y_2}\ar@/^/[lll]^{w_{3,r}} \\
{\phantom{\infty}}\\
&&V_{\scriptstyle m-1}\ar@/_/[rr]\ar@/^/[uur]^{w_{m-1,r}}&&\ldots\ar@/_1pc/[uurr]\ar@/^/[uul]&&
}
\end{align*}
where $v_{i,r}=V_{b_{i,r}}$, $w_{i,r}=V_{b_{i,r}^*}$, $r=1,\ldots,\zeta_i$.

Let us note that in the case of cyclic quiver there exists a unique cycle $p\in\mathsf P_i$ of the length $m$, denote it by $p_i$. The possible lengths of cycles is divided by $m$ and the cycle $p\in\mathsf P_i$ of the length $mk$ is unique for each $k$, so that $\mathsf P_i=\{p_i^k\mid k\in\ZZ_{\ge0}\}$ (where $p_i^0=1_i$). The cycles $p_i^k\in\mathsf P_i$ and $p_j^k\in\mathsf P_j$ differ by a cyclic permutation. This means that the set of Hamiltonians~\eqref{Hp} is reduced to the family
\begin{align} \label{Hmk}
 &H_{mk}(V):=\tr(V_{p_0^k})=\tr(V_{p_0}^k)=\tr(Y_0Y_1\cdots Y_{mk-1}), &&k\in\ZZ_{\ge1}.
\end{align}
Analogously, we obtain $\mathsf P_{ij}=\{p_{ij}p_i^k=p_j^kp_{ij}\mid k\in\ZZ_{\ge0}\}$ where $p_{ij}$ is the shortest path belonging to $\mathsf P_{ij}$. Thus,
\begin{multline} \label{HlrC}
 H_{\ell,r}(V)=-\hspace{-2pt}\sum_{i\in I}\tr(E_rw_{i-\ell}V_{p_{i,i-\ell}}V_{p_i}^kv_i)=-\hspace{-2pt}\sum_{i\in I}w_{i-\ell,r}Y_{i-\ell}\cdots Y_{i-2}Y_{i-1}v_{i,r}, \\
 \ell\in\ZZ_{\ge0},\quad r\le\min(\zeta_i,\zeta_{i-\ell}),\quad
\end{multline}
where $k=(\ell-|p_{i,i-\ell}|)/m$ is the integer part of the ratio $\ell/m$. By summing the Hamiltonians $H_{mk,r}$ over $r=1,\ldots,\max\limits_{i\in I}\{\zeta_i\}$ and using~\eqref{VPiXY} we derive
\begin{multline*}
 \sum_{r\ge1}H_{mk,r}(V)=-\hspace{-2pt}\sum_{i\in I}\tr(w_iV_{p_i}^kv_i)=-\hspace{-2pt}\sum_{i\in I}\tr(V_{p_i}^kv_iw_i)
=\sum_{i\in I}\tr(V_{p_i}^kX_{i-1}Y_{i-1}-V_{p_i}^kY_iX_i-\lambda_iV_{p_i}^k)= \\
=\sum_{i\in I}\tr(V_{a_i^*}V_{p_{i+1}}^kX_i)-\hspace{-2pt}\sum_{i\in I}\tr(V_{p_i}^kV_{a_i^*}X_i)-\hspace{-2pt}\sum_{i\in I}\lambda_i\tr(V_{p_i}^k)=-|\lambda|\,H_{mk}(V),
\end{multline*}
where we used $a_i^*p_{i+1}^k=p_{i+1,i}p_{i+1}^k=p_i^kp_{i+1,i}=p_i^ka_i^*$. Since $|\lambda|\ne0$, the Hamiltonians~\eqref{Hmk} are expressed through the Hamiltonians~\eqref{HlrC} as
\begin{align} \label{HmkHmkr}
 H_{mk}=-\frac1{|\lambda|}\sum_{r\ge1}H_{mk,r}.
\end{align}

The elements $A\in\mathsf L_\zeta$ have the components $A_{(i,\ell)}:=A_p$, where $i\in I$, $\ell\in\ZZ_{\ge0}$ and $p\in\mathsf P_{i,i-\ell}$ is the path of the length $|p|=\ell$. In this terms the commutator in $\mathsf L_\zeta$ has the form
\begin{align}
[A,B]_{(i,\ell)}=\sum\limits_{\ell'=0}^\ell(A_{(i-\ell',\ell-\ell')}B_{(i,\ell')}-B_{(i-\ell',\ell-\ell')}A_{(i,\ell')}).
\end{align}
The integrals~\eqref{IA} take the form
\begin{align} \label{IACycl}
 I_A(V)=-\hspace{-2pt}\sum_{i=0}^{m-1}\sum_{\ell=0}^\infty\tr(A_{(i,\ell)}w_{i-\ell}Y_{i-\ell}\cdots Y_{i-1}v_i).
\end{align}

The following notations will be used in Section~\ref{secApplKP}. Let $\mathbf V:=\bigoplus\limits_{i=0}^{m-1}V_i$ be the direct sum with the canonical embeddings $\mu_i\colon V_i\hookrightarrow\mathbf V$ and projections $\pi_i\colon\mathbf V\twoheadrightarrow V_i$. Denote
\begin{align} 
 &\mathbf X=\sum_{i=0}^{m-1}\mu_{i+1}\circ X_i\circ\pi_i, &&\mathbf Y=\sum_{i=0}^{m-1}\mu_i\circ Y_i\circ\pi_{i+1}, \label{XYDef} \\
 &\mathbf v_i=\mu_iv_i, &&\mathbf w_i=w_i\pi_i, \label{vwDef}
\end{align}
where $i\in I$. The matrices
\begin{align}
 &\mathbf X,\mathbf Y\in\End(\mathbf V), &&\mathbf v_i\in\Hom(\CC^{\zeta_i},\mathbf V), &&\mathbf w_i\in\Hom(\mathbf V,\CC^{\zeta_i}) \label{XYvw}
\end{align}
have the form~\eqref{XYDef}, $\eqref{vwDef}$ for some $X_i,Y_i,v_i,w_i$ if and only if
\begin{align}
 &\pi_j\mathbf X\mu_i=0, &&\pi_i\mathbf Y\mu_j=0, &&\text{for $j\ne i+1$;} \label{XYform} \\
 &\pi_j\mathbf v_i=0, &&\mathbf w_i\mu_j=0, &&\text{for $j\ne i$.} \label{vwform}
\end{align}
The relation~\eqref{VPiXY} is equivalent to
\begin{align}
 [\mathbf X,\mathbf Y]+\hspace{-2pt}\sum_{i\in I}\mathbf v_i \mathbf w_i&=\sum_{i\in I}\lambda_i\mu_i\pi_i.  \label{VPiXYbf}
\end{align}
Thus, the point of $M_\lambda(\alpha,\zeta)$ is given by $2m+2$ matrices~\eqref{XYvw} satisfying~\eqref{XYform}, \eqref{vwform}, \eqref{VPiXYbf} modulo the transformations
\begin{align}
  &\mathbf X\mapsto\mathbf g\mathbf X\mathbf g^{-1},
 &&\mathbf Y\mapsto\mathbf g\mathbf Y\mathbf g^{-1},
 &&\mathbf v\mapsto\mathbf g\mathbf v,
 &&\mathbf w\mapsto\mathbf w\mathbf g^{-1}
\end{align}
by matrices $\mathbf g=\sum\limits_{i\in I}\mu_ig_i\pi_i\in\GL(|\alpha|,\CC)$, where $g_i\in\GL(\alpha_i,\CC)$. The functions~\eqref{Hmk}, \eqref{HlrC}, \eqref{IACycl} in these notations have the form
\begin{align}
 &H_{mk}(V)=\frac1m\tr_{\mathbf V}(\mathbf Y^{mk}), \\
 &H_{\ell,r}(V)=-\hspace{-2pt}\sum_{i\in I}\tr(E_r\mathbf w_{i-\ell}\mathbf Y^\ell\mathbf v_i), \\
 &I_A(V)=-\hspace{-2pt}\sum_{i\in I}\sum_{\ell=0}^\infty\tr(A_{(i,\ell)}\mathbf w_{i-\ell}\mathbf Y^\ell\mathbf v_i).
\end{align}

\subsection{Calogero--Moser spaces}
\label{secCM}

Let $m=1$. In this case the vectors $\zeta$ and $\alpha$ are one-dimensional: we have $\zeta_0=d$, $\alpha_0=n$ for some $d,n\in\ZZ_{\ge1}$ (we do not consider the trivial cases when $n=0$ or $d=0$). In this section we start with the case $d=1$. The quivers $Q$ and $Q_\zeta$ in this case have the form
\begin{align*}
&\xymatrix{ 0\ar@(ul,ur)^{a_0}
}
&&\xymatrix{\infty\ar[rr]^{b_{0,1}} && 0\ar@(ul,ur)^{a_0}
}
\end{align*}
The matrices $X=X_0,Y=Y_0\in\End(\CC^n)$, $v=v_{0,1}\in\CC^n$, $w=w_{0,1}\in(\CC^n)^*$ satisfy the condition $[X,Y]+vw=\lambda\id_{\CC^n}$. Since $\lambda\ne0$, one can reach $\lambda=1$ by rescaling $Y$ and $w$. The $4$-tuple $(X,Y,v,w)$ satisfying the condition
\begin{align}
 [X,Y]+vw=\id_{\CC^n}, \label{CMS}
\end{align}
gives a point of the variety $\Rep\big(\Pi^\llambda(Q_\zeta,\aalpha)\big)$ where $\llambda=(-n,1)$, $\zeta=1$, $\aalpha=(1,n)$. The group $\GL(\alpha)=\GL(n,\CC)$ acts on this variety as
\begin{align} \label{gXYvw}
 &g\colon (X,Y,v,w)\mapsto(gXg^{-1},gYg^{-1},gv,wg^{-1}), &&g\in\GL(n,\CC).
\end{align}
The quotient of the variety of $4$-tuples $(X,Y,v,w)\in\big(\End(\CC^n)\big)^2\times(\CC^n)\times(\CC^n)^*$ satisfying~\eqref{CMS} over the action~\eqref{gXYvw} is called the {\it Calogero--Moser space}~\cite{W} and is denoted by $\C_n$.
In this way, the quiver variety in the considered case is $M_\lambda(\alpha,\zeta)=M_\lambda(n,1)\simeq M_1(n,1)=\C_n$. Due to the formula~\eqref{HmkHmkr} the Hamiltonian system on $\C_n$ is given by the functions $H_k(V)=\tr(Y^k)$. Each Hamiltonian $H_k$ defines the flows
\begin{align}
 &X(t_k)=X-kt_k Y^{k-1}, &&Y(t_k)=Y, &&v(t_k)=v, &&w(t_k)=w,
\end{align}
where $(X,Y,v,w)$ is a $4$-tuple corresponding to the initial point. Since the Hamiltonians $H_k$ Poisson commute with each other, the equations they give are compatible. The solution of the obtained system has the form
\begin{align} \label{Xt}
 &X(t)=X-\hspace{-2pt}\sum_{k\ge1}kt_k Y^{k-1}, &&Y=const, &&v=const, &&w=const,
\end{align}
where $t=(t_1,t_2,\ldots)\in\bigoplus\limits_{k\ge1}\CC$.

Let $\C'_n\subset\C_n$ be the subset of points for which the matrix $X$ is diagonalisable. After a transformation~\eqref{gXYvw} of a $4$-tuple $(X,Y,v,w)$ corresponding to a point of $\C'_n$ one obtains
\begin{align} \label{CMSX}
 &X=\begin{pmatrix}x_1 & &0\\  &\ddots &  \\ 0 &  & x_n \end{pmatrix},
\end{align}
where $x_1,\ldots,x_n\in\CC$. Then, the condition~\eqref{CMS} takes the form
\begin{align}
 &(x_a-x_b)Y_{ab}+v_aw_b=\delta_{ab}, &&a,b=1,\ldots,n. \label{xaxbY}
\end{align}
For $a=b$ this gives $v_a\ne0$ and $w_a=1/v_a$. By applying the transformation~\eqref{gXYvw} with the matrix $g=\diag(v_1^{-1},\ldots,v_n^{-1})$ we obtain $(X,Y,v,w)$ with the same $X$ and
\begin{align} \label{CMSvw}
 &v=\begin{pmatrix}1 \\\vdots \\1\end{pmatrix}, && w=(1,\ldots,1).
\end{align}
Now the equation~\eqref{xaxbY} for $a\ne b$ implies that the eigenvalues $x_1,\ldots,x_n$ are pairwise different and $Y_{ab}=-\frac1{x_a-x_b}$, hence
\begin{align} \label{CMSY}
 &Y=\begin{pmatrix}p_1 & &-(x_a-x_b)^{-1}\\  &\ddots & \\ -(x_b-x_a)^{-1} & & p_n \end{pmatrix},
\end{align}
where $p_1,\ldots,p_n\in\CC$ are arbitrary. From~\eqref{dimKolMn} we derive $\dim\C_n=\dim M_\lambda(n,1)=2n$. In particular, this implies that $\{x_1,\ldots,x_n;p_1,\ldots,p_n\}$ are local coordinates on $\C_n$. By substituting the formulae~\eqref{CMSX}, $\eqref{CMSvw}$ and \eqref{CMSY} to~\eqref{omegaF} one yields
\begin{align}
 \omega_{\llambda,\aalpha}=\tr(\d Y\wedge\d X)+\d w\wedge\d v=\sum_{a=1}^n\d p_a\wedge\d x_a.
\end{align}
This means that $\{x_1,\ldots,x_n;p_1,\ldots,p_n\}$ are Darboux coordinates.

The Hamiltonians (or, more precisely, their restrictions to $\C'_n$) in these coordinates take the form
\begin{align}
 &H_1=\sum_{a=1}^np_a, &
 &H_2=\sum_{a=1}^np_a^2-2\sum_{a<b}(x_a-x_b)^{-2}, &
 &H_k=\sum_{a=1}^np_a^k+\ldots
\end{align}
where dots mean lower terms with respect to degrees of $p_1,\ldots,p_n$. Algebraic independence of the functions $\sum\limits_{a=1}^np_a^k$, $k=1,\ldots,n$, implies that $H_k$, $k=1\ldots,n$ are also algebraically independent. Since the Hamiltonians $H_k$ Poisson commute, they define an integrable system. It is called {\it Calogero--Moser system} (of type $A_{n-1}$). The matrix~\eqref{CMSY}, which gives Hamiltonians as traces of its powers, is called {\it Lax matrix} of this system.

The phase space of the Calogero--Moser system is the cotangent bundle $T^*\CC^n_{reg}$ to the affine variety $\CC^n_{reg}=\{(x_1,\ldots,x_n)\in\CC^n\mid x_i\ne x_j\;\forall\,i\ne j\}$.
Due to symmetricity of the Hamiltonians one can consider that the phase space of the system is the quotient $T^*\CC^n_{reg}/S_n$ by the symmetric group $S_n$.

Since the eigenvalues $x_1,\ldots,x_n$ are pairwise different and defined up to a permutation, we obtain $\C'_n\simeq T^*\CC^n_{reg}/S_n$. Thus the Calogero--Moser space $\C_n$ is a completion of the phase space of the Calogero--Moser system, where the Hamiltonians of the Calogero--Moser system define complete flows.

The elements of $\C_n\backslash\C'_n$, by which one completes the phase space, correspond to the cases of non-diagonalisable $X$. The matrix $X$ can be always reduced to a Jordan form. Any two Jordan blocks corresponding to the same eigenvalue must have sizes differing by at least 2~\cite{W}.

The points of $\C_n\backslash\C'_n$ are interpreted as collisions of particles in the complex plane. We will see below that the consideration of these points has application in the theory of KP equation and its hierarchy (see Section~\ref{secKP}).

\begin{primer}
 Consider the case $n=2$. Then all the points of $\C_2\backslash\C'_2$ are
\begin{align}
 &X=\begin{pmatrix}x_1 & 1 \\ 0 & x_1 \end{pmatrix}, &&Y=\begin{pmatrix}a & b \\ 1 & a \end{pmatrix}, &&v=\begin{pmatrix}0 \\ 2 \end{pmatrix}, && w=(0,1), \label{Coll2a} \\
 &X=\begin{pmatrix}x_1 & 1 \\ 0 & x_1 \end{pmatrix}, &&Y=\begin{pmatrix}a & b \\ -1 & a \end{pmatrix}, &&v=\begin{pmatrix}1 \\ 0 \end{pmatrix}, && w=(2,0), \label{Coll2b}
\end{align}
where the numbers $x_1,a,b$ take all the complex values.
\end{primer}

\subsection{Gibbons--Hermsen systems}
\label{secGH}

Let still $m=1$ and $\alpha_0=n\in\ZZ_{\ge1}$, but $\zeta_0=d$ be an arbitrary positive integer. Then the quiver $Q_\zeta$ has the form
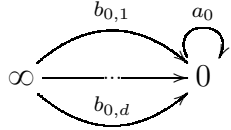
\begin{align*}
&\xymatrix{\infty\ar@/^1.5pc/[rr]^{b_{0,1}}\ar@/^0pc/[rr]|{\ldots}\ar@/_1.5pc/[rr]^{b_{0,d}} && 0\ar@(ul,ur)^{a_0}
}
\end{align*}
The matrices $X=X_0,Y=Y_0\in\End(\CC^n)$, $v_{0,r}\in\CC^n$, $w_{0,r}\in(\CC^n)^*$ satisfy the condition
\begin{align} \label{XYvwd}
 [X,Y]+\hspace{-2pt}\sum_{r=1}^dv_{0,r}w_{0,r}=\lambda\id_{\CC^n}.
\end{align}

Since $\mathsf P=\mathsf P_0=\{(a_0^*)^k\mid k\in\ZZ_{\ge0}\}$, the elements $A\in\mathsf L_\zeta$ has the form $A=(A_{(k)})_{k\ge0}$ where $A_{(k)}:=A_{(0,k)}\in\End(\CC^d)$. The map $A\mapsto\sum\limits_{k=0}^\infty A_{(k)}t^k$ gives an isomorphism of Lie algebras $\mathsf L_\zeta\simeq\mathfrak{gl}(d,\CC)\otimes\CC[t]$. If $\mathfrak h$ is a maximal commutative subalgebra of $\mathfrak{gl}(d,\CC)$, then $\mathfrak h\otimes\CC[t]$ is a maximal commutative subalgebra of $\mathfrak{gl}(d,\CC)\otimes\CC[t]$. Up to an automorphism we can choose the maximal commutative subalgebra $\mathfrak h\subset\mathfrak{gl}(d,\CC)$ to be the algebra of the diagonal matrices. Then $\mathfrak h\otimes\CC[t]$ corresponds to the subalgebra of $\mathsf L_\zeta$, spanned by $E^{(k)}_r$, $k\ge0$, $r=1,\ldots,d$. This explains the choice of the Hamiltonians~\eqref{Hkr} among the integrals $I_A$ for this case.

Let $\varphi_a\in\CC^d$, $\psi_a\in(\CC^d)^*$ be vectors and covectors enumerated by $a=1,\ldots,n$ with components
\begin{align}
 &(\varphi_a)_r=(v_{0,r})_a, &&(\psi_a)_r=(w_{0,r})_a, &&r=1,\ldots,d.
\end{align}
Then a point of the quiver variety $M_\lambda(\alpha,\zeta)=M_\lambda(n,d)$ is given by a family of matrices $(X,Y,\varphi_1,\ldots,\varphi_n,\psi_1,\ldots,\psi_n)$ satisfying the condition~\eqref{XYvwd}.
Let $U\subset M_\lambda(n,d)$ be the subset of points with diagonalisable~$X$:
\begin{align} \label{CMSXd}
 &X=\begin{pmatrix}x_1 & &0\\  &\ddots &  \\ 0 &  & x_n \end{pmatrix}.
\end{align}
The condition~\eqref{XYvwd} for these points takes the form
\begin{align} \label{xaxbYd}
 (x_a-x_b)Y_{ab}+\psi_b\varphi_a=\lambda\delta_{ab}.
\end{align}
Let $U'\subset U$ be subset (also dense) where $(\varphi_a)_1\ne0$ for all $a=1,\ldots,n$. By reasoning as above we can consider for the points of $U'$ that $(\varphi_a)_1=1$ for all $a=1,\ldots,n$. By letting $a=b$ in~\eqref{xaxbYd} we obtain $(\psi_a)_1=\lambda-\hspace{-2pt}\sum\limits_{r=2}^d(\varphi_a)_r(\psi_a)_r$.
Analogously to the previous case we have
\begin{align} \label{CMSYd}
 &Y_{aa}=p_a, &&Y_{ab}=-\frac{\psi_b\varphi_a}{x_a-x_b},\quad a\ne b,
\end{align}
for some $p_1,\ldots,p_n\in\CC$. The formula~\eqref{dimKolMn} implies $\dim\C_n=\dim M_\lambda(n,d)=2nd$. The local coordinates
\begin{align}
 \big\{x_a,(\varphi_a)_2\ldots,(\varphi_a)_d;p_a,(\psi_a)_2\ldots,(\psi_a)_d\big\}_{a=1}^n
\end{align}
are Darboux coordinates:
\begin{align*}
 \omega_{\llambda,\aalpha}=\tr(\d Y\wedge\d X)+\hspace{-2pt}\sum_{r=1}^d\d w_{0,r}\wedge\d v_{0,r}
 =\sum_{a=1}^n\d p_a\wedge\d x_a+
\hspace{-1pt}\sum_{a=1}^n\sum_{r=2}^d\d(\psi_a)_r\wedge\d(\varphi_a)_r.
\end{align*}

In these coordinates the Hamiltonians (their restrictions to $U$) have the form
\begin{align}
 &H_1=\sum_{a=1}^np_a, &
 &H_2=\sum_{a=1}^np_a^2-2\sum_{a<b}\frac{(\psi_a\varphi_b)(\psi_b\varphi_a)}{(x_a-x_b)^2}, &
 &H_k=\sum_{a=1}^np_a^k+\ldots
\end{align}
The expression $(\psi_a\varphi_b)(\psi_b\varphi_a)$ can be interpreted as spin interaction: in terms of "spin"{} operators $s_a=\varphi_a\psi_a$ it takes the form $\tr(s_as_b)$. Thus, one can call the coordinates $(\varphi_a)_r$ and $(\psi_a)_r$ {\it spin variables}. The Hamiltonians $H_k$ define the integrable system introduced in the work~\cite{GH}. It is called {\it Gibbons--Hermsen system} or {\it spin Calogero--Moser system} of type $A_{n-1}$. The variety $M_\lambda(n,d)$ is the completed phase space of this system.

In the paper~\cite{GH} authors also consider more general integrals of motion
\begin{align}
 &J_k(T)=\tr(T w_0 Y^kv_0), &&T\in\End(\CC^d).
\end{align}
They span the Lie algebra of the integrals~\eqref{IA} in this case:
\begin{align}
 &I_A=-\hspace{-2pt}\sum_{k\ge0}J_k(A_{(k)}), &&A\in\mathsf L_\zeta.
\end{align}

\subsection{The framing $\zeta=(d,0,\ldots,0)$}
\label{secde0}

Now consider the case of general $m\ge1$. Suppose first that all the framing edges come to only one vertex $i\in I$. Without loss of generality we can assume that $i=0$, that is $\zeta=d\varepsilon_0$ for some $d\in\ZZ_{\ge1}$. The framed quiver $Q_\zeta=Q_{d\varepsilon_0}$ looks as
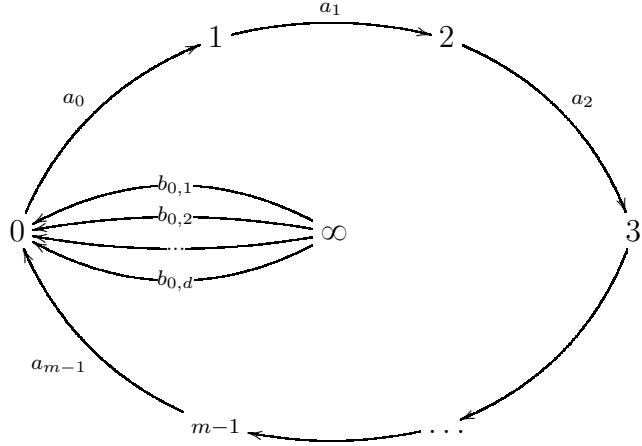
\begin{align*}
\xymatrix{&& 1\ar@/^/[rr]^{a_1} && 2\ar@/^1pc/[ddrr]^{a_2} && \\
{\phantom{\infty}}\\
0\ar@/^1pc/[uurr]^{a_0} &&&\infty\ar@/_1.5pc/[lll]|{b_{0,1}}\ar@/_0.5pc/[lll]|{b_{0,2}}\ar@/^0.5pc/[lll]|{\ldots}\ar@/^1.5pc/[lll]|{b_{0,d}}&&&3\ar@/^1pc/[ddll] \\
{\phantom{\infty}}\\
&&{\scriptstyle m-1}\ar@/^1pc/[uull]^{a_{m-1}}&&\ldots\ar@/^/[ll]&&
}
\end{align*}

For an element $A\in\mathsf L_{d\varepsilon_0}$ we have $A_p\ne0$ only if $p\in\mathsf P_0=\{p_0^k\mid k\in\ZZ_{\ge0}\}$. Hence the elements of the Lie algebra $\mathsf L_{d\varepsilon_0}$ have the components $A_{(k)}:=A_{(0,mk)}\in\End(\CC^d)$, so that $\mathsf L_{d\varepsilon_0}\simeq\mathfrak{gl}(d,\CC)\otimes\CC[t]$ and the previous reasoning gives an explanation of the choice of~\eqref{Hkr} in this case as well.

Non-zero Hamiltonians are
\begin{align}
 H_{mk,r}=-w_{0,r}(Y_0\cdots Y_{m-1})^kv_{0,r}. \label{Hmkr}
\end{align}

Consider the case $\alpha=n\delta$ where $n\in\ZZ_{\ge1}$ and $\delta=\sum\limits_{i=0}^{m-1}\varepsilon_i$. We have $\aalpha=(1,n\delta)\in\Delta_{im}^+(Q_{d\varepsilon_0})$ and $\llambda=(-n|\lambda|,\lambda)$. From~\eqref{dimKolMn} we obtain $\dim M_\lambda(\alpha,\zeta)=\dim M_\lambda(n\delta,d\varepsilon_0)=2nd$. Local coordinates are constructed as in Section~\ref{secGH}. At a generic point the product $X_{m-1}\cdots X_1X_0$ is a non-degenerate diagonalisable matrix with eigenvalues $x_1^m,\ldots,x_n^m$ for some $x_1,\ldots,x_n\in\CC\backslash\{0\}$. Let us apply the transformation
\begin{align}
g=(g_0,g_1,\ldots,g_{m-1})\in\GL(n\delta)
\end{align}
such that $g_0\in\GL(n,\CC)$ diagonalises this matrix:
\begin{align} \label{XXXdiag}
 X_{m-1}\cdots X_1X_0\mapsto\diag(x_1^m,\ldots,x_n^m).
\end{align}
Since the matrices $X_{m-1},\ldots,X_1$ are non-degenerate, one can make them to be arbitrary non-degenerate matrices by choosing successively the elements $g_{m-1},\ldots,g_1\in\GL(n,\CC)$, in particular, we can make
\begin{align} \label{Xidiag}
 &X_i\mapsto\diag(x_1,\ldots,x_n), &i=1,\ldots,m-1.
\end{align}
Then~\eqref{XXXdiag} and \eqref{Xidiag} imply
\begin{align}
 &X_0\mapsto\diag(x_1,\ldots,x_n). &\phantom{i=1,\ldots,m-1.}
\end{align}

As above we introduce $\varphi_a\in\CC^d$ and $\psi_a\in(\CC^d)^*$ with components
\begin{align}
 &(\varphi_a)_r=(v_{0,r})_a, \qquad (\psi_a)_r=(w_{0,r})_a, &&r=1,\ldots,d. \label{v0w00mat}
\end{align}
By summing the $a$-th diagonal elements in~\eqref{VPiXY} over $i\in I$ (where $v_i=0, w_i=0$ for $i\ne0$), we obtain $\psi_a\varphi_a=|\lambda|$. At a generic point we have $(\varphi_a)_1\ne0$ $\;\forall\,a=1,\ldots,n$. The variables $x_a$ can be completed to the set of local coordinates
\begin{align} \label{xphippsi0}
 \big\{x_a,(\varphi_a)_2\ldots,(\varphi_a)_d;p_a,(\psi_a)_2\ldots,(\psi_a)_d\big\}_{a=1}^n.
\end{align}
In these coordinates we have
\begin{align}
 &(X_i)_{ab}=x_a\delta_{ab}, \label{Xkij0mat} \\
 &(Y_i)_{aa}=\frac1m p_a+\frac{1}{x_a}\Big(\sum_{l=1}^{m-1}\frac{m-l}{m}\lambda_l-\hspace{-2pt}\sum_{l=1}^{i}\lambda_l\Big), \label{Ykii0mat} \\
 &(Y_i)_{ab}=-\frac{x_a^ix_b^{m-i-1}}{x_a^m-x_b^m}\psi_b\varphi_a, &&a\ne b, \label{Ykij0mat} \\
 &(\varphi_a)_1=1, \qquad\qquad (\psi_a)_1=|\lambda|-\hspace{-2pt}\sum_{r=2}^d(\varphi_a)_r(\psi_a)_r, \label{phia1psia1}
\end{align}
where $i=0,\ldots,m-1$, $a,b=1,\ldots,n$ (the formulae~\eqref{Ykii0mat},\eqref{Ykij0mat} follows from~\eqref{VPiXY}, see~\cite{ChS}). The coordinates~\eqref{xphippsi0} are Darboux coordinates as well:
\begin{multline} \label{omegamd}
 \omega_{\llambda,\aalpha}=\sum_{i\in I}\tr(\d Y_i\wedge\d X_i)+\hspace{-1pt}\sum_{r=1}^d\d w_{0,r}\wedge\d v_{0,r}
 =\sum_{a=1}^n\d p_a\wedge\d x_a+\hspace{-1pt}\sum_{a=1}^n\sum_{r=2}^d\d(\psi_a)_r\wedge\d(\varphi_a)_r.
\end{multline}

\begin{Rem}
The substitutions $p_a\to p_a+c_ax_a^{-1}$ with constant $c_a$ do not change the symplectic form~\eqref{omegamd}, but they can not make the coefficient at $x_a^{-1}$ in the formula~\eqref{Ykii0mat} to be zero, because it depends on $i$. This coefficient is chosen in such way that $\sum\limits_{i\in I}(Y_i)_{aa}=p_a$. Under this choice the variables $p_a$ can be interpreted as momenta of particles with coordinates $x_a$ (see examples below). In particular, this choice implies that the Hamiltonians $H_{mk}$ have the form $m^{-mk}\sum\limits_{a=1}^np_a^{mk}+O(|p|^{mk-2})$, i.e. the terms of order $mk-1$ with respect to momenta vanish.
\end{Rem}

By using the coordinates~\eqref{xphippsi0} one can prove the following theorem~\cite{ChS}.

\begin{Th} \label{Thde0}
 The Hamiltonians
\begin{align}
 &H_{mk,r}\in\CC\big[M_{\lambda}(n\delta,d\varepsilon_0)\big], &&k=1,\ldots,n,\quad r=1,\ldots,d,
\end{align}
are functionally independent. Hence, they form an integrable system on the quiver variety $M_\lambda(n\delta,d\varepsilon_0)$.
\end{Th}

\begin{primer}
 For $m=1$ we obtain the Gibbons--Hermsen systems described in Section~\ref{secGH} in details.
\end{primer}

\begin{primer}
 For $m=2$ we obtain a $B_n$ analogue of the Gibbons--Hermsen system with quadratic Hamiltonian
\begin{multline} \label{H2Bn}
 -\frac1{|\lambda|}\sum_{r=1}^dH_{2,r}=H_2=\tr(Y_0Y_1)= \\
=\frac14\sum_{a=1}^n\Big(p_a^2-\frac{\lambda_1^2}{x_a^2}\Big)
-\frac12\sum_{a<b}\Big(\frac{1}{(x_a-x_b)^2}+\frac{1}{(x_a+x_b)^2}\Big) (\psi_a\varphi_b)(\psi_b\varphi_a).
\end{multline}
\end{primer}

\begin{primer}
 For general $m$ and $d=1$ we do not have (independent) spin variables. We obtain the Calogero--Moser systems for the generalised symmetric group $S_n\ltimes(\ZZ/m\ZZ)^n$. In particular, if $m=2$ then we obtain the Calogero--Moser system of type $B_n$ with the Hamiltonian~\eqref{H2Bn} where $(\psi_a\varphi_b)(\psi_b\varphi_a)=|\lambda|^2$, i.e. without spin interaction. The Calogero--Moser systems for $S_n\ltimes(\ZZ/m\ZZ)^n$ are completed to the corresponding {\it Calogero--Moser spaces}~\cite{EG} isomorphic to $M_\lambda(n\delta,\varepsilon_0)$. The flows defined by the Hamiltonians $H_{mk,1}$ are compete. They can be written explicitly:
\begin{align}
 &X_i(t')=X_i-|\lambda|\sum_{k\ge1}kt_{mk} Y_{i+1}Y_{i+2}\cdots Y_{i+mk-1}, \label{Xit} \\
 &Y_i(t')=Y_i, \qquad v_0(t')=v_0, \qquad w_0(t')=w_0, \label{Yit}
\end{align}
where $t'=(t_m,t_{2m},t_{3m},\ldots)\in\bigoplus\limits_{k\ge1}\CC$. In the notations~\eqref{XYDef}, \eqref{vwDef} the formulae~\eqref{Xit}, \eqref{Yit} have the form
\begin{align}
 &\mathbf X(t')=\mathbf X-|\lambda|\sum_{k\ge1}kt_{mk} \mathbf Y^{mk-1},  \label{XtYtvtwt} \\
 &\mathbf Y(t')=\mathbf Y, \qquad \mathbf v_0(t')=\mathbf v_0, \qquad \mathbf w_0(t')=\mathbf w_0. \label{Yitbf}
\end{align}
\end{primer}

By means of Theorem~\ref{ThKolMnRF} one can generalise Theorem~\ref{Thde0} for more general $\alpha$.

\begin{Cor} \label{CorThde0}
 Let $\aalpha'=(1,n\delta)$ and $\llambda'=(-n|\lambda'|,\lambda')$ for a regular $\lambda'\in\CC^m$. If $\aalpha=(1,\alpha)$ and $\llambda=(-\lambda\cdot\alpha,\lambda)$ for some $\alpha\in\ZZ_{\ge0}^I$, $\lambda\in\CC^m$, and the pair $(\llambda,\aalpha)$ is obtained from $(\llambda',\aalpha')$ by a chain of admissible reflections, then the Hamiltonians $H_{mk,r}\in\CC\big[M_{\lambda}(\alpha,d\varepsilon_0)\big]$ form an integrable system on the variety $M_{\lambda}(\alpha,d\varepsilon_0)$.
\end{Cor}

The condition of Corollary~\ref{CorThde0} implies that $\aalpha\in\Delta^+_{im}(Q_{d\varepsilon_0})$. The question arise: which imaginary roots $\aalpha=(1,\alpha)$ this corollary can be applied for? Let $\aalpha=(1,\alpha)=s_{i_1}s_{i_2}\cdots s_{i_\ell}\aalpha'$ for some $i_1,i_2,\ldots,i_\ell\in\Iinflf$. Let $\llambda=(-\lambda\cdot\alpha,\lambda)$. If $(r_{i_{k-1}}\cdots r_{i_1}\llambda)_{i_k}\ne0$ for all $k=1,\ldots,\ell$, then $(\llambda,\aalpha)$ is obtained from $(\llambda',\aalpha')$ by a chain of admissible reflections: $\llambda'=r_{i_\ell}\cdots r_{i_1}\llambda$. Since $\llambda'\cdot\aalpha'=\llambda\cdot\aalpha=0$ we have $\llambda'=(-n|\lambda'|,\lambda')$ for some $\lambda'\in\CC^I$. The requirement of the regularity of $\lambda'$ imposes the conditions $r_{i_\ell}\cdots r_{i_1}\llambda\cdot(0,\beta)\ne0$ $\;\forall\,\beta\in\Delta(Q)$. Thus the condition of Corollary~\ref{CorThde0} holds for any root $\aalpha=(1,\alpha)$ from $W$-orbit of the root $\aalpha'=(1,n\delta)$ and a generic vector $\lambda$, where $W$ is the group generated by reflections $s_i$ for the quiver $Q_{d\varepsilon_0}$.

Consider the case $d=1$. Note that in this case all the roots of the form $(0,\alpha)$ and $(1,\alpha)$ from the fundamental domain $F$ are $(0,n\delta)$, $n\ge1$, and $(1,n\delta)$, $n\ge2$, respectively. Moreover, $(1,\delta)=s_\infty(0,\delta)$.

\begin{Conj} \label{Conje0}
 Any root $\aalpha=(1,\alpha)\in\Delta^+_{im}(Q_{\varepsilon_0})$ belongs to $W$-orbit of a root $(1,n\delta)$ for some $n\in\ZZ_{\ge1}$.
\end{Conj}

If Conjecture~\ref{Conje0} is valid, then due to Corollary~\ref{CorThde0} the Hamiltonians $H_{mk,1}$ form an integrable system on the variety $M_\lambda(\alpha,\varepsilon_0)$ for a generic vector $\lambda\in\CC^m$ and any $\alpha\in\ZZ_{\ge0}^I$ such that $(1,\alpha)\in\Delta^+_{im}(Q_{\varepsilon_0})$.

\begin{Rem} \label{RemConj}
 In the case $d\ge2$ Conjecture~\ref{Conje0} is not valid literally (if $m\ge2$). In this case there are elements $(1,\alpha)\in F$ such that $\alpha\ne n\delta$ for any $n$. However one can conjecture that every imaginary root $(1,\alpha)$ belongs to an orbit of a root of the form $(0,\beta)$ or $(1,\beta)$ from $F$.
\end{Rem}

\subsection{The framing $\zeta=(d,d,\ldots,d)$}
\label{secddelta}

Consider another particular case: let $m\ge1$ be general again, but now $\zeta=d\delta$ for some $d\in\ZZ_{\ge1}$. The framed quiver $Q_{d\delta}$ is
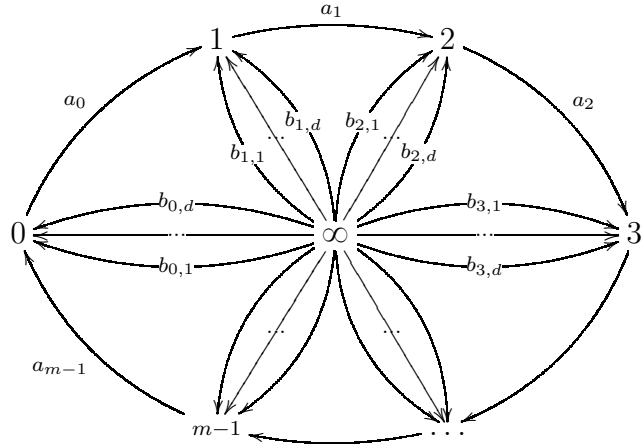
\begin{align*}
\xymatrix{&& 1\ar@/^/[rr]^{a_1} && 2\ar@/^1pc/[ddrr]^{a_2} && \\
{\phantom{\infty}}\\
0\ar@/^1pc/[uurr]^{a_0} &&&\infty\ar@/_1pc/[lll]|{b_{0,d}}\ar[lll]|{\ldots}\ar@/^1pc/[lll]|{b_{0,1}} \ar@/_1pc/[uul]|{b_{1,d}}\ar[uul]|{\ldots}\ar@/^1pc/[uul]|{b_{1,1}} \ar@/_1pc/[uur]|{b_{2,d}}\ar[uur]|{\ldots}\ar@/^1pc/[uur]|{b_{2,1}} \ar@/_1pc/[rrr]|{b_{3,d}}\ar[rrr]|{\ldots}\ar@/^1pc/[rrr]|{b_{3,1}} \ar@/_1pc/[ddr]\ar[ddr]|{\ldots}\ar@/^1pc/[ddr] \ar@/_1pc/[ddl]\ar[ddl]|{\ldots}\ar@/^1pc/[ddl] &&&3\ar@/^1pc/[ddll] \\
{\phantom{\infty}}\\
&&{\scriptstyle m-1}\ar@/^1pc/[uull]^{a_{m-1}}&&\ldots\ar@/^/[ll]&&
}
\end{align*}
In this case the components of elements $A\in\mathsf L_{d\delta}$ are square matrices $A_{(i,\ell)}\in\End(\CC^d)$. Consider the Lie subalgebra $\mathsf L'_{d\delta}\subset\mathsf L_{d\delta}$, consisting of elements $A$, for which $A_{(i,\ell)}$ do not depend on $i$. The corresponding Lie algebra of integrals is spanned by
\begin{align}
 &J_\ell(T)=\sum_{i\in I}\tr(T w_{i-\ell}Y_{i-\ell}\cdots Y_{i-1}v_i), 
\end{align}
where $\ell\in\ZZ_{\ge0}$, $T\in\End(\CC^d)$.
Subalgebra $\mathsf L'_{d\delta}$ is isomorphic to $\mathfrak{gl}(d,\CC)\otimes\CC[t]$, while the subspace spanned by $E^{(\ell)}_r$ is its maximal commutative subalgebra. The elements $E^{(\ell)}_r$ correspond to the Hamiltonians
\begin{align}
 H_{\ell,r}=-J_\ell(E_r)=-\hspace{-2pt}\sum_{i\in I} w_{i-\ell,r}Y_{i-\ell}\cdots Y_{i-1}v_{i,r}.
\end{align}

\begin{Prop} \label{PropIncl}
 The variety $M_\lambda(\alpha,d\varepsilon_0)$ can be identified with subvariety of $M_\lambda(\alpha,d\delta)$ defined by the equations
\begin{align}
 &v_i=0,\qquad w_i=0, &&i=1,\ldots,m-1. \label{viwi0eq}
\end{align}
The flows defined by the functions $J_\ell(T)$ preserve this subvariety if and only if $\ell$ is divided by $m$ or $T=0$. The restrictions of the Hamiltonians $H_{mk,r}=-J_{mk}(E_r)$ to $M_\lambda(\alpha,d\varepsilon_0)$ coincide with~\eqref{Hmkr}.
\end{Prop}

\noindent{\bf Proof.} The equations~\eqref{viwi0eq} define an embedding $\Rep\big(\Pi^\llambda(Q_{d\varepsilon_0}),\aalpha\big)\subset\Rep\big(\Pi^\llambda(Q_{d\delta}),\aalpha\big)$. The action of the group $\GL(\alpha)$ preserves the equations~\eqref{viwi0eq}. By taking quotient by this action we obtain the desired embedding of the quiver varieties. The rest is obtained by direct calculations. \qed

Let us consider the case $\alpha=n\delta$ in details again. The formula~\eqref{dimKolMn} gives the dimension: $\dim M_\lambda(n\delta,d\delta)=2nmd$.
Let $\varphi_a\in\Hom(\CC^m,\CC^d)$ and $\psi_a\in\Hom(\CC^d,\CC^m)$ be matrices with the components
\begin{align}
 &(\varphi_a)_{ri}=(v_{i,r})_a, \qquad (\psi_a)_{ir}=(w_{i,r})_a. \label{viwimat}
\end{align}
Similarly to Section~\ref{secde0}, we can express these matrices and the matrices $X_i$, $Y_i$ at a generic point via the local Darboux coordinates
\begin{align} \label{xphippsi}
 &x_a, (\varphi_a)_{ri}, &&p_a, (\psi_a)_{ir}, &&(i,r)\ne(0,1).
\end{align}
We have (see details in~\cite{ChS})
\begin{align}
 &(X_i)_{ab}=x_a\delta_{ab}, \label{Xkijmat} \\
 &(Y_i)_{aa}=\frac1m p_a+\frac{1}{x_a}\sum_{l=1}^{m-1}\frac{m-l}{m}\Big(\lambda_l-(\psi_a\varphi_a)_{ll}\Big)-\frac{1}{x_a}\sum_{l=1}^{i}\Big(\lambda_l-(\psi_a\varphi_a)_{ll}\Big), \notag \\
 &(Y_i)_{ab}=-\hspace{-2pt}\sum_{j=0}^{m-1}\frac{x_a^jx_b^{m-j-1}}{x_a^m-x_b^m}(\psi_b\varphi_a)_{i-j,i-j},\qquad\qquad a\ne b, \label{Ykijmat} \\
 &(\varphi_a)_{10}=1,  \label{phia10} \\
 &(\psi_a)_{01}=|\lambda|-\hspace{-2pt}\sum_{r=2}^d(\varphi_a)_{r0}(\psi_a)_{0r}-\hspace{-2pt}\sum_{i=1}^{m-1}\sum_{r=1}^d(\varphi_a)_{ri}(\psi_a)_{ir},
\end{align}
where $i=0,\ldots,m-1$, $a,b=1,\ldots,n$, and
\begin{align*}
 \omega_{\llambda,\aalpha}=\sum_{i\in I}\tr(\d Y_i\wedge\d X_i)+\hspace{-2pt}\sum_{i\in I}\sum_{r=1}^d\d w_{i,r}\wedge\d v_{i,r}
 =\sum_{a=1}^n\d p_a\wedge\d x_a+\hspace{-1pt}\sum_{a=1}^n\tr(\d\psi_a\wedge\d\varphi_a).
\end{align*}

\begin{Rem}
 Let $\mu:=e^{2\pi i/m}$ and $E=\mathrm{diag}(1, \mu, \mu^{2}, \dots, \mu^{m-1})$. Then, by using the identity%
\footnote{The formula~\eqref{xmymmu} is proven as follows. The multiplication of the left and right hand sides of~\eqref{xmymmu} by $m(x^m-y^m)=m\prod\limits_{i=0}^{m-1}(x-\mu^iy)$ gives the polynomials $f(x)=mx^{m-1-j}y^j$ and $g(x)=\sum\limits_{l=0}^{m-1}\mu^{-jl}\prod\limits_{i\ne l}(x-\mu^iy)$. The equality $\prod\limits_{i=1}^{m-1}(x-\mu^i)=(x^m-1)/(x-1)$ in the limit is $\prod\limits_{i=1}^{m-1}(1-\mu^i)=m$. Hence $f(\mu^ly)=g(\mu^ly)$. This is enough for the identity $f(x)\equiv g(x)$ since the degrees of the polynomials $f(x)$ and $g(x)$ are less or equal to $m-1$.
}
\begin{align} \label{xmymmu}
 &\frac{x^{m-j-1}y^j}{x^m-y^m}=\frac1m\sum_{l=0}^{m-1}\frac{\mu^{-jl}}{x-\mu^ly}, &&j=0,\ldots,m-1,
\end{align}
one can rewrite the expression~\eqref{Ykijmat} in the form
\begin{align*}
(Y_i)_{ab}=\frac1m\sum_{l=0}^{m-1} \,\frac{\mu^{-il}\tr\left(\varphi_a \, E^l \,\psi_b\,\right)}{x_b-\mu^{l}x_a}
\end{align*} 
(where $E^l$ is the power of the matrix $E$).
\end{Rem}

\begin{Rem}
 Consider the variety
\begin{align*}
 \mathsf Q_{m,d}=\big\{(\varphi,\psi)\in\Hom(\CC^m,\CC^d)\times\Hom(\CC^d,\CC^m)\mid \tr(\varphi\psi)=|\lambda|\big\}/\CC^\times, 
\end{align*}
where the group $\CC^\times=\CC\backslash\{0\}$ acts as $c.(\varphi,\psi)=(c\varphi,c^{-1}\psi)$. This variety is a Hamiltonian reduction for $M=T^*M_0$, $M_0=\Hom(\CC^m,\CC^d)$, $G=\CC^\times$. The points of $M_\lambda(n\delta,d\delta)$, at which the matrix $X_{m-1}\cdots X_0$ is diagonalisable and non-degenerate, are parametrised by coordinates $x_a,p_a$ and elements $q_a=(\varphi_a,\psi_a)\in\mathsf Q_{m,d}$. Analogously, generic points of $M_\lambda(n\delta,d\varepsilon_0)$ are parametrised by the coordinates $x_a,p_a$ and elements $q_a=(\varphi_a,\psi_a)\in\mathsf Q_{1,d}$, where $\mathsf Q_{1,d}=\big\{(\varphi,\psi)\in\CC^d\times(\CC^d)^*\mid \psi\varphi=|\lambda|\big\}/\CC^\times$ (see~\cite{W2}).
\end{Rem}

Theorem~\ref{Thde0} is generalised as follows~\cite{ChS}.

\begin{Th} \label{Thddelta}
 The Hamiltonians $H_{\ell,r}\in\CC\big[M_{\lambda}(n\delta,d\delta)\big]$, $\ell=1,\ldots,nm$, $r=1,\ldots,d$, are functionally independent and, hence, form an integrable system on the quiver variety $M_\lambda(n\delta,d\delta)$.
\end{Th}

\begin{Rem}
 The Hamiltonians $H_{\ell,r}$, $\ell=0,\ldots,nm-1$, $r=1,\ldots,d$, are dependent, since $\sum\limits_{r=1}^d H_{0,r}=-n|\lambda|$ (see Remark~\ref{RemHamil0}). One can obtain $nmd$ independent Hamiltonians, by replacing one of $H_{0,r}$ by $H_{mn}=-|\lambda|^{-1}\sum\limits_{r=1}^d H_{mn,r}$, for example.
\end{Rem}

At $m=1$ we obtain the Gibbons--Hermsen systems again. At grater $m$ the explicit form of the Hamiltonians is quite complicated even for $m=2$.

\begin{primer}~\cite{ChS}
 Let $m=2$ and $d=1$. Then the system is defined by the Hamiltonians $H_{\ell,1}$, $\ell\ge1$. Denote
$Z=\begin{pmatrix}1 & 0 \\ 0 & -1\end{pmatrix}$, $F_\pm=\begin{pmatrix}0 & \pm1 \\ 1 & 0\end{pmatrix}$. The first two Hamiltonians have the form
\begin{multline*}
 H_{1,1}=-w_0Y_0v_{1}-w_1Y_1v_{0}
=-\frac12\sum_{a=1}^n\Big((\varphi_aF_+\psi_a)p_a+\frac1{x_a}(\varphi_a F_-\psi_a)\big(\lambda_1+(\varphi_a)_1(\psi_a)_1\big)\Big)- \\ 
-\frac12\sum_{a\ne b}\Big(\frac{(\varphi_a F_+\psi_b)(\varphi_b\psi_a)}{x_a-x_b}
+\frac{(\varphi_a F_-\psi_b)(\varphi_b Z\psi_a)}{x_a+x_b}\Big),
\end{multline*}
\begin{multline*}
 H_2=-\frac1{|\lambda|}H_{2,1}=\tr(Y_0Y_1)=\frac14\sum_{a=1}^n\Big(p_a^2-\frac1{x_a^2}\big(\lambda_1-(\varphi_a)_1(\psi_a)_1\big)^2\Big)+\\
 +\frac14\sum_{a\ne b}\Big(\frac{(\varphi_a\psi_b)(\varphi_b\psi_a)}{(x_a-x_b)^2}
+\frac{(\varphi_a Z\psi_b)(\varphi_b Z\psi_a)}{(x_a+x_b)^2}\Big)+ \\
+\frac12\sum_{a\ne b}\frac{(\varphi_a)_1(\psi_b)_1(\varphi_b)_0(\psi_a)_0-(\varphi_a)_0(\psi_b)_0(\varphi_b)_1(\psi_a)_1}{x_a^2-x_b^2},
\end{multline*}
where $\varphi_a\in(\CC^2)^*$, $\psi_a\in\CC^2$ with components $(\varphi_a)_i=(\varphi_a)_{1i}$, $(\psi_a)_i=(\psi_a)_{i1}$, $i=0,1$. At a generic point (where $(\varphi_a)_1\ne0$ $\;\forall\,a$) one can choose $(\varphi_a)_0=1$, $(\psi_a)_0=|\lambda|-(\varphi_a)_1(\psi_a)_1$, so we have two independent spin variables $(\varphi_a)_1$ and $(\psi_a)_1$.
\end{primer}

As before, Theorem~\ref{ThKolMnRF} allows to generalise Theorem~\ref{Thddelta} for more general $\alpha$.

\begin{Cor} \label{CorThddelta}
Let $\aalpha'=(1,n\delta)$ and $\llambda'=(-n|\lambda'|,\lambda')$ for a regular $\lambda'\in\CC^m$. If $\aalpha=(1,\alpha)$ and $\llambda=(-\lambda\cdot\alpha,\lambda)$ for some $\alpha\in\ZZ_{\ge0}^I$ and $\lambda\in\CC^m$ such that the pair $(\llambda,\aalpha)$ is obtained from $(\llambda',\aalpha')$ by a chain of admissible reflections, then the Hamiltonians $H_{mk,r}\in\CC\big[M_{\lambda}(\alpha,d\delta)\big]$ form an integrable system on the variety $M_{\lambda}(\alpha,d\delta)$.
\end{Cor}

As before the condition of Corollary~\ref{CorThddelta} is valid for a generic vector $\lambda$ and a vector $\alpha$ such that $\aalpha=(1,\alpha)$ belongs to an $W$-orbit of a root $\aalpha'=(1,n\delta)$, where $W$ is the group generated by the reflections $s_i$ for the quiver $Q_{d\delta}$.

\section{Generalised KP hierarchies and their rational solutions}
\label{secApplKP}

In the case of the cyclic quiver the Hamiltonian dynamics on the quiver varieties, considered in Section~\ref{secCykl}, can be applied to construct solutions of the KP hierarchy and its generalisations. The reflection functor also plays some role here.

\subsection{Rational solutions of the usual KP hierarchy}
\label{secKP}

First we recall the definition of the KP hierarchy and present its rational solutions in terms of Calogero--Moser spaces.

The KP hierarchy is a system of infinitely many non-linear PDEs with infinite number of unknown functions and infinite number of unknown variables these functions depend on. It is convenient to regard the unknown functions as coefficients of a pseudo-differential operator and to write the equations in an "operator"{} form~\cite{Miwa}, \cite{Dikij}.

A {\it pseudo-differential operator} is an expression of the form
\begin{align}
 F=\sum_{k=-\infty}^N f_k(x)\partial^k, \label{PDO}
\end{align}
where $N\in\ZZ$, $f_k(x)\in\CC(x)$, $\partial=\partial_x$ is the operator of differentiation in $x$ and $\partial^{-1}$ is its formal inverse (one can consider the coefficients $f_{k}(x)$ in a larger algebra of functions which is preserved by $\partial$, but we will be interested in rational solutions only, so the function algebra $\CC(x)$ is enough for our purposes). Despite its name, the pseudo-differential operators do not act anywhere. They are defined purely algebraically by rules of commutation: one postulates commutativity of functions with each other and the relations
\begin{align}
 &[\partial,f(x)]=f'(x), && \partial\partial^{-1}=\partial^{-1}\partial=1.
\end{align}
These relations together with the condition of associativity imply
\begin{align}
 &\partial^kf(x)=\sum_{l=0}^\infty\left(k\atop l\right) f^{(l)}(x)\partial^{k-l}, && k\in\ZZ, \label{PDOmult}
\end{align}
where $f^{(l)}(x)$ is the $l$-th derivative. Formally, the pseudo-differential operators are elements of an associative algebra consisting of the expressions of the form~\eqref{PDO} with the multiplication defined by the formula~\eqref{PDOmult}. The correctness of this definition and the associativity of this multiplication can be checked by straightforward calculations. Purely differential and purely pseudo-differential parts of an element~\eqref{PDO} are denoted as
\begin{align}
 &F_+=\sum_{k=0}^N f_k(x)\partial^k &&\text{and}  &&F_-=\sum_{k=-\infty}^{-1} f_k(x)\partial^k 
\end{align}
respectively.

\begin{Rem}
 In the algebraic language the algebra of pseudo-differential operators is a microlocalisation~\cite{AVdBVO} of the algebra of the differential operators $\sum\limits_{k=0}^Nf_k(x)\partial^k$.
\end{Rem}

Consider a pseudo-differential operator
\begin{align} \label{L}
 L=\partial+\hspace{-1pt}\sum_{l=1}^\infty f_l(x)\partial^{-l}.
\end{align}
We will assume that the functions $f_l(x)$ depend also on an infinite number of "time"{} variables: $f_l(x)=f_l(x;t)$ where $t=(t_1,t_2,t_3,\ldots)$ (sometimes $t_1$ and $x$ are identified). The {\it KP hierarchy} is the systems of the compatible equations
\begin{align} \label{KP}
 &\frac{\partial}{\partial t_k}L=[(L^k)_+,L], &&k=1,2,3,\ldots
\end{align}
(see~\cite{Miwa},~\cite{Dikij}).

One can show that any pseudo-differential operator of the form~\eqref{L} can be written as
\begin{align} \label{LM}
 L=M\partial M^{-1}
\end{align}
for some
\begin{align} \label{M}
 M=1+\hspace{-1pt}\sum_{l=1}^\infty u_l(x)\partial^{-l}.
\end{align}
It is convenient to search a solution of the KP hierarchy in terms of $M$: if~\eqref{M} satisfies the system of equations
\begin{align}
 &\frac{\partial}{\partial t_k}M=-\big(M\partial^k M^{-1}\big)_-\cdot M, &&k=1,2,3,\ldots,
\end{align}
then~\eqref{LM} solves~\eqref{KP}.

Let $(X,Y,v,w)$ be a $4$-tuple corresponding to a point of the Calogero--Moser space $\C_n$ described in Section~\ref{secCM}. Consider the pseudo-differential operator
\begin{align} \label{MwXYv}
 M=1-w(X-x)^{-1}(Y-\partial)^{-1}v,
\end{align}
where $(Y-\partial)^{-1}=-\hspace{-2pt}\sum\limits_{l=0}^\infty Y^l\partial^{-l-1}$ and $(X-x)^{-1}$ is the matrix over the field $\CC(x)$, which is inverse to the matrix $X-x=X-x\id_{\CC^n}$. The operator~\eqref{MwXYv} has the form~\eqref{M} and does not depend on a representative $(X,Y,v,w)$ of the point of $\C_n$. Now let this point move under the Hamiltonian flows given by the equations $\frac{\partial}{\partial t_k}f=\{H_k,f\}$. Then the formula~\eqref{MwXYv} defines a pseudo-differential operator $M=M(t)$ with coefficients depending on the variables $t_1,t_2,t_2,\ldots$. It is obtained by replacement of $X$ by \eqref{Xt} in the formula~\eqref{MwXYv}. It was proved in the paper~\cite{W} that $L(t)=M(t)\partial M(t)^{-1}$ is a solution of the KP hierarchy~\eqref{KP} and that this exhausts all its rational solutions. Each such equation is given by the initial point $(X,Y,v,w)$. Thus, the rational solutions are in one-to-one correspondence with the points of the set $\bigsqcup\limits_{n\ge0}\C_n$.

\begin{Rem}
 This bijection is a part of the {\it Calogero--Moser correspondence}. The latter is one-to-one correspondence between the following elements~\cite{CH}, \cite{W}, \cite{BCE}:
\begin{itemize}
 \item the points of $\bigsqcup\limits_{n\ge0}\C_n$,
 \item the elements of adelic Grassmannians,
 \item the rational solutions of the KP hierarchy,
 \item the isomorphism classes of right ideals (as of right modules) of the {\it Weyl algebra} $A_1(\CC):=\Big\{\sum\limits_{l=0}^Nf_l(x)\partial^k\mid f_l(x)\in\CC[x], N\ge0\Big\}$.
\end{itemize}
\end{Rem}

The {\it KP equation} is the PDE
\begin{align} \label{KPeq}
 3\frac{\partial^2u}{\partial t_2^2}=\frac{\partial}{\partial x}\Big(4\frac{\partial u}{\partial t_3}-6u\frac{\partial u}{\partial x}-\frac{\partial^3 u}{\partial x^3}\Big),
\end{align}
where $u=u(x,t_2,t_3)$. It is obtained from~\eqref{KP} by putting $u=2f_1$~\cite{Miwa}. Substitution of~\eqref{CMSX}, \eqref{CMSvw} to~\eqref{MwXYv} gives $u_1=-\hspace{-2pt}\sum\limits_{a=1}^n\frac1{x-x_a}$ and
\begin{align} \label{uRatSol}
 u=2f_1=-2\frac{\partial u_1}{\partial x}=-\hspace{-2pt}\sum_{a=1}^n\frac2{(x-x_a)^2},
\end{align}
where $x_1=x_1(t),\ldots,x_n=x_n(t)$ is a solution of the Calogero--Moser system%
\footnote{The functions $f_1$ and $u_1$ are expressed via the coordinates $x_a$ only, while $f_l$ and $u_l$ for general $l$ are expressed via both the coordinates $x_a$ and the momenta $p_a$.
}
\begin{align}
 &\frac{\partial x_a}{\partial t_2}=\{H_2,x_a\},
 &&\frac{\partial p_a}{\partial t_2}=\{H_2,p_a\}, \\
 &\frac{\partial x_a}{\partial t_3}=\{H_3,x_a\},
 &&\frac{\partial p_a}{\partial t_3}=\{H_3,p_a\}. \label{H3flow}
\end{align}
The solutions~\eqref{uRatSol} of the equation~\eqref{KPeq} were found in the works~\cite{ChCh}, \cite{Kr}.

The points of the phase space $T^*\CC^n_{reg}$ do not give all the rational solutions, since there exists solutions corresponding to the points of $\C_n$ with non-diagonalisable $X$. They corresponds to the cases of "collisions"{} of particles.

\begin{Rem}
 Similar solutions of the {\it KdV equation} 
\begin{align}
 4\frac{\partial u}{\partial t_3}=6u\frac{\partial u}{\partial x}+\frac{\partial^3 u}{\partial x^3},
\end{align}
where $u=u(x,t_3)$, were found earlier in~\cite{AMM}. They have the same form~\eqref{uRatSol}, where $x_a=x_a(t_3)$ satisfy the equations $\{H_2,x_a\}=\{H_2,p_a\}=0$ and~\eqref{H3flow}.
\end{Rem}

\subsection{Generalised KP hierarchy and its solutions from cyclic quiver}
\label{secGKP}

We describe here the generalised KP hierarchy defined in~\cite{ChS} and its rational solutions found there.

Consider the group $\Gamma=\{1,\sigma,\sigma^2,\ldots,\sigma^{m-1}\}\simeq\ZZ/m\ZZ$ where $\sigma^m=1$.
Its group algebra $\CC\Gamma$ has the following basis consisting of the idempotents
\begin{align} \label{id}
 &\epsilon_k=\frac1m\sum_{j=0}^{m-1}\mu^{-kj}\sigma^j, &&k\in I,
\end{align}
where $\mu=e^{2\pi i/m}$ and $I=\{0,\dots, m-1\}=\ZZ/m\ZZ$. In order to generalise KP hierarchy we consider the {\it Cherednik algebra} $\mathcal H_\lambda$ for the group $\Gamma$. It is defined by generators $x,y,\sigma$ and relations
\begin{align}
 &yx-xy=\sum_{k=0}^{m-1}\lambda_k\epsilon_k, &&\sigma^m=1, &&\sigma x=\mu^{-1} x\sigma, &&\sigma y=\mu y\sigma,
\end{align}
where $\lambda_k$ are complex parameters.%
\footnote{One can simply say that the right hand side of the first relation is an arbitrary element of $\CC\Gamma$, but we intentionally decompose it by the basis~\eqref{id} to identify the coefficients with the parameters of the corresponding quiver variety.
}
In terms of the basis~\eqref{id} the last three relations have the form
\begin{align}
 &\epsilon_i\epsilon_j=\delta_{ij}\epsilon_i, &&\epsilon_ix=x\epsilon_{i+1}, && \epsilon_iy=y\epsilon_{i-1}. \label{epsilonxy}
\end{align}
As any Cherednik algebra, $\mathcal H_\lambda$ has a {\it PBW basis}:
\begin{align}
 &x^i\sigma^jy^k, &&i,k\in\ZZ_{\ge0},\quad j\in I.
\end{align}

Let us extend $\mathcal H_\lambda$ to the algebra $\overline{\mathcal H}_\lambda$ consisting of elements
\begin{align}
 F=\sum_{k=-\infty}^N \sum_{j=0}^{m-1}f_{jk}(x)\sigma^jy^k, \label{PDOy}
\end{align}
where $f_{lj}(x)\in\CC(x)$
and $y^{-1}$ is the inverse of $y$: that is $yy^{-1}=y^{-1}y=1$. The associativity of the algebra $\overline{\mathcal H}_\lambda$ follows from the theory of microlocalisation~\cite{AVdBVO} (see detail in~\cite{ChS}). The subalgebra spanned by $f(x)\sigma^j$, where $f(x)\in\CC(x)$, is denoted by $\CC(x)\#\Gamma$ (or $\CC(x)*\Gamma$). The elements~\eqref{PDOy} have the form
\begin{align} \label{PDOxy}
 &F=\sum_{k=-\infty}^N f_k(x)y^k,
\end{align}
for some $f_k(x)\in\CC(x)\#\Gamma$. For the elements $F\in\overline{\mathcal H}_\lambda$ of the form~\eqref{PDOxy} we introduce notations the following notations (analogous to those for the pseudo-differential operators):
\begin{align}
 &F_+=\sum_{k=0}^N f_k(x)y^k &&\text{and}  &&F_-=\sum_{k=-\infty}^{-1} f_k(x)y^k.
\end{align}

Consider the elements $L\in\overline{\mathcal H}_\lambda$ of the form
\begin{align} \label{LG}
 L=y+\hspace{-1pt}\sum_{l=0}^\infty f_l(x)y^{-l}
\end{align}
with coefficients $f_l(x)=f_l(x;t)\in\CC(x)\#\Gamma$, where $t=(t_1,t_2,t_3,\ldots)\in\bigoplus\limits_{\ell=1}^\infty\CC$ (if $m\ge2$ then the variable  $t_1$ is not identified with $x$). The {\it generalised KP hierarchy} is the system of compatible equations
\begin{align} \label{GKP}
 &\frac{\partial}{\partial t_\ell}L=[(L^\ell)_+,L], &&\ell=1,2,3,\ldots
\end{align}
The element~\eqref{LG} can be always written as
\begin{align}
 &L=MyM^{-1}, &M=1+\hspace{-1pt}\sum_{l=1}^\infty u_l(x)y^{-l}, \label{LMyM}
\end{align}
where $u_l(x)\in\mathcal F\#\Gamma$ for some algebra of (multivalued) functions $\mathcal F$ containing $\CC(x)$ (any $M$ of this form is invertible). It is easy to deduce that the equations~\eqref{GKP} follow from
\begin{align} \label{GKPM}
 &\frac{\partial}{\partial t_\ell}M=-\big(My^\ell M^{-1}\big)_-\cdot M, &&\ell=1,2,3,\ldots
\end{align}

Let $\lambda=(\lambda_0,\ldots,\lambda_{m-1})$ where $\lambda_k$ are parameters of the algebra $\overline{\mathcal H}_\lambda$ (as before, we suppose that $\lambda$ is regular with respect to the root system of the cyclic quiver unless otherwise indicated).

Let the matrices~\eqref{XYvw} define a point of the variety $M_\lambda(\alpha,\delta)$ (this is the case $d=1$ described in Section~\ref{secddelta}). The Hamiltonians $H_{\ell,1}$ give the flows $\frac{\partial}{\partial t_\ell}$ on $M_\lambda(\alpha,\delta)$: $\mathbf X=\mathbf X(t)$, $\mathbf Y=\mathbf Y(t)$, $\mathbf v_i=\mathbf v_i(t)$, $\mathbf w_i=\mathbf w_i(t)$.

\begin{Prop} {\normalfont~\cite{ChS}}
The matrix
\begin{align} \label{MwXYvG}
 M(t)=1-\hspace{-4pt}\sum_{i,j\in I}\epsilon_i\mathbf w_i(\mathbf X-x)^{-1}(\mathbf Y-y)^{-1}\mathbf v_j\varepsilon_j
\end{align}
satisfies the equations~\eqref{GKPM}.
\end{Prop}

Consequently, the formulae~\eqref{LMyM}, \eqref{MwXYvG} give a solution of~\eqref{GKP}. In this way we obtain a solution of the generalised KP hierarchy from the flows on $M_\lambda(\alpha,\delta)$ with initial values $\mathbf X(0)$, $\mathbf Y(0)$, $\mathbf v_i(0)$, $\mathbf w_i(0)$.

If one considers~\eqref{LG} as deformation of $y$, it is natural to fix the commutation relation with the elements of the group:
\begin{align} \label{Lequiv}
 &\sigma L=\mu L\sigma &&\text{or, equivalently,} &&\epsilon_i L=L\epsilon_{i-1},\quad i\in I.
\end{align}
The condition~\eqref{Lequiv} is compatible with those and only those equations of the system~\eqref{GKP} which correspond to $\ell$ divisible by $m$:
\begin{align} \label{KPSph}
 &\frac{\partial}{\partial t_{mk}}L=[(L^{mk})_+,L], &&k=1,2,3,\ldots
\end{align}
The system~\eqref{KPSph} with the condition~\eqref{Lequiv} is called the {\it spherical sub-hierarchy}.

Consider the variety $M_\lambda(\alpha,\varepsilon_0)$ (this is the case $d=1$ of Section~\ref{secde0}). The Hamiltonians $H_{km,1}=-|\lambda|H_{km}$ define the flows $\mathbf X=\mathbf X(t')$, $\mathbf Y=\mathbf Y(t')$, $\mathbf v_0=\mathbf v_0(t')$, $\mathbf w_0=\mathbf w_0(t')$, where $t'=(t_m,t_{2m},t_{3m},\ldots)$ (see the formulae~\eqref{XtYtvtwt}, \eqref{Yitbf}). Then, by applying Proposition~\ref{PropIncl} to the formula~\eqref{MwXYvG} we obtain a solution of the spherical sub-hierarchy: $L(t')=M(t')yM(t')^{-1}$ where
\begin{align} \label{MtSph}
 M(t')=1-\epsilon_0\mathbf w_0(\mathbf X-x)^{-1}(\mathbf Y-y)^{-1}\mathbf v_0\varepsilon_0.
\end{align}

\subsection{Generalised matrix KP hierarchy}
\label{secGMKP}

In the case of general $d$ the varieties $M_\lambda(\alpha,d\varepsilon_0)$, $M_\lambda(\alpha,d\delta)$ and the Hamiltonian systems on them give solutions of generalised matrix KP hierarchies and corresponding sub-hierarchies~\cite{ChS}. The usual matrix KP hierarchy is described, for example, in~\cite{Dikij}. We define it in the general from (for general $m$).

The role of the algebra $\overline{\mathcal H}_\lambda$ in this case is played by the algebra tensor product $\End(\CC^d)\otimes\overline{\mathcal H}_\lambda$, that is by the algebra of $d\times d$ matrices with coefficients in $\overline{\mathcal H}_\lambda$. The operations $(\cdot)_+$ and $(\cdot)_-$ on this algebra of matrices is defined entrywise. In the tensor notations they have the form $(A\otimes F)_\pm=A\otimes F_\pm$ where $A\in\End(\CC^d)$, $F\in\overline{\mathcal H}_\lambda$.

Consider the elements of the algebra $\End(\CC^d)\otimes\overline{\mathcal H}_\lambda$ of the form
\begin{align}
 &L=y+\hspace{-1pt}\sum_{l=0}^\infty F_ly^{-l}, \label{Lmat} \\
 &R_r=E_r+\hspace{-1pt}\sum_{l=1}^\infty R_{l,r}y^{-l}, &&r=1,\ldots,d, \label{Ralpha}
\end{align}
where $F_l,R_{l,r}\in\End(\CC^d)\otimes\big(\mathbb C(x)\#\Gamma\big)$ and the matrices $E_r\in\End(\CC^d)$ are defined in Section~\ref{secHamKolMn}. Impose the conditions
\begin{align}
 &[L,R_r]=0\,, &&R_r R_s=\delta_{rs}R_r, &&\sum_{r=1}^d R_r=1. \label{LRalpha}
\end{align}
Define the {\it generalised matrix KP hierarchy} as compatible system of equations on $L,R_1,\ldots,R_d$:
\begin{align}
 &\frac{\partial L}{\partial t_{\ell,r}}=\big[(L^\ell R_r)_+,L\big], \label{tkbetaL} \\
 &\frac{\partial R_s}{\partial t_{\ell,r}}=\big[(L^\ell R_r)_+,R_s\big]. \label{tkbetaRalpha}
\end{align}
This system defines a dependence of the elements $F_l,R_{l,s}$ on the variables $t_{\ell,r}$,  $\ell\in\ZZ_{\ge0}$, $r=1,\ldots,d$. The variables $t_{0,r}$ (or, more precisely, their flows) are dependent, since the third formula~\eqref{LRalpha} implies $\sum\limits_{r=1}^d \dfrac{\partial}{\partial t_{0,r}}=0$.

Associate the variables $t_{\ell,r}$ with the Hamiltonians $H_{\ell,r}$ on the quiver variety $M_\lambda(\alpha,d\delta)$. These Hamiltonians define the flows $\mathbf X=\mathbf X(t)$, $\mathbf Y=\mathbf Y(t)$, $\mathbf v_i=\mathbf v_i(t)$, $\mathbf w_i=\mathbf w_i(t)$ on $M_\lambda(\alpha,d\delta)$, where $t=(t_{\ell,r})$. The dependence $\sum\limits_{r=1}^d \dfrac{\partial}{\partial t_{0,r}}=0$ for these flows follows from $\sum\limits_{r=1}^dH_{0,r}=-\lambda\cdot\alpha$ (see Remark~\ref{RemHamil0}). Note that
\begin{gather*}
 \mathbf v_j\in\Hom(\CC^d,\mathbf V), \qquad (\mathbf Y-y)^{-1}=-\hspace{-2pt}\sum\limits_{l=0}^\infty Y^ly^{-l-1}\in\End\mathbf V\otimes\overline{\mathcal H}_\lambda, \\
 (\mathbf X-x)^{-1}\in\End\mathbf V\otimes\mathbb C(x)\subset\End\mathbf V\otimes\overline{\mathcal H}_\lambda, \qquad \mathbf w_i\in\Hom(\mathbf V,\CC^d). 
\end{gather*}
Hence, the matrix $M$ defined by the formula~\eqref{MwXYvG} with this flows on $M_\lambda(\alpha,d\delta)$ belongs to $\End(\CC^d)\otimes\overline{\mathcal H}_\lambda$ and has the form~\eqref{LMyM} where $u_l(x)\in\End(\CC^d)\otimes\big(\mathcal F\#\Gamma\big)$.

\begin{Prop} {\normalfont~\cite{ChS}}
\begin{itemize}
 \item[(1)] The matrix $M=M(t)$ satisfies the system of equations
\begin{align} \label{GMKPM}
 &\frac{\partial}{\partial t_{\ell,r}}M=-\big(My^\ell E_r M^{-1}\big)_-\cdot M, &&\ell\in\ZZ_{\ge0},\; r=1,\ldots,d.
\end{align}
 \item[(2)] The equations~\eqref{GMKPM} imply that the matrices
\begin{align}
 &L=MyM^{-1}, &&R_r=ME_rM^{-1} \label{LRM}
\end{align}
solve the system~\eqref{tkbetaL}, \eqref{tkbetaRalpha}.
\end{itemize}
\end{Prop}

Since the matrices~\eqref{LRM} have the form~\eqref{Lmat}, \eqref{Ralpha} and satisfies the conditions~\eqref{LRalpha}, they gives a solution of the generalised KP hierarchy.

{\it Matrix spherical sub-hierarchy} is the system
\begin{align}
 &\frac{\partial L}{\partial t_{mk,r}}=\big[(L^{mk} R_r)_+,L\big], \label{tkbetaLSph} \\
 &\frac{\partial R_s}{\partial t_{mk,r}}=\big[(L^{mk} R_r)_+,R_s\big], \label{tkbetaRalphaSph}
\end{align}
where $k\in\ZZ_{\ge0}$, $r=1,\ldots,d$ and $L,R_r$ are matrices of the form~\eqref{Lmat}, \eqref{Ralpha} satisfying the conditions~\eqref{LRalpha} and
\begin{align}
 &\sigma L=\mu L\sigma, &&\sigma R_r=R_r\sigma.
\end{align}
Let $t'=(t_{mk,r})$ and $M=M(t')$ be the matrix defined by the formula~\eqref{MtSph} with the flows $\mathbf w_0=\mathbf w_0(t')$, $\mathbf X=\mathbf X(t')$, $\mathbf Y=\mathbf Y(t')$, $\mathbf v_0=\mathbf v_0(t')$ on the quiver variety $M_\lambda(\alpha,d\varepsilon_0)$. Substitution of $M=M(t')$ to~\eqref{LRM} gives a solution of the matrix spherical sub-hierarchy.

\begin{Rem} \label{RemMatA}
 To follow~\cite{Dikij} more strictly one should consider more general hierarchies. Namely, one should replace the formula~\eqref{Lmat} by
\begin{align}
 &L=Ay+\hspace{-2pt}\sum_{l=0}^\infty F_ly^{-l}, \label{LmatA}
\end{align}
where $A=\diag(a_1,\ldots,a_d)$ is a constant diagonal matrix with elements $a_r\in\CC\backslash\{0\}$. The solution of this hierarchy corresponding to the flows on $M_\lambda(\alpha,d\delta)$ is obtained as follows. One needs to replace the variables $t_{\ell,r}$ by $a_r^{-\ell}t_{\ell,r}$ in the same matrix $M=M(t)$ and to substitute the obtained $M$ to $L=MAyM^{-1}$ and $R_r=ME_rM^{-1}$. The solutions of the matrix spherical sub-hierarchy corresponding to the flows on $M_\lambda(\alpha,d\varepsilon_0)$ is obtained in the same way.
\end{Rem}

\begin{primer}
 Let $m=1$. In this case the parameter $\lambda$ is one-dimensional and does not vanish, so one can normalise it: $\lambda=1$. Then the elements of the algebra $\End(\CC)\otimes\overline{\mathcal H}_\lambda$ are matrices over the algebra of pseudo-differential operators~\eqref{PDO}. In this case the system~\eqref{tkbetaL}, \eqref{tkbetaRalpha} with the conditions~\eqref{Lmat}--\eqref{LRalpha} (and the spherical matrix sub-hierarchy) is the usual matrix (multicomponent) KP hierarchy~\cite{Dikij} (modulo Remark~\ref{RemMatA}). Its solutions are given by the Gibbons--Hermsen systems on the varieties $M_1(n,d)$ considered in Section~\ref{secGH}. The connection of the Gibbons--Hermsen systems with the matrix KP hierarchy was considered in the works~\cite{KBBT}, \cite{W2}, \cite{BP}.
\end{primer}

The following table shows the correspondence between hierarchies and quiver varieties: a rational solution of a hierarchy is constructed by means of the corresponding quiver varieties. Each hierarchy corresponds to a set of varieties associated to different $\alpha$ (or $n$). The vertical arrows means the transition to the particular case $d=1$, while the horizontal ones --- to the case $m=1$, so that to the case of one-loop quiver (here one can normalise $\lambda=1$).

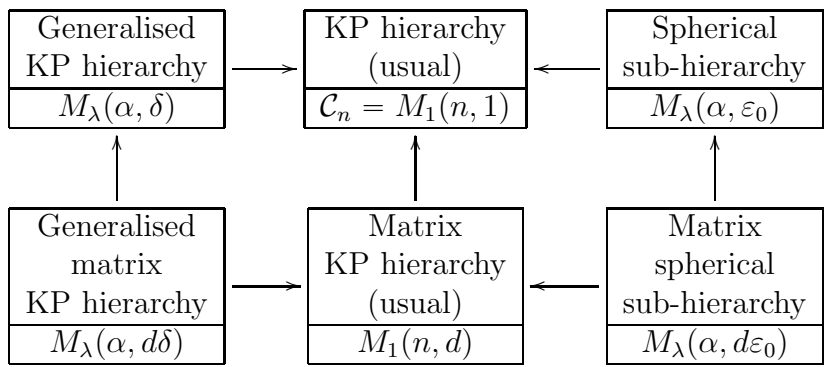
\begin{align*}
\xymatrix{
{\begin{tabular}{|c|}
\hline
\text{Generalised} \\
\text{KP hierarchy} \\
\hline
\text{$M_\lambda(\alpha,\delta)$} \\
\hline
\end{tabular}}\ar[r]
&
{\begin{tabular}{|c|}
\hline
\text{KP hierarchy} \\
\text{(usual)} \\
\hline
\text{$\C_n=M_1(n,1)$} \\
\hline
\end{tabular}}
&
{\begin{tabular}{|c|}
\hline
\text{Spherical} \\
\text{sub-hierarchy} \\
\hline
\text{$M_\lambda(\alpha,\varepsilon_0)$} \\
\hline
\end{tabular}}\ar[l] \\
{\begin{tabular}{|c|}
\hline
\text{Generalised} \\
\text{matrix} \\
\text{KP hierarchy} \\
\hline
\text{$M_\lambda(\alpha,d\delta)$} \\
\hline
\end{tabular}}\ar[r]\ar[u]
&
{\begin{tabular}{|c|}
\hline
\text{Matrix} \\
\text{KP hierarchy} \\
\text{(usual)} \\
\hline
\text{$M_1(n,d)$} \\
\hline
\end{tabular}}\ar[u]
&
{\begin{tabular}{|c|}
\hline
\text{Matrix} \\
\text{spherical} \\
\text{sub-hierarchy} \\
\hline
\text{$M_\lambda(\alpha,d\varepsilon_0)$} \\
\hline
\end{tabular}}\ar[l]\ar[u]
} 
\end{align*}

To obtain a solution of a hierarchy from this table one needs to take a solution of the Hamiltonian system on the corresponding quiver variety and to substitute it to the corresponding formulae written above. At $m\ge2$ we have a special class of solutions corresponding to $\alpha=n\delta$. In this case the matrices $\mathbf X(t)$, $\mathbf Y(t)$, $\mathbf v_i(t)$, $\mathbf w_i(t)$ are obtained from a solution of a "spin"{} generalisation of the Calogero--Moser system, described in Section~\ref{secde0} or \ref{secddelta}. If $\alpha,\lambda$ satisfy the condition of Corollary~\ref{CorThde0} or~\ref{CorThddelta}, then the solutions of the corresponding Hamiltonian system are obtained by application of reflection functors to the solutions of the "spin"{} Calogero--Moser systems on $M_\lambda(n\delta,\zeta)$, where $\zeta=d\delta$ or $d\varepsilon_0$.
For instance, if $d=1$ in the spherical case and Conjecture~\ref{Conje0} is valid, then the problem is reduced to the application of the reflection functors to the solutions~\eqref{XtYtvtwt}, \eqref{Yitbf}.

If $\aalpha\in\Delta^+_{re}(Q_\zeta)$, then $\dim M_\lambda(\alpha,\zeta)=0$, i.e. $\mathbf w_i,\mathbf X,\mathbf Y,\mathbf v_j$ are constant. Hence, we obtain in this case a stationary solution of the hierarchy.

\begin{Rem} \label{RemIrr}
One can define the matrix $M$ depending on a point of $M_\lambda(\alpha,d\delta)$ (or $M_\lambda(\alpha,d\varepsilon_0)$) by the formula~\eqref{MwXYvG} (or \eqref{MtSph}) for an arbitrary $\alpha\in\ZZ_{\ge0}^I$ (even if $\aalpha=(1,\alpha)$ is not a root). Let $M$ be constructed via the matrices $\mathbf X$, $\mathbf Y$, $\mathbf v_i$, $\mathbf w_i$, which gives a reducible $\Pi^\llambda(Q_\zeta)$-module of the dimension $\aalpha=(1,\alpha)$, where $\zeta=d\delta$ (or $\zeta=d\varepsilon_0$). Then $\aalpha=\sum_l\aalpha^{(l)}$, where $\aalpha^{(l)}\in\ZZ_{\ge0}^I$ are dimensions of the composition factors $V^{(l)}$ for this module. One has $\aalpha^{(l)}_\infty=\delta_{l,l_0}$ for some~$l_0$. Substituting the matrices $\wt{\mathbf X}$, $\wt{\mathbf Y}$, $\wt{\mathbf v_i}$, $\wt{\mathbf w_i}$ corresponding to the module $V^{(l_0)}$, to~\eqref{MwXYvG} (or~\eqref{MtSph}) one obtains the same matrix $M$. Since the $\Pi^\llambda(Q_\zeta)$-module $V^{(l_0)}$ is simple, its dimension $\aalpha^{(l_0)}$ is a root~\cite[Theorem~1.2]{CB01}. Therefore, one can suppose that $\aalpha$ is a positive root and that the initial point $\big(\mathbf X(0),\mathbf Y(0),\mathbf v_i(0),\mathbf w_i(0)\big)$ gives a simple module.
\end{Rem}

\begin{Rem}
 If the vector $\lambda\in\CC^I$ is not regular then the quiver varieties $M_\lambda(\alpha,\zeta)$ should be defined as varieties $N_\llambda(\aalpha)$ consisting from isomorphism classes of the semi-simple modules of the dimension $\aalpha=(1,\alpha)$ (see Remark~\ref{RemCQ}). In this case the construction described above also gives a solution of the corresponding hierarchy for regular initial points $\big(\mathbf X(0),\mathbf Y(0),\mathbf v_i(0),\mathbf w_i(0)\big)$ of the variety $M_\lambda(\alpha,\zeta)$. By virtue of Remark~\ref{RemIrr} it is enough to take simple modules (which are regular points) as initial points of the flows on $M_\lambda(\alpha,\zeta)$.
\end{Rem}

\section*{Open questions}
\addcontentsline{toc}{section}{Open questions}

There are still a lot of unclear questions on the subjects concerned above. Here we present some of them (see also the list in the end of~\cite{ChS}).

\begin{enumerate}
 \item The integrability of the system~\eqref{Hp} is proved for cases of the cyclic quiver only. In the general case it is not integrable, but possibly there exists an orientation for each graph such that the system with the Hamiltonians $H_p$ is integrable. It is interesting to find out when the Hamiltonian systems defined in Section~\ref{secHamKolMn} are integrable or, at least, to prove integrability for a wider class of quivers.
 \item Even for the cyclic quiver not all the cases have been investigated. First, not all the roots $(1,\alpha)$ from the fundamental domain $F$ are considered (see Remark~\ref{RemConj}). Second, the case of general $\zeta$ is not considered even for $\alpha=n\delta$. In particular, the case $\zeta=d\sum\limits_{i=0}^{m'}\varepsilon_i$, $0\le m'\le m-1$, should be described in complete analogy with to the cases $\zeta=d\varepsilon_0$ and $\zeta=d\delta$. There is also a question: which hierarchies correspond to another framing vector $\zeta$?
 \item In the case of cyclic quiver it is also unclear, can one obtain all the positive imaginary roots $\aalpha=(1,\alpha)$ by applying the reflection functor to the roots $(1,\beta)\in F$. This is not proved even for the simplest case $\zeta=\varepsilon_0$ (Conjecture~\ref{Conje0}).
 \item By considering a trigonometric or elliptic version of the Cherednik algebra, one can define a corresponding generalised KP hierarchy. To obtain their solutions one presumably needs to consider more general preprojective algebras.
\end{enumerate}



\appendix

\refstepcounter{section}
\label{appPoisson}
\section*{Appendix~\thesection.\\ Poisson brackets in tensor notations} 
\addcontentsline{toc}{section}{Appendix \thesection. Poisson brackets in tensor notations}

In this section we prove the formula~\eqref{PalphaV} for the moment map $P_\alpha\colon\Rep(\overline Q,\alpha)\to\mathfrak g^*$ and deduce the formulae~\eqref{IAIB}, \eqref{wiIAFlow}. Before that, we introduce a tensor formalism to calculate the Poisson brackets in the matrix form.

Let $A\in\Hom(\CC^n,\CC^{n'})$ and $B\in\Hom(\CC^m,\CC^{m'})$ be arbitrary matrices. Their tensor product is the matrix $A\otimes B\in\Hom(\CC^n,\CC^{n'})\otimes\Hom(\CC^m,\CC^{m'})=\Hom(\CC^n\otimes\CC^m,\CC^{n'}\otimes\CC^{m'})$ with  the entries $(A\otimes B)_{ij,kl}=A_{ik}B_{jl}$. Introduce the notations $A^{(1)}=A\otimes \id_{\CC^m}$ and $B^{(2)}=\id_{\CC^n}\otimes B$ (we suppose $m=m'$ in the former and $n=n'$ in the latter). Note that $A\otimes B=A^{(1)}B^{(2)}$ (for any $n,n',m,m'$). In the case of square matrices ($n=n'$, $m=m'$) we have $\tr(A^{(1)}B^{(2)})=\tr(A)\tr(B)$. Define the {\it transposition matrix} $\P\colon\CC^n\otimes\CC^m\to\CC^m\otimes\CC^n$ by $\P_{ij,kl} =\delta_{il}\delta_{jk}$. It has the following properties:
\begin{align}
 &\tr( A^{(1)}\P B^{(2)})=\tr(AB), &&A\in\Hom(\CC^m,\CC^n),\quad B\in\Hom(\CC^n,\CC^m); \label{trPAB} \\
 &\tr_1( A^{(1)}\P)=A, &&A\in\Hom(\CC^n,\CC^n), \qquad (n=m); \label{tr2AP}
\end{align}
(in the left hand side of~\eqref{trPAB} the trace is over $\CC^n\otimes\CC^n$ and in the right hand side of~\eqref{trPAB} is over $\CC^n$; in the left hand side of~\eqref{tr2AP} the trace is over the fist tensor factor $\CC^n$).

The Poisson brackets on $\Rep(\overline Q,\alpha)$ corresponding to the symplectic form~\eqref{omega} are given by the formula $\{(V_a)_{kl},(V_{b^*})_{pq}\}=(-1)^a\delta_{ab}\delta_{pl}\delta_{kq}$ where $a,b\in\overline Q$. In the matrix form the brackets have the form $\{V_a^{(1)},V_{b^*}^{(2)}\}=(-1)^a\delta_{ab}\P$ (the size of $\P$ depends on $a\colon i\to j$, namely, $n=\alpha_i$ and $m=\alpha_j$). The Leibniz rule in the matrix form can be written as $\{A^{(1)}B^{(1)},C^{(2)}\}=A^{(1)}\{B^{(1)},C^{(2)}\}+\{A^{(1)},C^{(2)}\}B^{(1)}$ where $A,B,C$ are matrix-valued functions.

Let us derive the moment map for the action of the group $G=G(\alpha)$ on the variety $M=\Rep(\overline Q,\alpha)$ (see Section~\ref{secHamRed}). The element $[g]\in G(\alpha)$ (where $g\in\GL(\alpha)$) acts by the formula~\eqref{GLaction} (where $a\in\overline Q$). The Lie algebra~$\mathfrak g$ of the group $G(\alpha)$ consists of $\theta=[(\theta_i)_{i\in I}]\in\End(\alpha)/(\CC\cdot1)$, where $\theta_i\in\End(\CC^{\alpha_i})$ and $\CC\cdot1\subset\End(\alpha)$ is the Lie subalgebra corresponding to the subgroup $\CC^\times\subset\GL(\alpha)$. The Lie brackets in $\mathfrak g$ have the form $[\theta,\eta]=\big[\big([\theta_i,\eta_i]\big)_{i\in I}\big]$. Infinitesimal action of the group $G=G(\alpha)$ is given by the vector fields $\mathcal V_\theta\in\Gamma(M,TM)$, defined as $(\mathcal V_\theta f)(V)=\frac{\d}{\d t}f\big(\exp(-t\theta).V\big)\big|_{t=0}$. Denote by $\partial_{V_a}$ the matrix with entries $(\partial_{V_a})_{ll'}=\dfrac{\partial}{\partial(V_a)_{l'l}}$. Then one can write $\mathcal V_\theta=\sum\limits_{a\colon i\to j}\tr\big((V_a\theta_i-\theta_jV_a)\partial_{V_a}\big)$. Let us prove that they are Hamiltonian vector fields with Hamiltonians $H_\theta(V)=\sum\limits_{a\colon i\to j}(-1)^a\tr(V_aV_{a^*}\theta_j)$ and that the latter ones satisfy $\{H_\theta,H_\eta\}=H_{[\theta,\eta]}$. First note that the Hamiltonians are correctly defined, since the simultaneous shift $\theta_i\to\theta_i+c$ gives the additional term $c\sum\limits_{a\in\overline Q}(-1)^a\tr(V_aV_{a^*})$, which vanishes due to cyclicity of the trace. Then, we have
\begin{multline*}
 \{H_\theta,f\}(V)=\sum_{b\in\overline Q}\{H_\theta,(V_b)_{kl}\}\partial f/\partial(V_b)_{kl}=\sum_{b\in\overline Q}\tr\big(\{H_\theta,V_b\}\partial_{V_b}\big)f= \\
=\sum_{a,b\in\overline Q\atop a\colon i\to j}(-1)^a\tr\big(\{V_a^{(1)}V_{a^*}^{(1)}\theta_j^{(1)},V_b^{(2)}\}\partial_{V_b}^{(2)}\big)f= \\
=\sum_{a,b\in\overline Q\atop a\colon i\to j}(-1)^a\tr\big(V_{a^*}^{(1)}\theta_j^{(1)}\{V_a^{(1)},V_b^{(2)}\}\partial_{V_b}^{(2)}\big)f+\\
+\hspace{-6pt}\sum_{a,b\in\overline Q\atop a\colon i\to j}(-1)^a\tr\big(\theta_j^{(1)}V_a^{(1)}\{V_{a^*}^{(1)},V_b^{(2)}\}\partial_{V_b}^{(2)}\big)f= \\
=\sum_{a\colon i\to j}\tr\big(V_{a^*}^{(1)}\theta_j^{(1)}\P\partial_{V_{a^*}}^{(2)}\big)f-\hspace{-4pt}\sum_{a\colon i\to j}\tr\big(\theta_j^{(1)}V_a^{(1)}\P\partial_{V_a}^{(2)}\big)f=\\
=\sum_{a\colon i\to j}\tr\big((V_a\theta_i-\theta_j V_a)\partial_{V_a}\big)f=(\mathcal V_\theta f)(V),
\end{multline*}
where we used~\eqref{trPAB}. In the same way we obtain
\begin{multline*}
\{H_\theta,H_\eta\}(V)=\sum_{a\colon i\to j\atop b\colon k\to l}(-1)^a(-1)^b\tr\big(\{V_a^{(1)}V_{a^*}^{(1)}\theta_j^{(1)},V_b^{(2)}V_{b^*}^{(2)}\eta_l^{(2)}\}\big)= \\
=\sum_{a\colon i\to j}(-1)^a\tr\big(-V_{a^*}\theta_jV_a\eta_i-\theta_jV_aV_{a^*}\eta_j+V_{a^*}\theta_j\eta_jV_a+\theta_jV_a\eta_iV_{a^*}\big)
=H_{[\theta,\eta]}(V).
\end{multline*}
Thus we have a Poisson action of the group $G(\alpha)$ given by the Hamiltonians $H_\theta$. From the definition of the moment map, $P(V)(\theta)=H_\theta(V)$, one yields $P(V)=\Big(\sum\limits_{i\in I\atop a\colon i\to j}(-1)^aV_aV_{a^*}\Big)_{j\in I}\in\End(\alpha)_0$.

Now are going to prove~\eqref{IAIB} (not assuming that $\aalpha_\infty=1$: see the end of Section~\ref{secHamKolMn}). Due to Proposition~\ref{propPoisson}, the Poisson brackets on $N_\llambda(\aalpha)$ can be calculated via Poisson brackets of the corresponding invariant functions on $\Rep\big(\Pi^\llambda(Q_\zeta),\aalpha\big)$. In the notations of Section~\ref{secHamKolMn} we have $\{v_i^{(1)},w_j^{(2)}\}_{rl,l's}=\{(v_i)_{rl'},(w_j)_{ls}\}=\{(V_{b_{ir}})_{l'},(V_{b_{js}^*})_l\}=\delta_{ij}\delta_{rs}\delta_{ll'}$, that is
 $\{v_i^{(1)},w_j^{(2)}\}=\delta_{ij}\P$,  $\;\forall\,i,j\in I$. By taking into account $\{V^{(1)}_{p''},V^{(2)}_{p'}\}=0$ and that $V_{p''}V_{p'}=V_{p''p'}$ for any $p',p''\in\mathsf P$ such that $p''p'\in\mathsf P$, we obtain
\begin{multline*}
 \{I_A,I_B\}(V)=
\sum_{i,j\in I\atop k,l\in I}\sum_{p''\in\mathsf P_{ij}\atop p'\in\mathsf P_{kl}} \tr\Big(A_{p''}^{(1)}\{w_j^{(1)}V_{p''}^{(1)}v_i^{(1)},w_l^{(2)}V_{p'}^{(2)}v_k^{(2)}\}B_{p'}^{(2)}\Big)= \\
=\sum_{i,j\in I\atop k,l\in I}\sum_{p''\in\mathsf P_{ij}\atop p'\in\mathsf P_{kl}} \tr\Big(A_{p''}^{(1)}\big(w_j^{(1)}V_{p''}^{(1)}\{v_i^{(1)},w_l^{(2)}\}V_{p'}^{(2)}v_k^{(2)}+ \\
+w_l^{(2)}V_{p'}^{(2)}\{w_j^{(1)},v_k^{(2)}\}V_{p''}^{(1)}v_i^{(1)}\big)B_{p'}^{(2)}\Big)= \\
=\sum_{i,j,k\in I}\sum_{p''\in\mathsf P_{ij}\atop p'\in\mathsf P_{ki}} \tr\Big(A_{p''}w_jV_{p''}V_{p'}v_kB_{p'}\Big)
-\hspace{-6pt}\sum_{i,j,l\in I}\sum_{p''\in\mathsf P_{ij}\atop p'\in\mathsf P_{jl}} \tr\Big(V_{p''}v_iA_{p''}B_{p'}w_lV_{p'}\Big)= \\
=\sum_{j,k\in I\atop p\in\mathsf P_{kj}}\sum_{p'',p'\in\mathsf P\atop p''p'=p} \tr\Big(B_{p'}A_{p''}w_jV_pv_k\Big)
-\hspace{-4pt}\sum_{i,l\in I\atop p\in\mathsf P_{il}}\sum_{p'',p'\in\mathsf P\atop p'p''=p} \tr\Big(A_{p''}B_{p'}w_lV_pv_i\Big).
\end{multline*}
So that, we derive $\{I_A,I_B\}(V)=I_{[A,B]}(V)$.

Finally, let us calculate the Poisson bracket in the formula~\eqref{wiIAFlow}. By using~\eqref{tr2AP} one yields
\begin{align*}
 \{I_A,w_i\}=-\hspace{-2pt}\sum_{j\in I}\sum_{\;\;p\in\mathsf P_{ij}}\tr_1\big(A_p^{(1)}w_j^{(1)}V_p^{(1)}\{v_i^{(1)},w_i^{(2)}\}\big)=
-\hspace{-2pt}\sum_{j\in I}\sum_{\;\;p\in\mathsf P_{ij}}A_pw_jV_p.
\end{align*}

\refstepcounter{section}
\label{appRRep}
\section*{Appendix~\thesection\\ Reflection functor and the varieties $\Rep\big(\Pi^\lambda(Q),\alpha\big)$} 
\addcontentsline{toc}{section}{Appendix~\thesection. Reflection functor and the varieties $\Rep\big(\Pi^\lambda(Q),\alpha\big)$}

Here we describe explicitly how the reflection functor can be represented on the varieties $\Rep\big(\Pi^\lambda(Q),\alpha\big)$. Then we apply this to prove the regularity of $\varrho^{\lambda,\alpha}_k$ without using Proposition~\ref{PropTr} (see Theorem~\ref{ThIsom}). In this appendix we use notations of Section~\ref{secRF} and suppose that $\lambda_k\ne0$ (for some fixed $k\in\Ilf$ and $\lambda\in\CC^I$).

Consider a point $V\in \Rep\big(\Pi^\lambda(Q),\alpha\big)$. Its affine coordinates are matrix entries of the operators $V_a$. They are defined via the standard bases of the spaces $V_j=\CC^{\alpha_j}$, we denote them by $\{e^{(j)}_l\}_{l=1}^{\alpha_j}$. Let us construct a representative $V'\in\Rep\big(\Pi^{r_k\lambda}(Q),s_k\alpha\big)$ isomorphic to the module $F^{\lambda}_k(V)$ by fixing bases in the spaces $V'_j$, $j\in I$. For $j\ne k$ the basis of $V'_j=V_j$ is the same $\{e^{(j)}_l\}_{l=1}^{\alpha_j}$. Note that the vectors $e^\oplus_{al}:=\mu_a(e^{(j)}_l)$, $a\colon j\to k$, $j\in I$, form a basis of the space $V_\oplus=\bigoplus\limits_{a\colon j\to k}V_j$. Its subspace $V'_k=\Ker\pi=\Im\mu'=\Im\mu'\pi'=\Im(1-\mu\pi)$ is spanned by the vectors $e'_{al}:=(1-\mu\pi)(e^\oplus_{al})$. The components of the vectors $e'_{al}$ in the basis $\{e^\oplus_{al}\}$ form a matrix of the rank $p_0:=\dim V'_k=(s_k\alpha)_k$. Its entries are determined from the formula $(1-\mu\pi)(e^\oplus_{a'l'})= \sum\limits_{a,l} (1-\mu\pi)_{al,a'l'} e^\oplus_{al}$. Explicitly,
\begin{multline*}
\mu\pi(e^\oplus_{a'l'})=\frac1{\lambda_k}\sum\limits_{a\in H}\mu_aV_{a^*}\sum\limits_{b\in H}(-1)^bV_b\pi_b(e^\oplus_{a'l'})= \\
=\frac1{\lambda_k}\sum\limits_{a,b\in H}(-1)^b(\mu_aV_{a^*}V_b)(\delta_{ba'}e_{l'})
=\frac1{\lambda_k}(-1)^{a'}\sum\limits_{a,l}\mu_a(e_l)(V_{a^*}V_{a'})_{l,l'}= \\
=\frac1{\lambda_k}(-1)^{a'}\sum\limits_{a,l}(V_{a^*}V_{a'})_{ll'}e^\oplus_{al},
\end{multline*}
(the index $j$ in $e^{(j)}_l$ is omitted). Thus, the entries $(1-\mu\pi)_{al,a'l'}=\delta_{aa'}\delta_{ll'}-\frac1{\lambda_k}(-1)^{a'}(V_{a^*}V_{a'})_{ll'}$ are polynomials of the coordinates $(V_a)_{ll'}$. For a given $V$ one can choose the basis $\{e'_p\}_{p=1}^{p_0}$ of the space $V'_k$ from the set $\{e'_{al}\}$. In particular, $e'_p=\sum\limits_{a,l}n_{al,p}e'_{al}$ for some $n_{al,p}\in\{0,1\}$. This gives us an isomorphism $V'_k\simeq\CC^{p_0}$ that identifies $V'$ with an element of $\Rep\big(\Pi^{r_k\lambda}(Q),s_k\alpha\big)$. The linear idependence of $e'_1,\ldots,e'_{p_0}$ is equivalent to non-vanishing of the corresponding $p_0\times p_0$ minor of the matrix $(1-\mu\pi)$ at the point $V$. Since this minor is a polynomial of $(V_a)_{ll'}$, it does not vanish on a Zariski open set $U\subset\Rep\big(\Pi^\lambda(Q),\alpha\big)$ containing the given point $V$. Hence we can choose the basis $e'_p=\sum\limits_{a,l}n_{al,p}e'_{al}$ for all $V\in U$ (where $n_{al,p}$ does not depend on $V$). The vectors $e'_{al}$ can be (uniquely) decomposed in the basis $e'_p$, i.e. $e'_{al}=\sum\limits_{p=1}^{p_0}t_{p,al}e'_p$, where $t_{p,al}$ are rational functions of $(V_a)_{ll'}$, regular on $U$. Then, we have
\begin{align*}
 &\mu'(e'_p)=e'_p=\sum_{a',l'}n_{a'l',p}e'_{a'l'}=\sum_{a',l'}n_{a'l',p}(1-\mu\pi)(e^\oplus_{a'l'})= \\
&\qquad\qquad\qquad\qquad\qquad\qquad\qquad=\sum_{a,a',l,l'}n_{a'l',p}(1-\mu\pi)_{al,a'l'}e^\oplus_{al}, \\
 &\pi'(e^\oplus_{al})=\mu'\pi'(e^\oplus_{al})=(1-\mu\pi)(e^\oplus_{al})=e'_{al}=\sum_p t_{p,al}e'_p.
\end{align*}
For $a\in H$ we derive
\begin{align*}
 &V'_a(e_l)=-\lambda_k(-1)^a\pi'(e^\oplus_{al})=-\lambda_k(-1)^a\sum_p t_{p,al}e'_p, \\
 &V'_{a^*}(e'_p)=\pi_a\big(\mu'(e'_p)\big)=\sum_{a',l,l'}n_{a'l',p}(1-\mu\pi)_{al,a'l'}e_l.
\end{align*}
So that, for $V\in U$ the element $V'\in\Rep\big(\Pi^{r_k\lambda}(Q),s_k\alpha\big)$ defined above has the coordinates
\begin{align*}
 &(V'_a)_{ll'}=(V_a)_{ll'} &&\text{for $a,a^*\notin H$}, \\
 &(V'_a)_{pl}=-\lambda_k(-1)^a t_{p,al} &&\text{for $a\in H$}, \\
 &(V'_{a^*})_{lp}=\sum_{a',l'}n_{a'l',p}(1-\mu\pi)_{al,a'l'} &&\text{for $a\in H$}.
\end{align*}
Thus, the variety $\Rep\big(\Pi^\lambda(Q),\alpha\big)$ is covered by open sets $U^{(\rho)}$ where we have the regular maps $$\varphi^{(\rho)}\colon U^{(\rho)}\to\Rep\big(\Pi^{r_k\lambda}(Q),s_k\alpha\big),$$ which send $V\in U^{(\rho)}$ to $V'$ constructed above. On the intersections $U^{(\rho)}\cap U^{(\sigma)}$ we have $\varphi^{(\rho)}(V)=g_{\rho\sigma}\varphi^{(\sigma)}(V)$ for some $g_{\rho\sigma}\in G(s_k\alpha)$ independent of $V\in U^{(\rho)}\cap U^{(\sigma)}$. The choice of the maps $\varphi^{(\rho)}$ is not unique, they can be replaced by $\tilde\varphi^{(\rho)}(V)=g_\rho\varphi^{(\rho)}$ for some $g_\rho\in G(s_k\alpha)$ (note that $\varphi^{(\rho)}$ is not injective in general since $\varphi^{(\rho)}(gV)=\varphi^{(\rho)}(V)$ for any $V\in U^{(\rho)}$ and $g\in\GL(\alpha_k)\subset \GL(\alpha)$ such that $gV\in U^{(\rho)}$).

Let $\cl_{\lambda}(\alpha)$ be subset of semi-simple $V\in\Rep\big(\Pi^{\lambda}(Q),\alpha\big)$. Then $N_\lambda(\alpha)=\cl_{\lambda}(\alpha)/G(\alpha)$ is an affine variety and the algebra of regular functions on  this variety is $\CC[N_\lambda(\alpha)]=\CC\big[\Rep\big(\Pi^{\lambda}(Q),\alpha\big)\big]^{G(\alpha)}$ (see Remark~\ref{RemCQ}). If any $V\in\Rep\big(\Pi^{\lambda}(Q),\alpha\big)$ is semi-simple, then $N_\lambda(\alpha)=\Rep\big(\Pi^{\lambda}(Q),\alpha\big)/G(\alpha)$. In particular, it is so if the condition of Theorem~\ref{ThSmooth} is valid, in this case the following proposition is a part of Theorem~\ref{ThIsom}. In general case it is proved in the same way (by applying Lemma~\ref{LemTr} and Proposition~\ref{PropTr}). We formulate it to show an alternative proof.

\begin{Prop} \label{PropIsomApp}
 The reflection functor $F^{\lambda}_k$ induces an isomorphism of affine varieties $\varrho_k^{\lambda,\alpha}\colon N_\lambda(\alpha)\to N_{r_k\lambda}(s_k\alpha)$.
\end{Prop}

\noindent{\bf Proof.} The bijectivity of $\varrho^{\lambda,\alpha}_k$ is argued in Remark~\ref{RemCQIsom}. Its regularity is equivalent to the claim that $f\circ\varrho^{\lambda,\alpha}_k\in\CC\big[N_{\lambda}(\alpha)\big]$ for any $f\in\CC\big[N_{r_k\lambda}(s_k\alpha)\big]$.

Under the isomorphism $\CC[N_{r_k\lambda}(s_k\alpha)]=\CC\big[\Rep\big(\Pi^{r_k\lambda}(Q),s_k\alpha\big)\big]^{G(s_k\alpha)}$ the function $f$ is mapped to $\bar f\in\CC\big[\Rep\big(\Pi^{r_k\lambda}(Q),s_k\alpha\big)\big]^{G(s_k\alpha)}$, such that $\bar f(V)=f([V])$ for any $V\in\cl_{r_k\lambda}(s_k\alpha)$. Define a $G(\alpha)$-invariant function $\bar h\colon\Rep\big(\Pi^{\lambda}(Q),\alpha\big)\to\CC$ by $\bar h(V)=\bar f(V')$ where $V'$ is any element $\Rep\big(\Pi^{r_k\lambda}(Q),s_k\alpha\big)$ isomorphic to $F^{\lambda}_k(V)$ as a $\Pi^{r_k\lambda}(Q)$-module. On each domain $U^{(\rho)}$ we have $\bar h\big|_{U^{(\rho)}}=\bar f\circ\varphi^{(\rho)}$. The regularity of the functions $\bar f$ and $\varphi^{(\rho)}$ for any $U^{(\rho)}$ implies $\bar h\in\CC\big[\Rep\big(\Pi^{\lambda}(Q),\alpha\big)\big]^{G(\alpha)}$. We have $\bar h(V)=\bar f(V')=f([V'])=f\big(\varrho^{\lambda,\alpha}_k([V])\big)$ for any $V\in\cl_\lambda(\alpha)$, hence the function $\bar h$ is mapped to $f\circ\varrho^{\lambda,\alpha}_k$ under the isomorphism $\CC\big[\Rep\big(\Pi^{\lambda}(Q),\alpha\big)\big]^{G(\alpha)}=\CC\big[N_{\lambda}(\alpha)\big]$. Hence $f\circ\varrho^{\lambda,\alpha}_k\in\CC\big[N_{\lambda}(\alpha)\big]$. \qed


\end{document}